%% file: template.tex
\documentclass[journal]{vgtc}                

\ifpdf
  \pdfoutput=1\relax                   
  \usepackage{graphicx}                
  \DeclareGraphicsExtensions{.pdf,.png,.jpg,.jpeg} 
\else
  \usepackage{graphicx}                
  \DeclareGraphicsExtensions{.eps}     
\fi%

\graphicspath{{figures/}{pictures/}{images/}{./}} 
\usepackage{color}
\usepackage{array}
\usepackage{multirow}
\usepackage{rotating}
\usepackage[table,xcdraw]{xcolor}
\usepackage{microtype}                 
\usepackage{makecell}
\PassOptionsToPackage{warn}{textcomp}  
\usepackage{textcomp}                  
\usepackage{mathptmx}                  
\usepackage{times}                     
\usepackage{cite}                      
\usepackage{tabu}                      
\usepackage{booktabs}                  

\onlineid{0}

\title{FFEINR: Flow Feature-Enhanced Implicit Neural Representation for Spatio-temporal Super-Resolution}

\author{Chenyue Jiao, Chongke Bi and Lu Yang}
\authorfooter{
\item
 Chenyue Jiao and Chongke Bi are with Tianjin University. E-mail: cherryjiao98@163.com, bichongke@tju.edu.cn.
\item
 Lu Yang is with Tianjin University of Technology. E-mail: yanglu8206@163.com.
}

\input{body/0-Abstract}

\begin{document}

\input{body/1-Introduction}
\input{body/2-RelatedWork}
\input{body/3-FINR}
\input{body/4-FEINR}

\input{body/5-ResultsAndDiscusssion}
\input{body/6-Conclusions}
\input{body/7-Acknowledgments}

\bibliographystyle{abbrv-doi}

\bibliography{template}
\input{body/8-Appendices}

\end{document}

%% file: body/0-Abstract.tex
\abstract{
Large-scale numerical simulations are capable of generating data up to terabytes or even petabytes.
As a promising method of data reduction, super-resolution (SR) has been widely studied in the scientific visualization community. 
However, most of them are based on deep convolutional neural networks (CNNs) or generative adversarial networks (GANs) 
and the scale factor needs to be determined before constructing the network.
As a result, a single training session only supports a fixed factor and has poor generalization ability. 
To address these problems, this paper proposes a Feature-Enhanced Implicit Neural Representation (FFEINR) for spatio-temporal super-resolution of flow field data. 
It can take full advantage of the implicit neural representation in terms of model structure and sampling resolution. 
The neural representation is based on a fully connected network with periodic activation functions, 
which enables us to obtain lightweight models. 
The learned continuous representation can decode the low-resolution flow field input data  
to arbitrary spatial and temporal resolutions, allowing for flexible upsampling. 
The training process of FFEINR is facilitated 
by introducing feature enhancements for the input layer, 
which complements the contextual information of the flow field.
To demonstrate the effectiveness of the proposed method, 
a series of experiments are conducted on different datasets by setting different hyperparameters. 
The results show that FFEINR achieves significantly better results than the trilinear interpolation method.
}

\keywords{Implicit neural representation, spatio-temporal super-resolution, flow field data.}

%% file: body/1-Introduction.tex

\firstsection{Introduction}
\maketitle 

The study of flow field evolution laws is of great importance in the fields of aerospace and atmospheric physics. 
With the rapid development of high performance computing technology, it has brought great help to experts in these fields.
According to the statistics of November 2022\footnote{ https://www.top500.org/lists/top500/2022/11/}, the theoretical peak performance of supercomputers has reached 1685.65 PFlop/s. 
The powerful computing power supports domain experts to execute simulations in more fine-grained grids, with higher temporal resolution and at faster iteration speed. 
The size of the simulation results can reach terabytes or even petabytes. 
{However, such large-scale data is difficult to transfer and store due to the limitations of I/O speed and storage resource.} 
It is not feasible to transfer the data from the high-performance computing device to the workstation for research at the original resolution. 
Researchers have to reduce data by temporal and spatial downsampling. 
\textcolor{black}{Nevertheless, sparse and discrete sampling results in a significant data loss, 
which indirectly leads to loss of potential research value of the data, 
missed opportunities to discover physical laws 
and even hinders the domain experts from making accurate judgments about design and law.
This poses a {challenge for the visualization and data analysis in the scientific workflow.}}

In order to achieve {data reduction} more efficiently, researchers have proposed a number of solutions, including: 1. truly lossless compression; 2. near lossless compression; 3. lossy compression; 4. mesh reduction and 5. derived representations~\cite{li_data_2018}. 
The methods mentioned above achieve excellent data reduction results. 
With the rapid development of ML4Science, DL4Science~\cite{wang_dl4scivis_2022,karniadakis_physics-informed_2021,pandey_perspective_2020}, 
an increasing amount of research focuses on leveraging the powerful data fitting capabilities of neural networks.
Among them, super-resolution has been widely used for reconstructing from low-resolution images, videos, and volume data. 
By definition, super-resolution achieves upsampling from a low-resolution source data to a high-resolution target data. 
The trained model can be used for data upscaling, so that we only need to store the low-resolution data and can readily generate higher resolution data by the model. 
\textcolor{black}{As a result, {super-resolution} can be considered as a {data reduction} method.}
According to different sampling dimensions, the super-resolution task is subdivided into spatial super-resolution, temporal super-resolution, and spatio-temporal super-resolution. 
\textcolor{black}{In general, spatio-temporal super-resolution allows for higher compression rates, due to the ability to interpolate in both temporal and spatial dimensions.}
However, from the perspective of network architecture, 
most of the {existing super-resolution networks \textcolor{black}{for data reduction}} are based on convolutional neural networks and generative adversarial networks, 
which {have two common problems}: 
Firstly, \textcolor{black}{some of }these networks are {complex} and need a {long training time}. 
In order to achieve better results, these works often use very complex and deep networks. 
For example, ESRGAN~\cite{wang_esrgan_2019} contains a generator and a discriminator.
The generator is composed of 23 Residual-in-Residual Blocks (RRDBs) based on residual connections, and each RRDB contains 15 convolutional layers, which greatly increases the training time. 
In addition, the network needs a two-stage training, performing 20,000 iterations for pre-training and 10,000 iterations for training, and the total time is more than ten hours~\cite{jiao_esrgan-based_2023}. 
Secondly, these networks have {poor generalization ability}. 
Since the scale factor, which is a hyperparameter of the network, needs to be set before training, 
the model obtained by training more than ten hours at a time only supports a fixed scale factor.
It is very time expensive to perform multiple training in order to obtain data with different resolutions. 
These problems reduce the efficiency of applying super-resolution to data reduction.

In recent years, {implicit representation} has achieved good results in solving scene representation~\cite{mildenhall_nerf_nodate}, data generation~\cite{lu_compressive_2021}, and physical partial differential equation solving~\cite{raissi_physics-informed_2019} . 
The basic implication is to exploit multilayer perceptron (MLP) networks to represent the data domain or problem domain. 
This model is structurally simple compared to the increasingly deeper convolutional neural networks. 
In addition, continuous implicit representation can generate values of interest at arbitrary locations within the data domain. 
Combining the implicit representation with the super-resolution task can well extend the factor-fixed interpolation space to the {space of arbitrary resolution}. 
\textcolor{black}{In order to improve the {generalization of super-resolution models} and achieve more effective data reduction,}
this paper proposes a continuous implicit representation scheme for spatio-temporal super-resolution in flow fields, 
which \textcolor{black}{combines the two main advantages of traditional convolutional networks in data feature extraction and implicit neural representation in infinite resolution}. 
This implicit representation can be regarded as a representation function where spatio-temporal coordinates are inputs and the corresponding physical field data are outputs.
In this way, the continuous flow field can be obtained by inputting continuous query coordinates, which enables more flexible scale factors. 
In addition, to achieve temporal super-resolution task, inputting low-resolution flow field data at both ends expands the perceptual field of the implicit network, allowing it to obtain data-context-dependent features in addition to data-agnostic spatio-temporal coordinates. 
\textcolor{black}{Through this implicit representation, we implement the spatio-temporal super-resolution problem of flow field data and ultimately provide a novel solution for {data reduction}.}

To summarize, our contributions include:
\begin{itemize}
\item We propose a \textbf{F}low \textbf{F}eature-\textbf{E}nhanced \textbf{I}mplicit \textbf{N}eural \textbf{R}epresentation (FFEINR) to solve the spatio-temporal super-resolution for \textcolor{black}{flow fields data reduction}. 
\item We use a fully connected network based on periodic activation functions for the super-resolution task, and the model can support \textcolor{black}{multiple} scale factors, 
\textcolor{black}{which has advantages in generalization ability}.
\item A series of experiments are conducted on different datasets by setting different hyperparameters to verify that FFEINR outperforms the baseline method.
\end{itemize}

%% file: body/2-RelatedWork.tex
\section{Related work}\label{sec:relatedwork}

\subsection{Implict Representation}\label{ssec:inr}
Based on MLP, implicit representation can encode data with the help of the powerful representational capabilities of neural networks. 
The data of interest, such as images, color and transparency values, physical field data, etc., are output by inputting temporal or/and spatial coordinates. 
This network has been widely used in the fields of computer vision, computer graphics, visualization,  computational fluid dynamics, and so on.

In the computer vision community, Chen et al.~\cite{chen_nerv_nodate} proposed a neural representation of video that encodes the video in a neural network.
Given a temporal coordinates or the frame index, the corresponding frame can be obtained. 
Using this representation of video helps to simplify downstream tasks, such as video compression, video denoising, etc. 
However, the frame index is input in this model to obtain both temporal and spatial information of the video. 
Therefore, there are a large number of redundant parameters in the network, which may cause the model to be very large. 
For this reason, Li et al.~\cite{avidan_e-nerv_2022} proposed E-NeRV, a network that achieves faster training with fewer parameters by decomposing the image-wise implicit neural representation into separate spatial and temporal contexts, while preserving the network's representational power. 
Most implicit representations only take fixed temporal or spatial coordinates as inputs. 
Reconstructing video frames from such content-agnostic information largely limits the generalization ability of video interpolation. 
This problem was optimized by Chen et al.~\cite{chen_hnerv_2023}. 
They proposed a content-adaptive embeddings related to video content and optimized the design of the network architecture, and experimentally demonstrated that this approach has competitive advantages in terms of reconstruction quality, convergence speed, and generalization ability.

In the field of computer graphics, neural radiation fields (NeRF)~\cite{mildenhall_nerf_nodate,nguyen-phuoc_snerf_2022,yu_pixelnerf_2021,chen_mvsnerf_2021,barron_mip-nerf_nodate} can be regarded as spatial implicit representations. 
NeRF~\cite{mildenhall_nerf_nodate} takes as input a single continuous 5D coordinate (spatial location ($x$, $y$, $z$) and viewing direction ($\theta$, $\phi$)) with the volume density and view-dependent emitted radiance at that spatial location as the output.
These outputs are used to generate the projected image at the given viewpoint by volume rendering, and supervised by sparse input view. 
The optimized radiance field can render realistic new views of scenes with complex geometry and appearance, outperforming the prior work on neural rendering and view synthesis.
To enhance the input features, \cite{yu_pixelnerf_2021,chen_mvsnerf_2021} combined CNN and MLP, using the features extracted from the convolutional layer as input of MLP networks, 
which can obtain higher rendering quality using fewer input views 
and significantly reduce the training time and improve the efficiency of NeRF.

In the direction of computational fluid mechanics, physics-informed neural networks (PINNs)~\cite{raissi_physics-informed_2019,raissi_hidden_2020,jin_nsfnets_2021,cai_physics-informed_2021,takamoto_pdebench_2022}  use the MLP architecture to form a new family of data-efficient spatio-temporal function approximators.
The network takes as input temporal and spatial coordinates or even simulation parameters and as output physical quantities.
Given laws of physics described by general nonlinear partial differential equations were incorporated into the design of the loss function, constraining the network to optimize in a physically interpretable direction.

In the visualization community, Lu et al.~\cite{lu_compressive_2021} used an implicit representation of volumetric scalar fields as a compression method. It enables a compressed representation of scalar fields when the number of weights of the neural network is smaller than the spatial scale of the input, and supports random access to field data. 
Inspired by the above-mentioned work of NeRF and PINN, Chu et al~\cite{chu_physics_2022} proposed a continuous space-time scene representation for solving smoke reconstruction. 
This network leveraged both the governing equations (Navier-stokes equtions) and sparse flow field video frame images as supervision.
This method used a hybrid architecture that separates static and dynamic content, and reconstructed dynamic fluid phenomena in an end-to-end optimization, achieving high-quality results with relaxed constraints and strong flexibility.
Han et al.~\cite{han_coordnet_2022} modeled different tasks in scientific visualization as a black-box function with inputting coordinates and outputting values. They proposed a unified framework based on implicit representation for solving data generation (i.e., temporal super-resolution and spatial super-resolution) and visualization generation (i.e., view synthesis and ambient occlusion prediction) tasks, achieving good quantitative and qualitative results.

\subsection{\textcolor{black}{Data Reduction} and Super-Resolution}\label{ssec:sr}
\textcolor{black}{The massive flow field data generated by large-scale data simulation calculations is difficult to directly visualize and analyze, which has prompted researchers to study various data reduction methods.
As mentioned earlier, it can be roughly divided into five categories: 
1. truly lossless compression; 2. near lossless compression; 3. lossy compression; 4. mesh reduction and 5. derived representations~\cite{li_data_2018}. 
Compression methods target the characteristics of the data itself and reduce data by encoding, transformation, truncation, quantization and so on. 
The object of mesh simplification is the mesh where data generate. 
By simplifying at the basic mesh, the scale of mesh is reduced, thereby reducing the data size. 
Derived representation discards the original data completely and designs alternative representations that match the data characteristics for different specific applications. 
The methods mentioned above achieve excellent data reduction results. }

\textcolor{black}{With the rapid development of deep learning in the visualization community, 
super-resolution has attracted more and more researchers' attention as a potential data reduction method. 
Super-resolution can leverage neural networks to upsample low-resolution inputs to high-resolution outputs, 
and thus we can only need to store low resolution data and achieve the goal of data reduction.} 
Specifically, it can be divided into three categories: spatial super-resolution~\cite{bashir_comprehensive_2021,han_ssr-tvd_2020,guo_ssr-vfd_2020}, 
temporal super-resolution~\cite{liu_video_2022,han_tsr-tvd_2020,han_tsr-vfd_2022}, 
and spatio-temporal super-resolution~\cite{goos_increasing_2002,xiang_zooming_2020,chen_videoinr_2022,han_stnet_2022}.

Spatial super-resolution is commonly used for upscaling low-resolution images, videos, and volume data. 
Temporal super-resolution is commonly applied to interpolation between video frames and intermediate time slices of time-varying data. 
Spatio-temporal super-resolution is a combination of the two, which can upscale low-resolution and low frame rate videos, and sparsely sampled time-varying data to higher spatio-temporal resolutions. 
It can be used in application scenarios such as medical image processing, remote sensing, and data compression. 
In this paper, we mainly focus on spatio-temporal super-resolution.
The concept of video spatio-temporal super-resolution was first proposed by~\cite{goos_increasing_2002}. 
They constructed a high spatio-temporal resolution video sequence by combining information from multiple low resolution video sequences from the same dynamic scene.
Xiang et al.~\cite{xiang_zooming_2020} proposed a single-stage spatio-temporal video super-resolution framework based on the idea of temporal feature interpolation and global information fusion, achieving faster inference speed. 
Chen et al.~\cite{chen_videoinr_2022} introduced implicit representation into super-resolution tasks, using continuous representation to encode the video space. 
The learned neural representation can be decoded into videos of any spatial resolution and frame rate, achieving competitive performance. 
In the scientific visualization community, Han et al.~\cite{han_stnet_2022} proposed a spatio-temporal super-resolution framework based on the generative adversarial networks. 
The generator realized the interpolation from the low-resolution volume at both ends to the spatio-temporal super-resolution volume at the middle and both ends, and designed a spatio-temporal discriminator for adversarial training.

%% file: body/3-FINR.tex
\section{Flow implict neural representation}\label{sec:FINR}

Due to the shortcomings of deep convolutional super-resolution networks in terms of training complexity and generalization ability, 
we attempt to find a lightweight model to solve the spatio-temporal super-resolution problem. 
Inspired by~\cite{han_coordnet_2022}, we focus on the coordinate-based implicit neural representation network. 
Implicit representation is a powerful method for continuous data representation. 
It takes arbitrary spatio-temporal indices as inputs and outputs the corresponding  values of interest. 
The model can be defined as:
\begin{equation}
D = F(x,t)
\end{equation}
where $D$ denotes the generated data, 
$x=\{(x_0,...,x_i)|x\in{N}\}$ represents the input vector in spatial dimension, 
$t$ represents the input in temporal dimension, 
and $F$ represents the implicit representation, which is generally based on MLP. 
In NeRF, $D=(c,\sigma)$ and is used by volume rendering algorithm and to generate images.
In PINN, $D$ can be the physical field data such as velocity and pressure, 
and is used to calculate the derivatives of $D$ with respect to $x$ and $t$, respectively. 
Finally, it is used in the loss function composed of the deformed partial differential equation to constrain the physical interpretability of the network. 
In the implicit representation of the flow field, 
$D$ is also physical field data, which is similar to the outputs of PINN, 
but we do not consider adding physical constraints to the model at the moment 
because it may drastically increase the difficulty of model training~\cite{wang_when_2022}.

Our goal is to train a continuous implicit representation to implement the spatiotemporal super-resolution of flow fields. 
We consider it as a representation function with inputting spatio-temporal indices and outputting corresponding flow field data.
Since the input coordinates are continuous, we can obtain flow field data that is theoretically continuous, 
and such data can be considered as infinite resolution compared to low-resolution data. 
Therefore, this method can overcome the problem of fixed resolution in the traditional convolutional super-resolution networks. 
In addition, the implicit representation often uses periodic activation functions, 
which can efficiently learn high-frequency data features with a simple network structure, 
thus compensating for the training difficulties of convolutional super-resolution networks.

%% file: body/4-FEINR.tex
\section{Feature-enhanced implict representation}\label{sec:FEINR}

\begin{figure*}[ht]
    \centering
    \setlength{\abovecaptionskip}{0.5cm}
    \includegraphics[width=1.95\columnwidth]{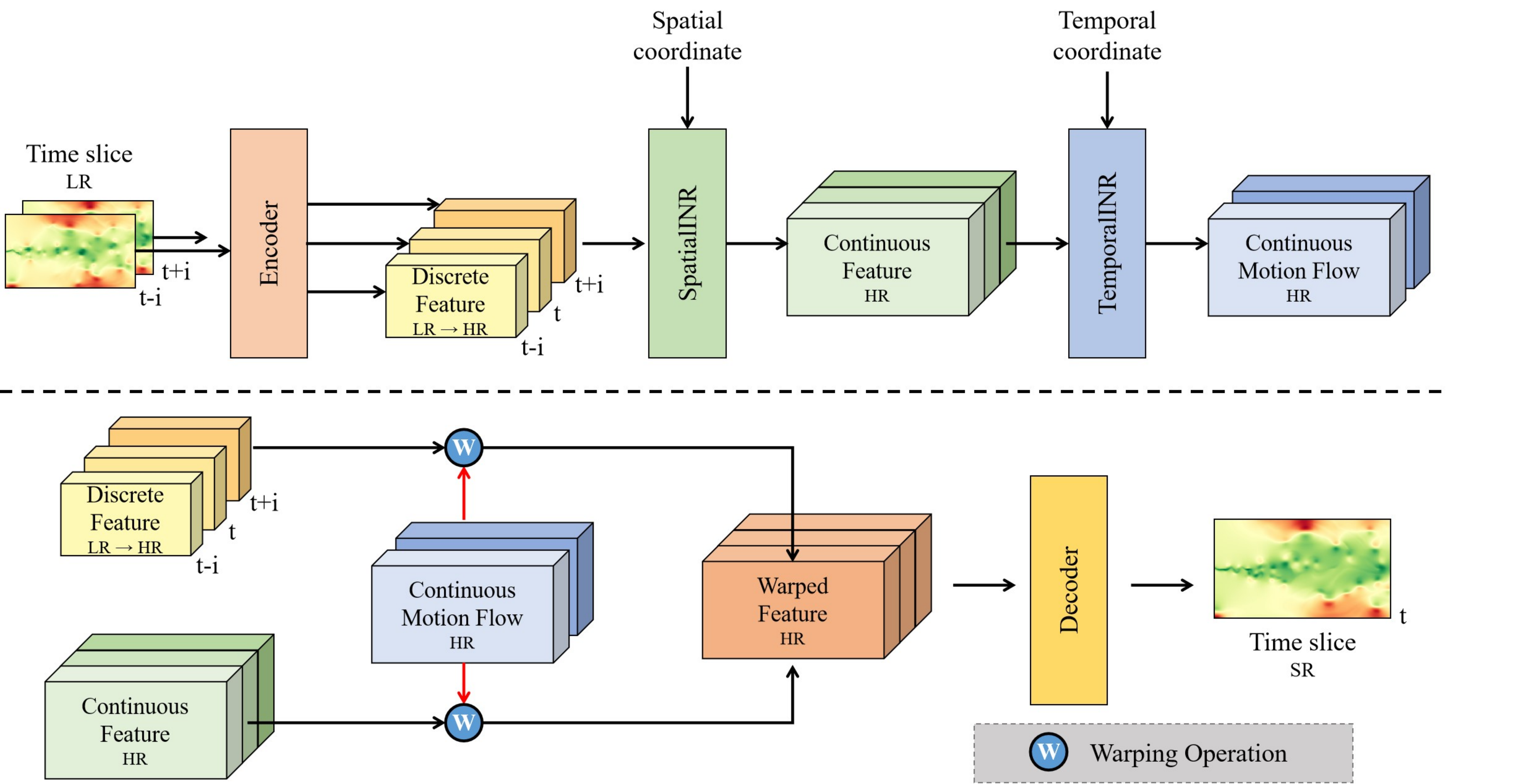}
    \caption{\textcolor{black}{Feature-enhanced implicit neural representation.} Two input time slices are encoded as discrete feature maps, which are then decoded by SpatialINR to obtain continuous spatial feature maps that provide spatial context information. These continuous feature maps are then decoded as continuous motion flow maps by TemporalINR to learn the changing patterns of flow fields in the time dimension. The resulting flow map is used to warp the spatial features, and the warped features are finally decoded into flow field data by the decoder. The entire framework achieves the task of inputting arbitrary coordinates and outputting the corresponding physical values. Since higher resolution coordinates can be used as inputs, super-resolution flow field data can be obtained.}
    \label{fig:model}
\end{figure*}

The model used in this paper to solve the spatio-temporal super-resolution for flow data is based on a framework for solving the video spatio-temporal super-resolution task proposed in~\cite{chen_videoinr_2022}. 
This network represents the video frame space as a decoupled spatial implicit representation and temporal implicit representation 
to achieve interframe interpolation of low frame rate video and upsampling of low-resolution video. 
We make some adjustments and optimizations to this network for the flow field spatio-temporal super-resolution task.

As shown in Fig.~\ref{fig:model}, the entire framework comprises four modules: 
input encoder, spatial implicit neural representation (SpatialINR), temporal implicit neural representation (TemporalINR), and decoder. 
Taking the two-dimensional data spatio-temporal super-resolution task as an example, 
we first obtain low-resolution time slices at both ends of time, 
which can be encoded by an input encoder~\cite{xiang_zooming_2020} . 
This encoder incorporates feature interpolation and ConvLSTM to obtain features in both local and global contexts. 
This process could be expressed as
\begin{equation}
    f_i = E(I^L)
\end{equation}
where $E$ is the encoder, $I^L$ is two consecutive input time slices, $f_i$ is encoded features.
Next, the extracted discrete features are fed into SpatialINR as contextual complementary information to help represent the spatial field. 
After decoding arbitrary query coordinates in space by SpatialINR, 
a spatially continuous feature embedding $f_s$ is obtained:
\begin{equation}
    f_s = F_s(x,f_i)
\end{equation}
where $F_s$ and $f_s$ represent SpatialINR and the spatial feature, respectively.
To learn temporally continuous features, we construct TemporalINR, which extends the feature domain from two-dimensional space to three-dimensional space and time by including the temporal coordinates of the desired query as input.  
We consider that the task of fusing all the features and generating the target results directly using only a single network may be difficult to achieve, 
as it requires learning contextual information in both spatio-temporal dimensions. 
Therefore the output of TemporalINR is designed as a motion flow:
\begin{equation}
    m_t = F_t(t,f_s)
\end{equation}
where $F_t$ and $m_t$ represent TemporalINR and the motion flow, respectively.
With this continuous motion flow field, we can obtain the pattern of flow field variation between time slices in the time dimension. This variation pattern is used to warp the feature embedding in the spatial dimension:
\begin{equation}
    x^*=x+m_t
\end{equation}
\begin{equation}
    f_{st}=F_s(x^*,f_i)
\end{equation}
The warped features are decoded by the decoder $D$, enabling us to obtain physical values in arbitrary spatio-temporal coordinates, i.e., higher spatio-temporal resolution time-slice data $O^H$: 
\begin{equation}
    O^H=D(f_{st})
\end{equation}
This entire process could be expressed as
\begin{equation}
    O^H=F_{st}(x,t,I^L)
\end{equation}
As mentioned before, the input to each implicit network includes feature embeddings in addition to coordinates in the time/space dimension. 
This increases the perceptual field of the network and enables the neural representation to learn features faster in the context of the problem domain.
We call this feature-enhanced implicit neural representation.

There are two key designs aspects of the framework that help us achieve better experimental results. 
Firstly, it is a spatio-temporal decoupled architecture. As mentioned in~\cite{fukami_machine-learning-based_2021}, it is very challenging to achieve spatio-temporal super-resolution by only one model. 
The use of spatio-temporal decoupling framework can alleviate the pressure on one model to achieve all tasks, reduce the difficulty of model training, and improve the training results.
Secondly, it is a feature-enhanced architecture. 
As the temporal super-resolution task can be understood as interpolation between frames, 
which requires data from both ends of the time dimension.
Such a task requires the model to understand the features of the input data and generate data at the intermediate time steps. 
While from the perspective of implicit representation, according to~\cite{chen_hnerv_2023,yu_pixelnerf_2021,chen_mvsnerf_2021}, supplementing content-related context in addition to spatio-temporal coordinates in the input layer of implicit representation helps to improve the representational power.

Compared with~\cite{chen_videoinr_2022}, we remove the two-stage training process, 
i.e., training the same network sequentially setting fixed and arbitrary scale factors. 
Although the two-stage training may theoretically allow the network to learn features at different scales, 
improving the ability of one model to support multiple factors of super-resolution. 
However, in our experiments, the attention of the model is distracted after training for an arbitrary super-resolution stage, which instead reduces the effectiveness of the model. 
Therefore, we improve the problem. In Section~\ref{ssec:results}, 
we will experimentally demonstrate that arbitrary resolution upsampling can also be supported by training only at fixed factors.

%% file: body/5-ResultsAndDiscusssion.tex
\section{Results and Discussion}\label{sec:resultsanddiscussion}

\subsection{Experiments Setup}\label{ssec:setup4experiments}

\subsubsection{Datasets}\label{sssec:datasets}
The flow field dataset comes from an open source visualization data website\footnote{https://cgl.ethz.ch/research/visualization/data.php}, as shown in Table~\ref{tab:datasets}. 
The simulations were all done using the Gerris flow solver~\cite{gerrisflowsolver}. 
Cylinder~\cite{Guenther17} was from the simulation of viscous 2D flow around a cylinder. 
Over the course of the simulation, the characteristic von-Karman vortex street is formed. 
HeatedCylinder~\cite{Guenther17} was from the simulation of a 2D flow generated by a heated cylinder. 
PipeCylinder~\cite{BaezaRojo19SciVisa} was from the simulation of a viscous 2D flow around two cylinders. 
Initially, a vortex street forms behind the first obstacle, 
which then flows around two corners. 
Behind each corner, a standing vortex forms. 
The latter one blocks half of the flow to the second obstacle, creating a one-sided vortex street. 

\begin{table}[tb]
\centering
\setlength{\abovecaptionskip}{0.5cm}
\setlength{\belowcaptionskip}{-0.8cm}
\caption{Datasets}
\label{tab:datasets}
\setlength\tabcolsep{3pt} 
\setlength{\abovecaptionskip}{0.cm}
\setlength{\belowcaptionskip}{-0.cm}
\begin{tabular}{c|c|c}
\cline{1-3}
\textbf{Dataset} &
  \textbf{\begin{tabular}[c]{@{}c@{}}Resolution \\ $x$, $y$, $t$\end{tabular}} &
  \textbf{Physical quantity}
   \\ \cline{1-3}
\makecell[c]{Cylinder}& 640, 80,1501   & $u_x$, $u_y$ \\ \cline{1-3}
\makecell[c]{HeatedCylinder}& 150, 450, 2001 & $u_x$, $u_y$ \\ \cline{1-3}
\makecell[c]{PipeCylinder}& 450, 150, 1501 & $u_x$, $u_y$ \\ \cline{1-3}
\end{tabular}%
\end{table}

\subsubsection{Metrics}\label{sssec:metrics}
From a data perspective, we use the Root Mean Squared Error (RMSE) \textcolor{black}{and peak signal-to-noise ratio (PSNR)} to measure the model's overall performance. From a visualization perspective, we evaluate the visualization results of our model (FFEINR), baseline (Trilinear), and ground truth (GT) by \textcolor{black}{structural similarity index (SSIM) and visual results}.

The PSNR is defined as follows:
\begin{equation}
    PSNR = 10 \cdot \log_{10}\left(\frac{MAX^2_I}{MSE}\right),
\end{equation}
where $MAX_I$ is \textcolor{black}{the maximum value of the data.}


SSIM is calculated as:
\begin{equation}
    SSIM = \frac{(2\mu_S\mu_H+c_1)(2\sigma_{S,H}+c_2)}{(\mu^2_S+\mu^2_H+c_1)(\sigma^2_S+\sigma^2_H+c_2)},
\end{equation}
where $\mu_S$ and $\mu_H$ are the average values of $I^S$ and $I^H$,
$\sigma_{S,H}$ is the covariance between $I^S$ and $I^H$, 
$\sigma^2_S$ and $\sigma^2_H$ are the variance of $I^S$ and $I^H$, 
and $c_1$ and $c_2$ are constants for avoiding instability.

\subsubsection{Implementation details}\label{ssec:impledetails}
Following the advice of ~\cite{chen_videoinr_2022}, the encoder is based on the structure of ~\cite{xiang_zooming_2020}, which integrates feature interpolation and ConvLSTM module.
The network architecture of SpatialINR, TemporalINR and decoder is based on SIREN~\cite{sitzmann_implicit_2020}, which is a fully connected neural network with periodic activation function and requires specific initialization. 
Because it can fit data containing high-frequency changes better, this activation function is widely used in implicit representation~\cite{han_coordnet_2022,chu_physics_2022}. 
In this paper, we retain the original SIREN structure instead of using SIREN with residual connections like~\cite{han_coordnet_2022}. 
Residual connections are often used in traditional super-resolution tasks based on convolution, 
but our goal is to find a simple and lightweight model to achieve super-resolution tasks, 
rather than pursuing deeper network. 
We use Adam as the optimizer, with ${\beta}_ 1=0.9$, ${\beta}_ 2=0.99$. 
For the loss function, we use the Charbonnier loss function for optimization. 
The models involved in the experiment are all trained on a single NVIDIA Tesla T4 16G GPU. 
In the following experiments, unless otherwise specified, 
we use the Cylinder dataset with batch size and patch size both being 16 $\times$ 16. 
Unlike~\cite{chen_videoinr_2022}, we do not use a two-stage training scheme. 
As mentioned above, this scheme first samples the data at a fixed spatio-temporal super-resolution scale and performs training, followed by upsampling training of the data at any scale within a certain range. 
In our experiment, using a two-stage training method will reduce the effectiveness of the model. 
For flow field data, the pursuit of data accuracy is essential, so we do not use this method. 
By contrast, we perform fixed downsampling on the flow field data and set a fixed scale factor for each training. 
Although the scale factor during the training phase is fixed, 
we test and discussed in Section~\ref{ssec:results} whether the training results of this mode support upsampling at any spatio-temporal resolution, 
proving that better results than baseline method can be obtained without the use of two-stage training mode. 
In the one-stage training, we train for 7500 iterations.
We use the trilinear interpolation (Trilinear) method as the baseline.

\subsection{Results}\label{ssec:results}

\begin{figure}[t]
\setlength{\abovecaptionskip}{0.5cm}
\setlength{\belowcaptionskip}{-0.5cm}
\centering
\begin{tabular}{ccc}
\centering
\includegraphics[width=0.28\columnwidth]{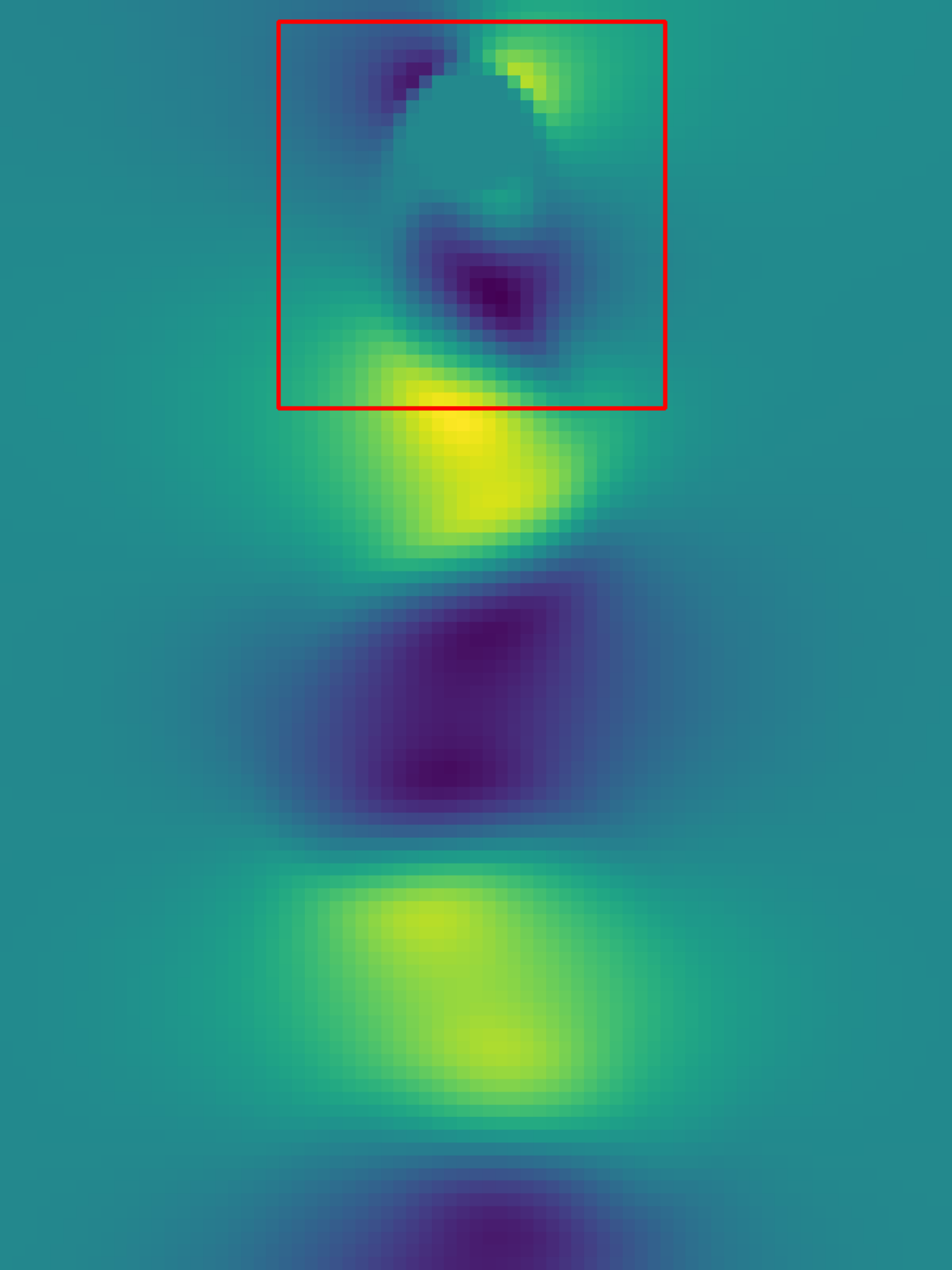} &     
\includegraphics[width=0.28\columnwidth]{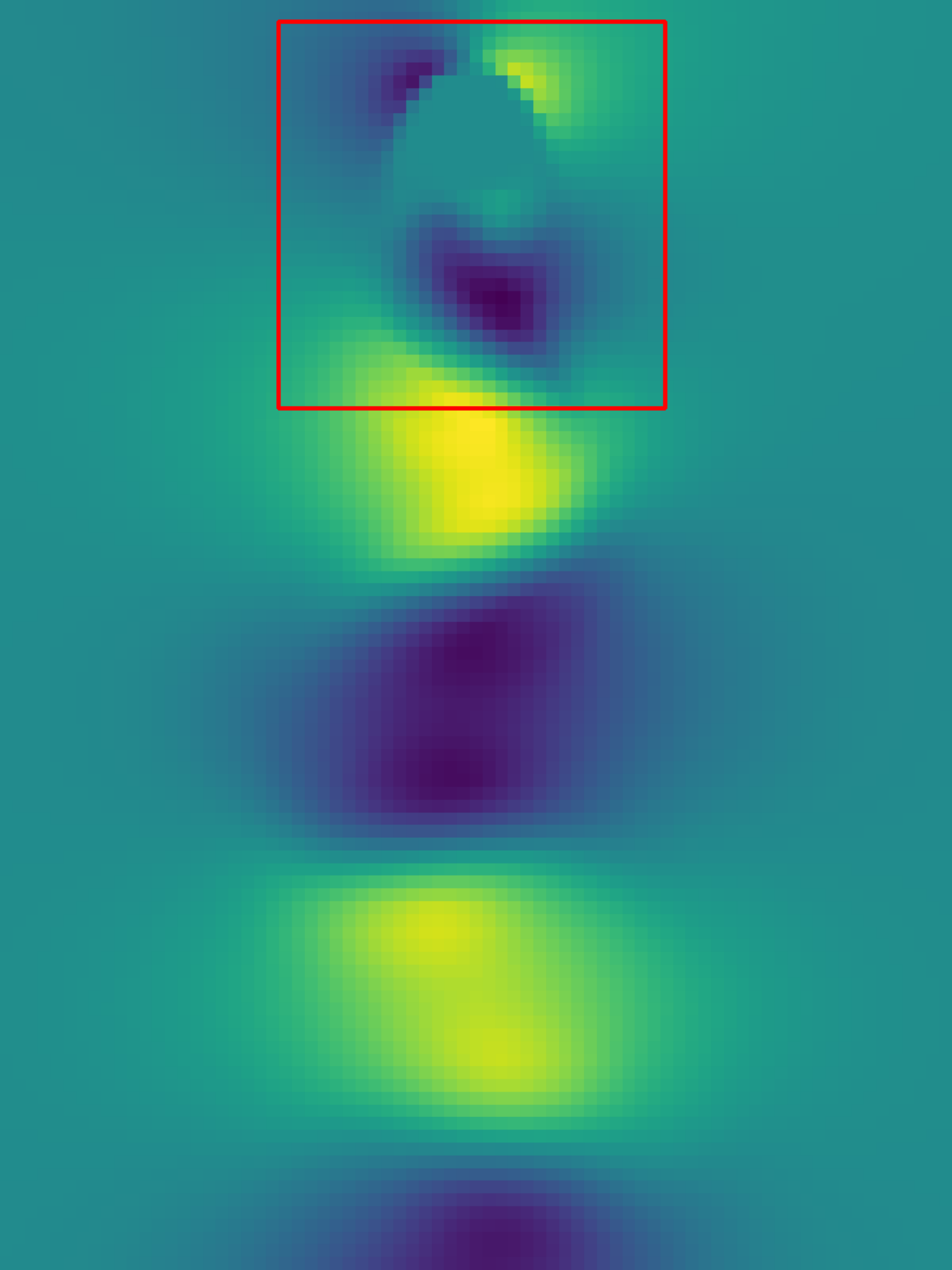} &     
\includegraphics[width=0.28\columnwidth]{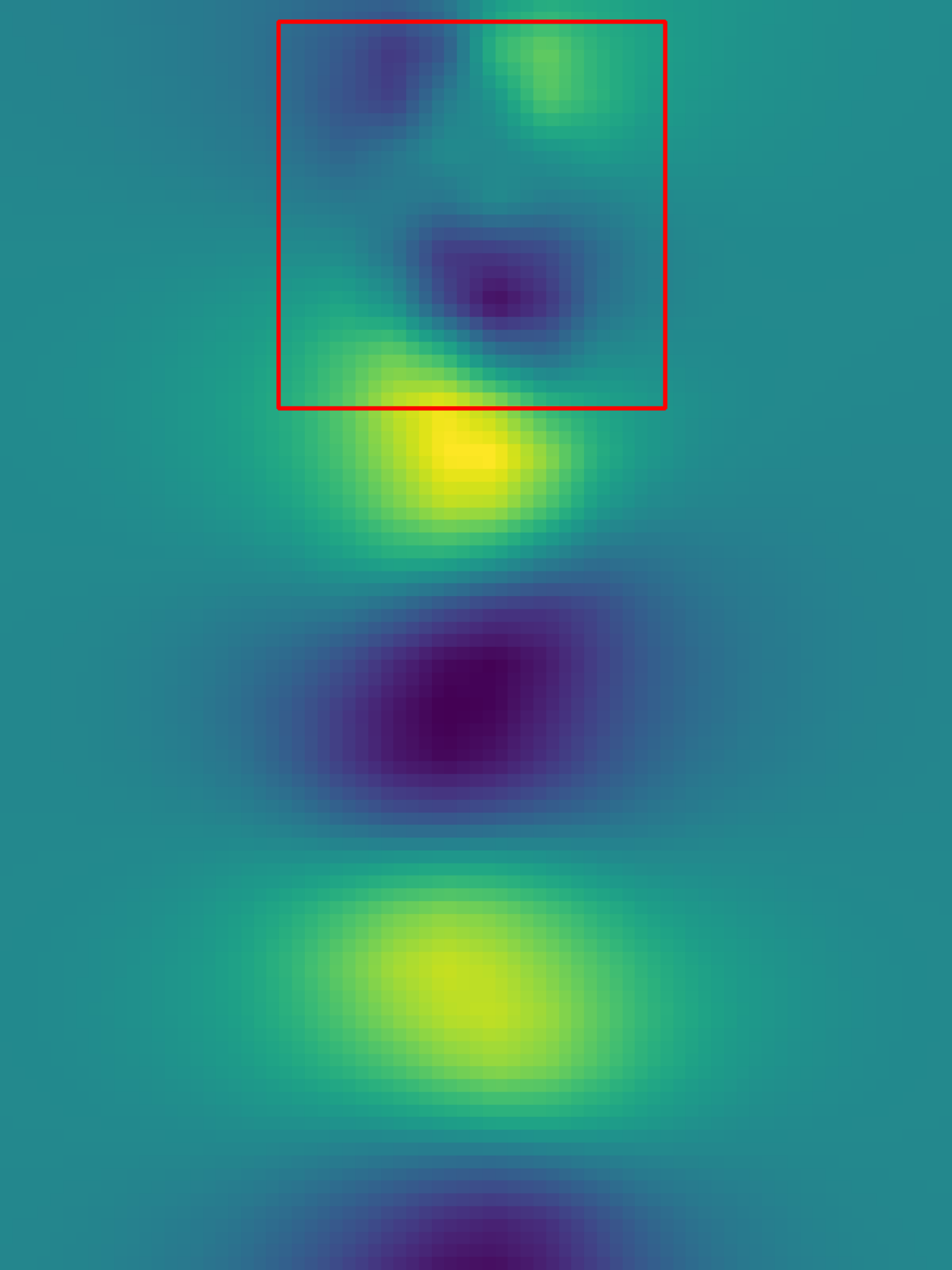} 
\\
\textcolor{black}{(1) Ground Truth} &  \textcolor{black}{(2) FFEINR} & \textcolor{black}{(3) Trilinear}
\\
\includegraphics[width=0.38\columnwidth, angle=-90]{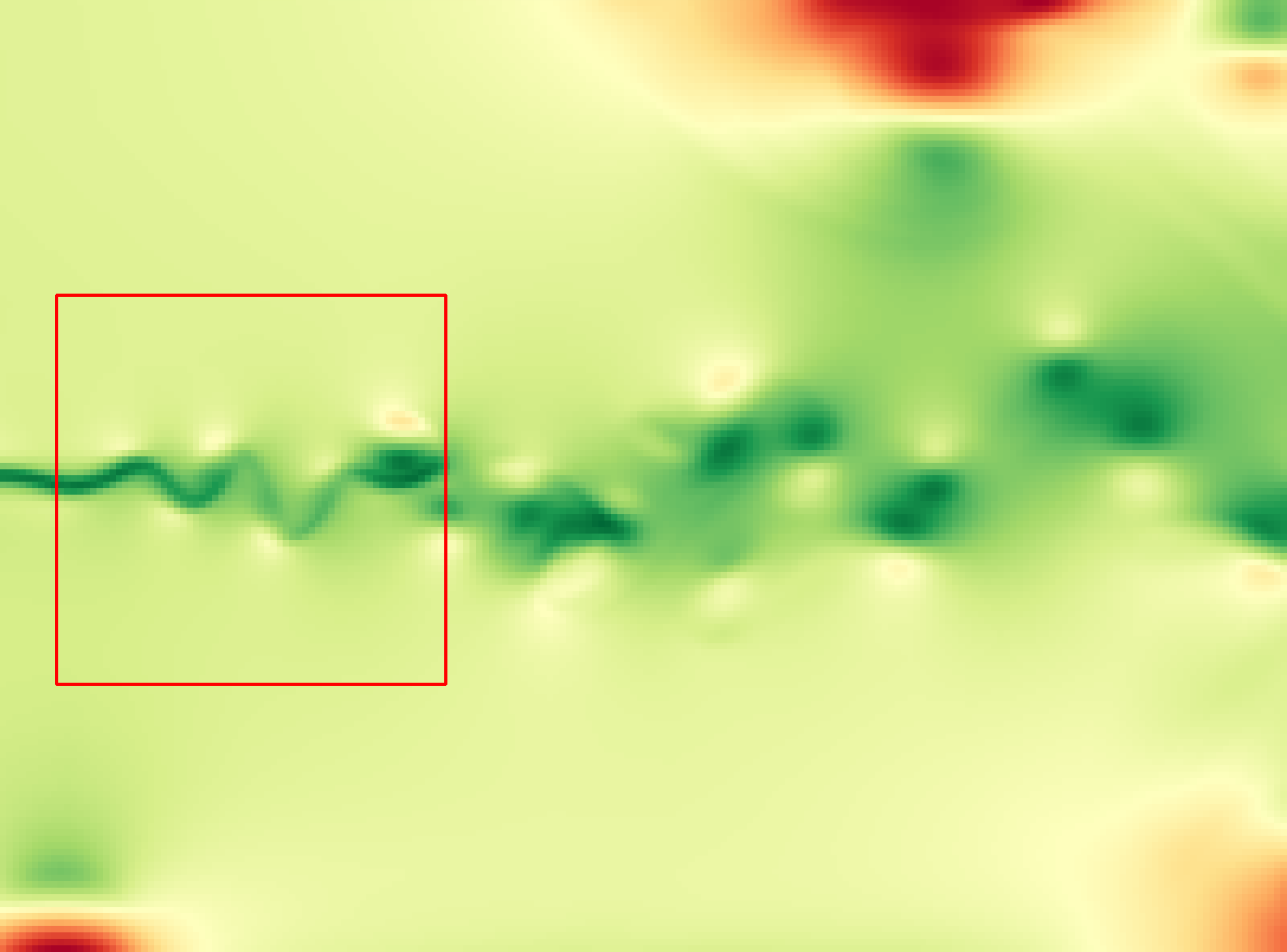} &     
\includegraphics[width=0.38\columnwidth, angle=-90]{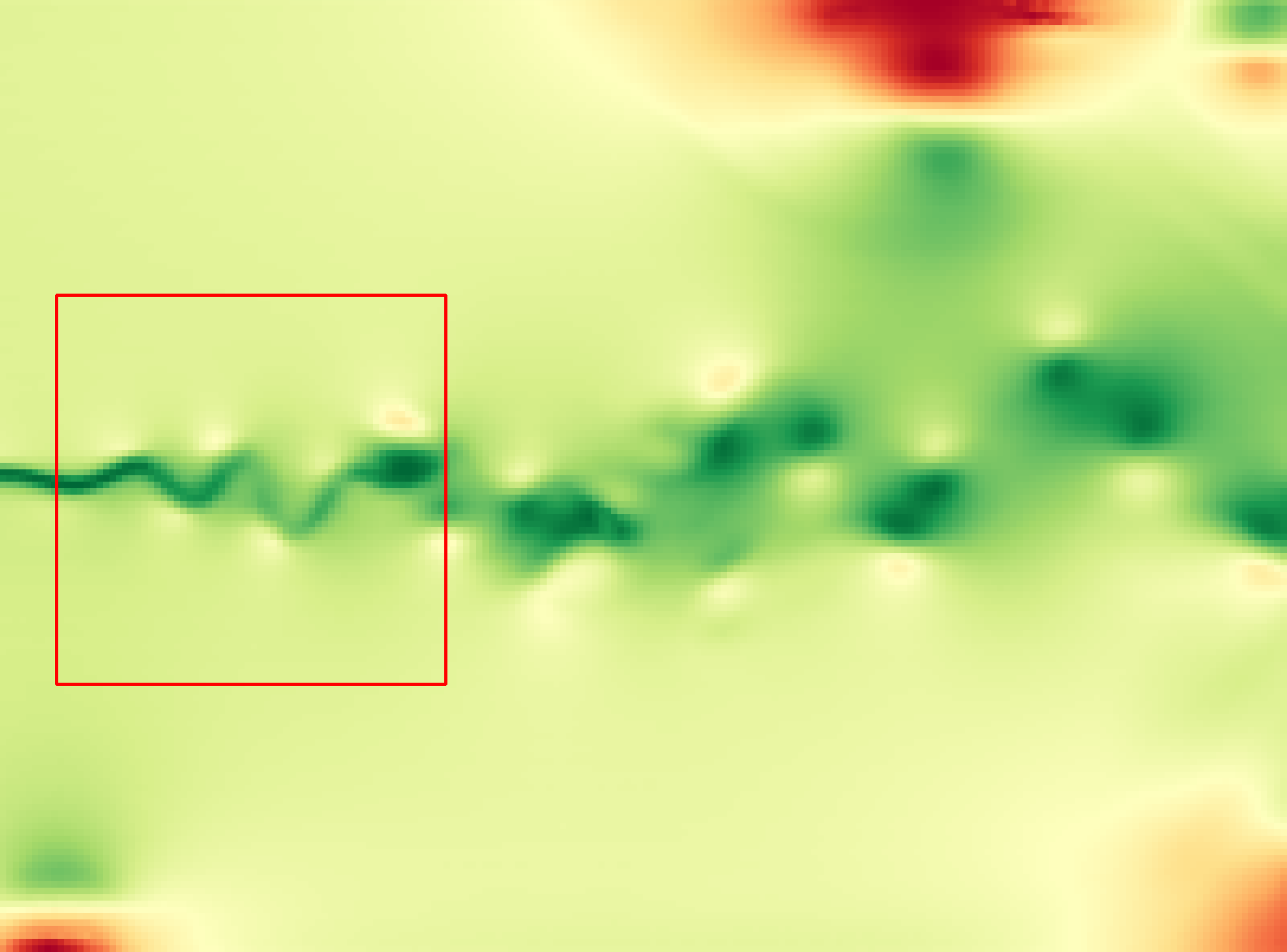} &     
\includegraphics[width=0.38\columnwidth, angle=-90]{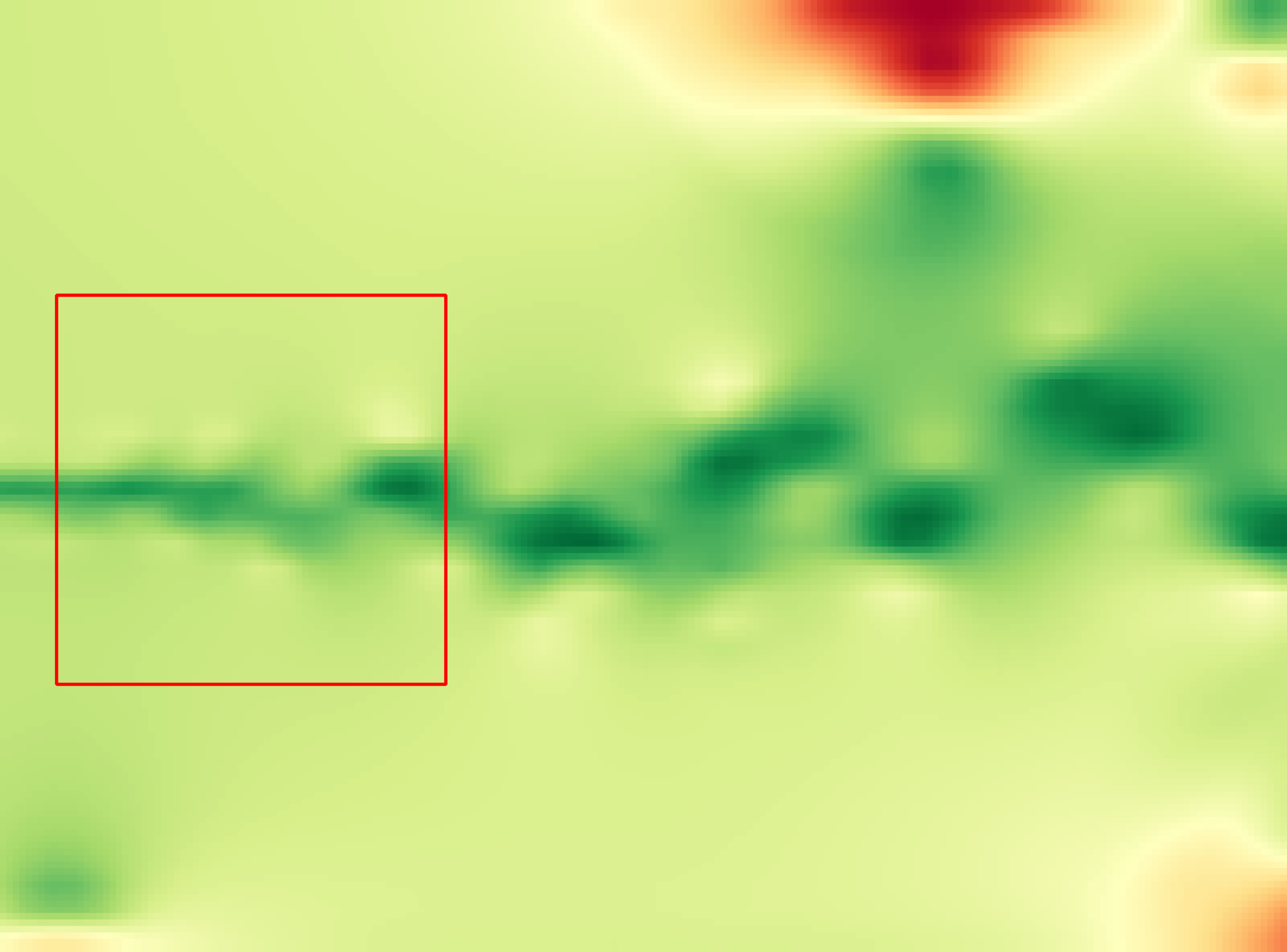} 
\\
\textcolor{black}{(1) Ground Truth} &  \textcolor{black}{(2) FFEINR} & \textcolor{black}{(3) Trilinear}
\\
\includegraphics[width=0.28\columnwidth]{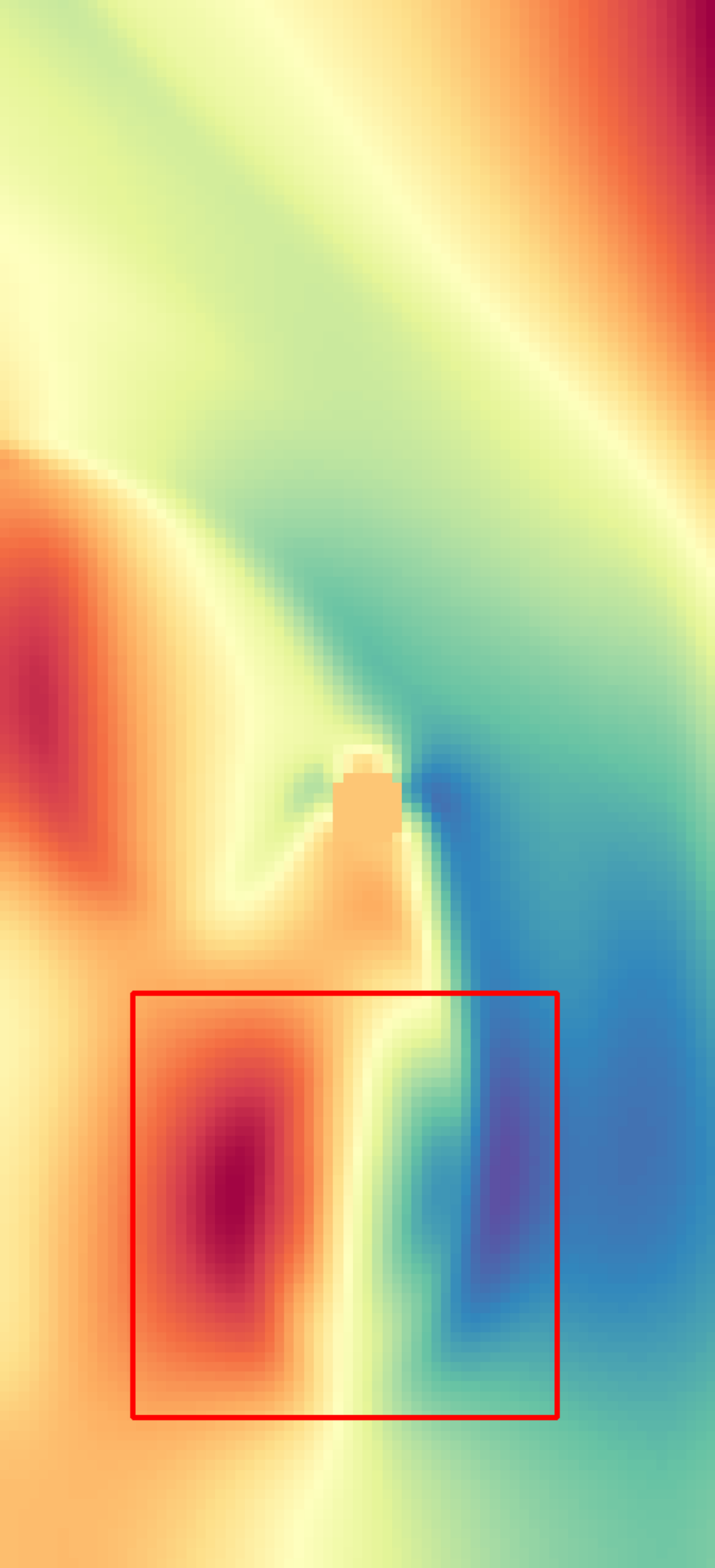} &     
\includegraphics[width=0.28\columnwidth]{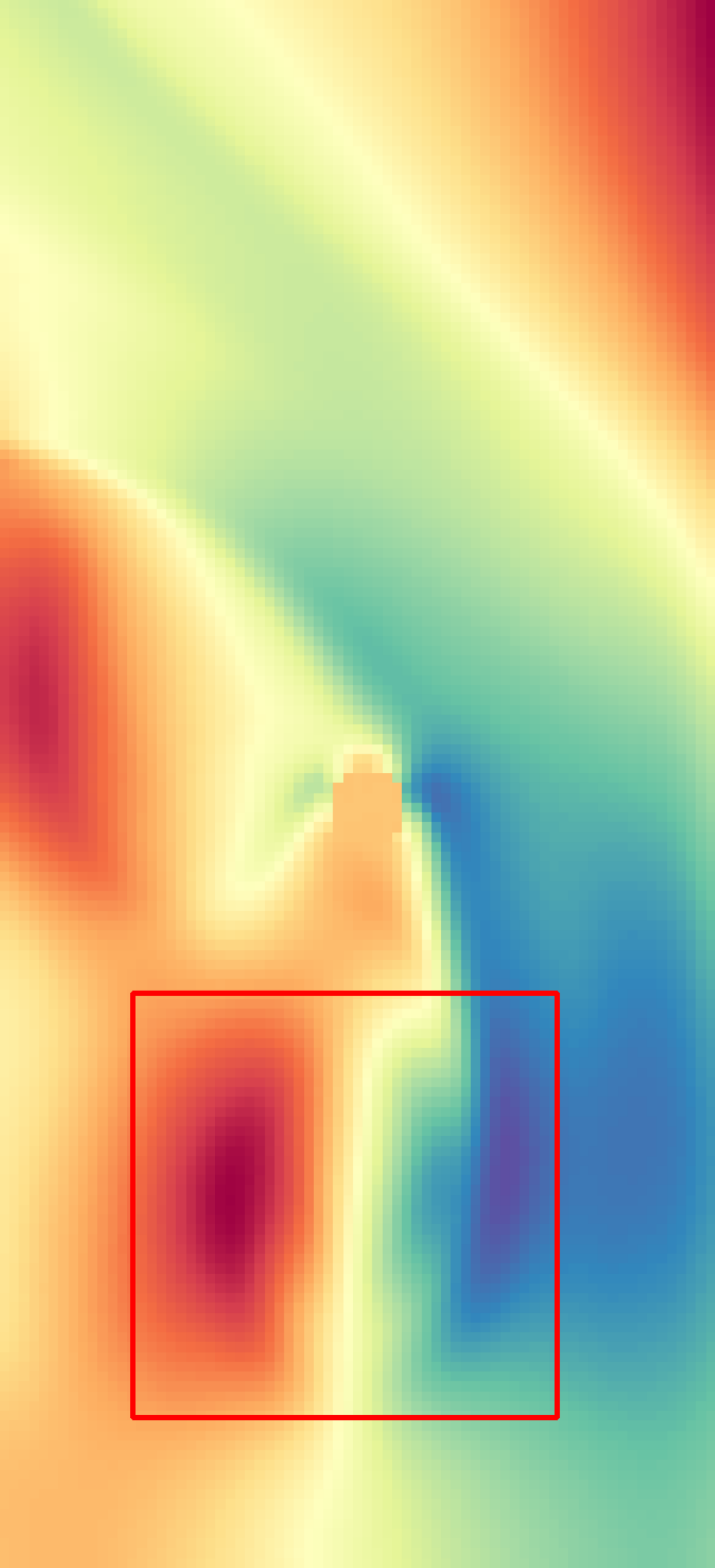} &     
\includegraphics[width=0.28\columnwidth]{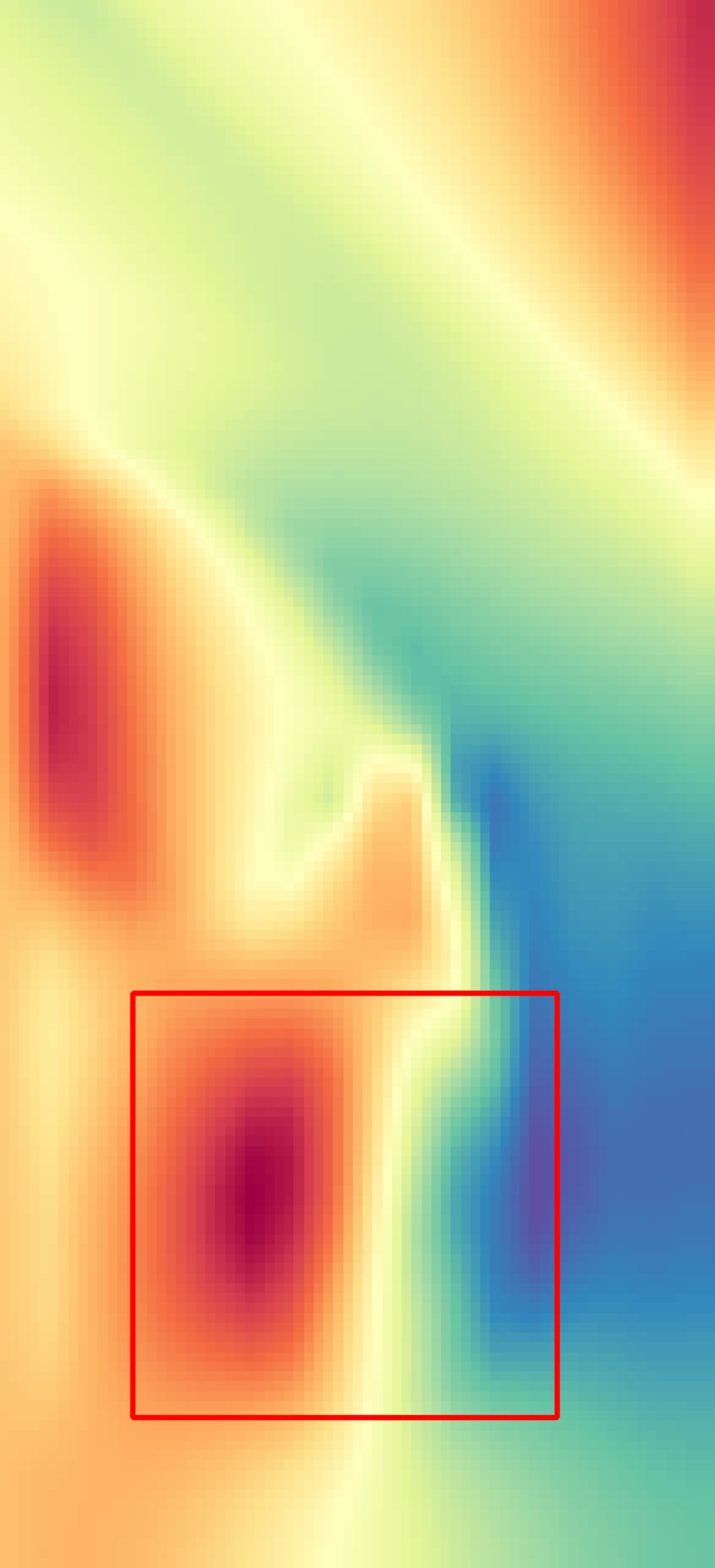} 
\\
(1) Ground Truth & (2) FFEINR & (3) Trilinear
\end{tabular}
\caption{Qualitative comparison on different datasets. Top to bottom: Cylinder, HeatedCylinder, PipeCylinder. It can be seen that FFEINR presents better visualization results than Trilinear in regions such as the edges of cylinder (Cylinder), the fluctuation region (HeatedCylinder) and the interior of the vortex (PipeCylinder). \textcolor{black}{Overall, FFEINR effectively preserves these sharp features, while the baseline Trilinear method blurs the edge information and results in poor visual effects.}}
\label{fig:qual4dataset}
\end{figure}

\textbf{Quantitative and qualitative analysis on different datasets.}
As shown in Table~\ref{tab:quan4dataset}, from a quantitative perspective, the model used in this paper (FFEINR) outperforms the baseline method in all three indicators. 
FFEINR achieves over 40dB and 0.98 for PSNR and SSIM on both the Cylinder dataset and PipeCylinder dataset with regular and gentle flow field variations, and its PSNR leads Trilinear by more than 4dB, or even 10dB. 
For the HeatedCylinder data with the most severe and complex flow field features, although the PSNR is below 40dB, it still outperforms the baseline method. 
In terms of data accuracy, the RMSEs of FFEINR are all lower than Trilinear, indicating a significant advantage in performance. 
From a qualitative perspective, Fig.~\ref{fig:qual4dataset} shows that FFEINR effectively preserves the sharp features of the obstacle edges in the Cylinder dataset, while the baseline Trilinear method blurs the edge information and results in poor visual effects. 
Similar situations occur in experiments on other datasets. 
On the HeatedCylinder dataset, the Trilinear does not effectively preserve the high-frequency variations of the flow field motion behind the obstacles. 
On the PipeCylinder dataset, the locally severe numerical changes in vortex streets are only successfully preserved in FFEINR. 
\textcolor{black}{In order to present the qualitative results more comprehensively, we show other visual results in Fig.~\ref{fig:qual4streamline}. From the streamlines in the HeatedCylinder dataset, it can be seen that the predicted streamlines obtained from the FFEINR are closer to the true streamline map. Specifically, the red box on the left shows the streamlines near the cylindrical obstacle. The predicted streamlines from FFEINR are the same as ground truth, bypassing both sides of the obstacle, while the streamlines obtained from Trilinear bypasses from the same side of the obstacle.
}
Overall, it can be seen that FFEINR has significant superiority in terms of performance across different datasets, especially in terms of providing better results than Trilinear in visualization results in regions with significant variation, such as edges and the interior of vortices.

\begin{table}[tb]
\setlength{\abovecaptionskip}{0.5cm}
\setlength{\belowcaptionskip}{-0.8cm}
\centering
\caption{Quantitative comparison on different datasets. We \textcolor{black}{have bolded} the test results of FFEINR when it is superior to the baseline trilinear interpolation. The same applies in the following.}
\label{tab:quan4dataset}
\renewcommand\arraystretch{1.0}
\setlength\tabcolsep{3pt} 

\begin{tabular}{c|c|ccc}
\cline{1-5}
\textbf{Dataset} & \textbf{Method} & \textbf{PSNR} & \textbf{SSIM} & \textbf{\begin{tabular}[c]{@{}c@{}}RMSE\\ $u_x$/$u_y$\end{tabular}} \\ \cline{1-5}
                    & FFEINR    & \textcolor{black}{\textbf{46.68}} & \textbf{\textcolor{black}{0.994}} & \textbf{\textcolor{black}{0.050/0.069}}          \\
\multirow{-2}{*}{Cylinder} & Trilinear & \textcolor{black}{35.59}                         & \textcolor{black}{0.986}                        & \textcolor{black}{0.194/0.222}                                \\ \cline{1-5}
                    & FFEINR    & \textbf{\textcolor{black}{39.32}} & \textbf{\textcolor{black}{0.963}}  & \textbf{\textcolor{black}{0.155/0.238}}          \\
\multirow{-2}{*}{HeatedCylinder} & Trilinear & \textcolor{black}{32.82}                         & \textcolor{black}{0.955}                        & \textcolor{black}{0.371/0.401}                                 \\ \cline{1-5}
                    & FFEINR    & \textbf{\textcolor{black}{40.21}} & \textbf{\textcolor{black}{0.983}} & \textbf{\textcolor{black}{0.523/0.150}}          \\
\multirow{-2}{*}{PipeCyliner} & Trilinear & \textcolor{black}{36.01}                        & \textcolor{black}{0.972}                        & \textcolor{black}{0.711/0.597}                                 \\ \cline{1-5}
\end{tabular}%
\end{table}

\begin{figure}[!h]
\setlength{\abovecaptionskip}{0.5cm}
\setlength{\belowcaptionskip}{-0.5cm}
\begin{tabular}{c}


\includegraphics[width=0.32\columnwidth, angle=-90]{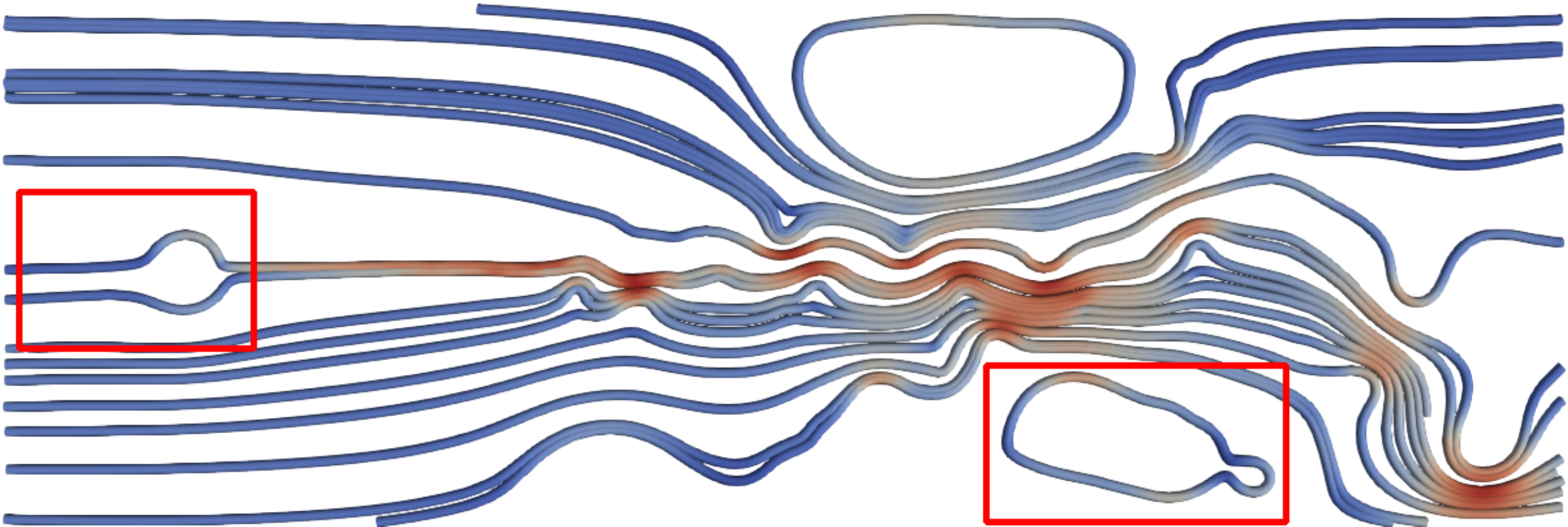} \\   
(1) Ground Truth \\
\includegraphics[width=0.32\columnwidth, angle=-90]{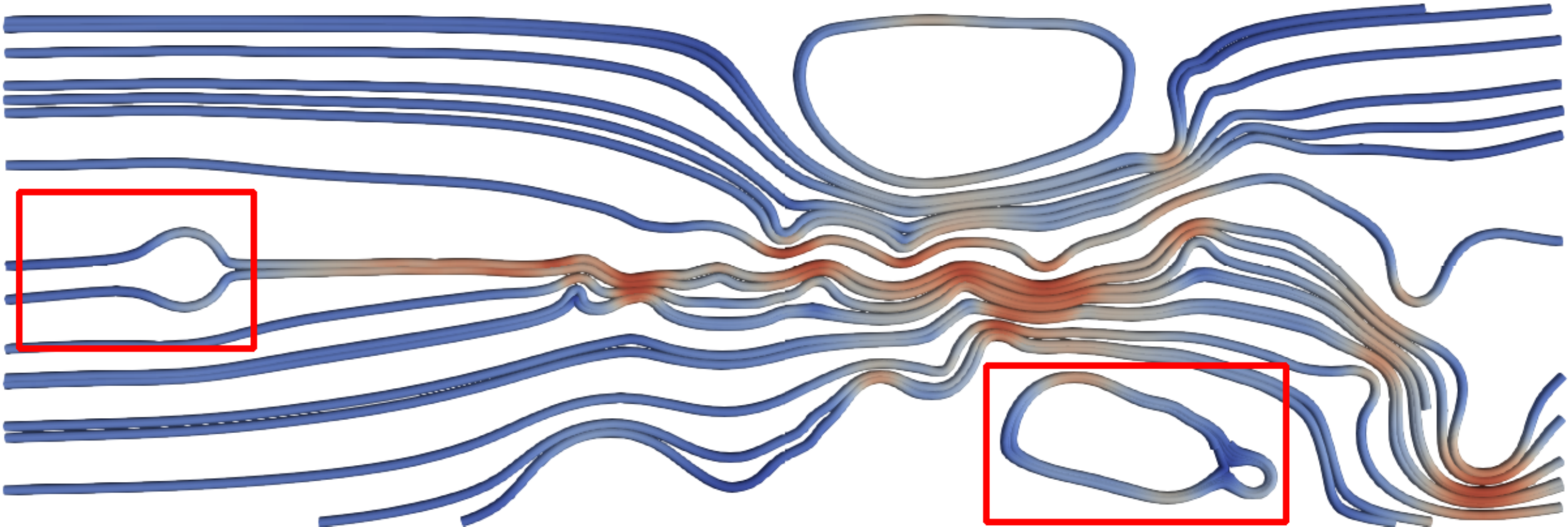}  \\    
(2) FFEINR \\
\includegraphics[width=0.32\columnwidth, angle=-90]{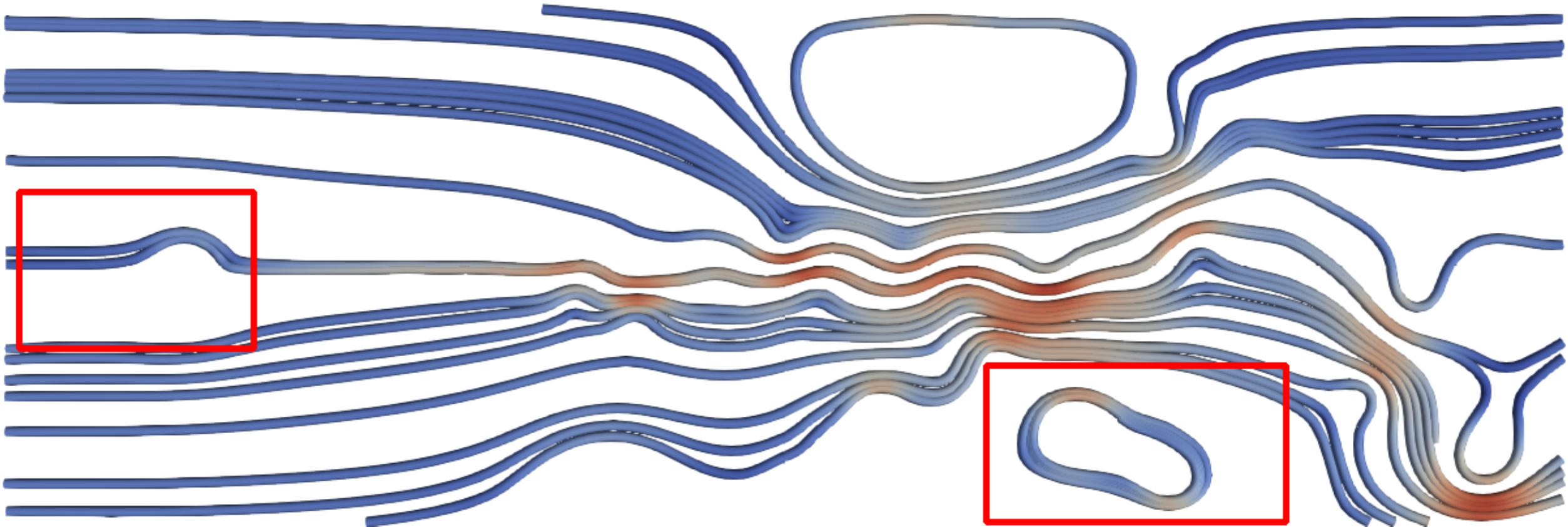} \\
(3) Trilinear
\end{tabular}
\caption{\textcolor{black}{Streamlines map (HeatedCylinder). We randomly select 20 seed points and render the streamline map. The red box on the left shows the streamlines near the cylindrical obstacle. The predicted streamlines from FFEINR are identical to the ground truth and bypass both sides of the obstacle, while the streamlines obtained from Trilinear bypass the same side of the obstacle. It can also be seen from the red box on the right that the predicted streamlines obtained from the FFEINR is closer to the true streamline map.}}
\label{fig:qual4streamline}
\end{figure}

\begin{table}[!h]
\setlength{\abovecaptionskip}{0.5cm}
\setlength{\belowcaptionskip}{-1.cm}
\caption{Quantitative comparison for extended resolution \textcolor{black}{on three datasets}. ESR means extended super-resolution. $(S\times4, T\times2)$ means $(Spatial\times4, Temporal\times2)$. For all datasets, we train the model at a factor of $(S\times4, T\times2)$. FFEINR achieves higher PSNR, SSIM and RMSE than the baseline method in most experiments.}
\label{tab:quan4extended}
\setlength\tabcolsep{3pt} 
    \begin{tabular}{c|c|c|ccc}
    \hline
\textbf{Dataset} & \textbf{Method} & \textbf{\begin{tabular}[c]{@{}c@{}}SR \& \\ESR\end{tabular}} & \textbf{PSNR} & \textbf{SSIM} & \textbf{\begin{tabular}[c]{@{}c@{}}RMSE\\ $u_x$/$u_y$\end{tabular}} \\ \hline
\multicolumn{1}{c|}{\multirow{8}{*}{Cylinder}} & FIFENR & \multicolumn{1}{c|}{\multirow{2}{*}{Sx4, Tx2}} & \textcolor{black}{\textbf{46.68 }} & \textcolor{black}{\textbf{0.994 }} & \textcolor{black}{\textbf{0.050/0.069}} \\
\cline{2-2}          & Trilinear &       & \textcolor{black}{35.59 } & \textcolor{black}{0.986 } & \textcolor{black}{0.194/0.222} \\
\cline{2-6}          & FIFENR & \multirow{2}{*}{Sx2, Tx2} & \textcolor{black}{\textbf{40.04 }} & \textcolor{black}{\textbf{0.993 }} & \textcolor{black}{0.152/\textbf{0.104}} \\
\cline{2-2}          & Trilinear &       & \textcolor{black}{38.82 } & \textcolor{black}{0.992 } & \textcolor{black}{0.104/0.178} \\
\cline{2-6}          & FIFENR & \multirow{2}{*}{Sx4, Tx4} & \textcolor{black}{\textbf{38.50 }} & \textcolor{black}{\textbf{0.984 }} & \textcolor{black}{\textbf{0.122/0.234}} \\
\cline{2-2}          & Trilinear &       & \textcolor{black}{34.08 } & \textcolor{black}{0.979 } & \textcolor{black}{0.209/0.286} \\
\cline{2-6}          & FIFENR & \multirow{2}{*}{Sx4, Tx8} & \textcolor{black}{\textbf{36.79 }} & \textcolor{black}{0.978 } & \textcolor{black}{\textbf{0.143}/0.282} \\
\cline{2-2}          & Trilinear &       & \textcolor{black}{34.92 } & \textcolor{black}{0.982 } & \textcolor{black}{0.200/0.253} \\
    \hline
    \multirow{8}{*}{HeatedCylinder} & FIFENR & \multicolumn{1}{c|}{\multirow{2}{*}{Sx4, Tx2}} & \textcolor{black}{\textbf{39.32 }} & \textcolor{black}{\textbf{0.963 }} & \textcolor{black}{\textbf{0.155/0.238}} \\
\cline{2-2}          & Trilinear &       & \textcolor{black}{32.82 } & \textcolor{black}{0.955 } & \textcolor{black}{0.371/0.401} \\
\cline{2-6}          & FIFENR & \multirow{2}{*}{Sx2, Tx2} & \textcolor{black}{\textbf{37.04 }} & \textcolor{black}{0.962 } & \textcolor{black}{\textbf{0.218}/0.264} \\
\cline{2-2}          & Trilinear &       & \textcolor{black}{36.19 } & \textcolor{black}{0.968 } & \textcolor{black}{0.301/0.229} \\
\cline{2-6}          & FIFENR & \multirow{2}{*}{Sx4, Tx4} & \textcolor{black}{\textbf{35.64 }} & \textcolor{black}{0.956 } & \textcolor{black}{\textbf{0.285/0.302}} \\
\cline{2-2}          & Trilinear &       & \textcolor{black}{33.08 } & \textcolor{black}{0.961 } & \textcolor{black}{0.372/0.403} \\
\cline{2-6}          & FIFENR & \multirow{2}{*}{Sx4, Tx8} & \textcolor{black}{\textbf{34.79 }} & \textcolor{black}{0.953 } & \textcolor{black}{\textbf{0.326/0.324}} \\
\cline{2-2}          & Trilinear &       & \textcolor{black}{33.45 } & \textcolor{black}{0.963 } & \textcolor{black}{0.347/0.395} \\
    \hline
    \multirow{8}{*}{PipeCylinder} & FIFENR & \multicolumn{1}{c|}{\multirow{2}{*}{Sx4, Tx2}} & \textcolor{black}{\textbf{40.21 }} & \textcolor{black}{\textbf{0.983 }} & \textcolor{black}{\textbf{0.523/0.150}} \\
\cline{2-2}          & Trilinear &       & \textcolor{black}{36.01 } & \textcolor{black}{0.972 } & \textcolor{black}{0.711/0.597} \\
\cline{2-6}          & FIFENR & \multirow{2}{*}{Sx2, Tx2} & \textcolor{black}{38.21 } & \textcolor{black}{0.978 } & \textcolor{black}{0.638/\textbf{0.237}} \\
\cline{2-2}          & Trilinear &       & \textcolor{black}{39.60 } & \textcolor{black}{0.987 } & \textcolor{black}{0.496/0.371} \\
\cline{2-6}          & FIFENR & \multirow{2}{*}{Sx4, Tx4} & \textcolor{black}{\textbf{38.75 }} & \textcolor{black}{\textbf{0.981 }} & \textcolor{black}{\textbf{0.563/0.306}} \\
\cline{2-2}          & Trilinear &       & \textcolor{black}{35.71 } & \textcolor{black}{0.970 } & \textcolor{black}{0.730/0.641} \\
\cline{2-6}          & FIFENR & \multirow{2}{*}{Sx4, Tx8} & \textcolor{black}{\textbf{38.28 }} & \textcolor{black}{\textbf{0.980 }} & \textcolor{black}{\textbf{0.581/0.355}} \\
\cline{2-2}          & Trilinear &       & \textcolor{black}{35.89 } & \textcolor{black}{0.971 } & \textcolor{black}{0.721/0.621} \\
    \hline
    \end{tabular}%
\end{table}

\begin{figure*}[!h]
\setlength{\abovecaptionskip}{0.5cm}
\setlength{\belowcaptionskip}{-0.4cm}
\centering
\begin{tabular}{ccccc}
\centering
    \multirow{-2}{*}{\rotatebox{90}{\textcolor{black}{FFEINR}}} &
    \includegraphics[width=0.32\columnwidth, angle=90]{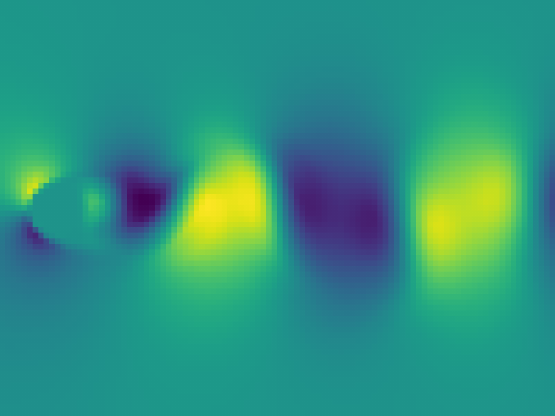} &     
    \includegraphics[width=0.32\columnwidth, angle=90]{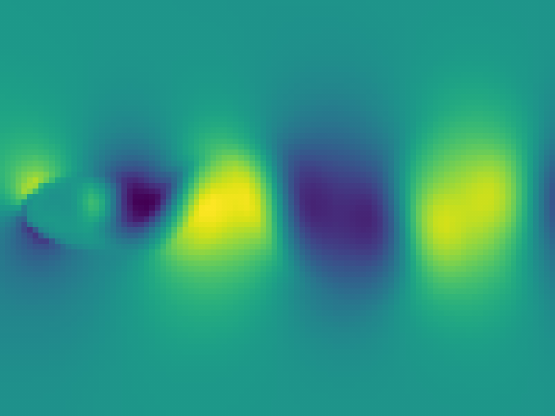} &     
    \includegraphics[width=0.32\columnwidth, angle=90]{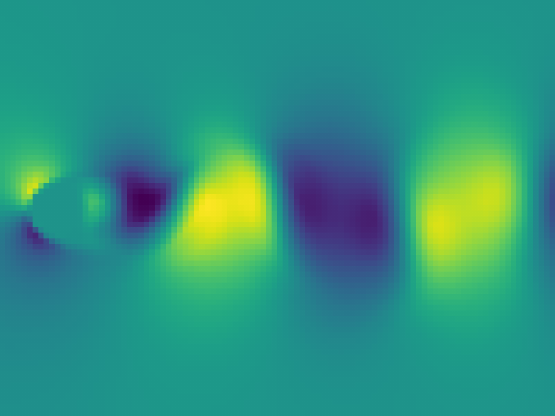} &     
    \includegraphics[width=0.32\columnwidth, angle=90]{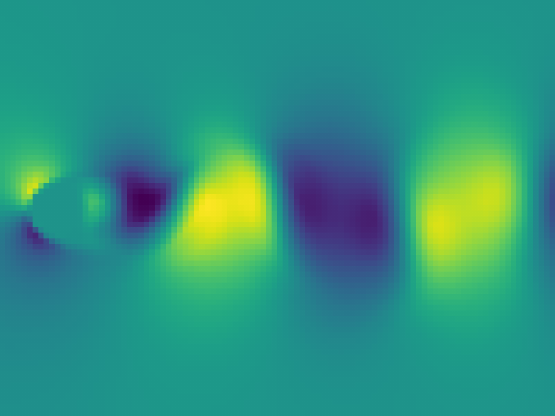}     
\\  
    &
    \includegraphics[width=0.32\columnwidth, angle=90]{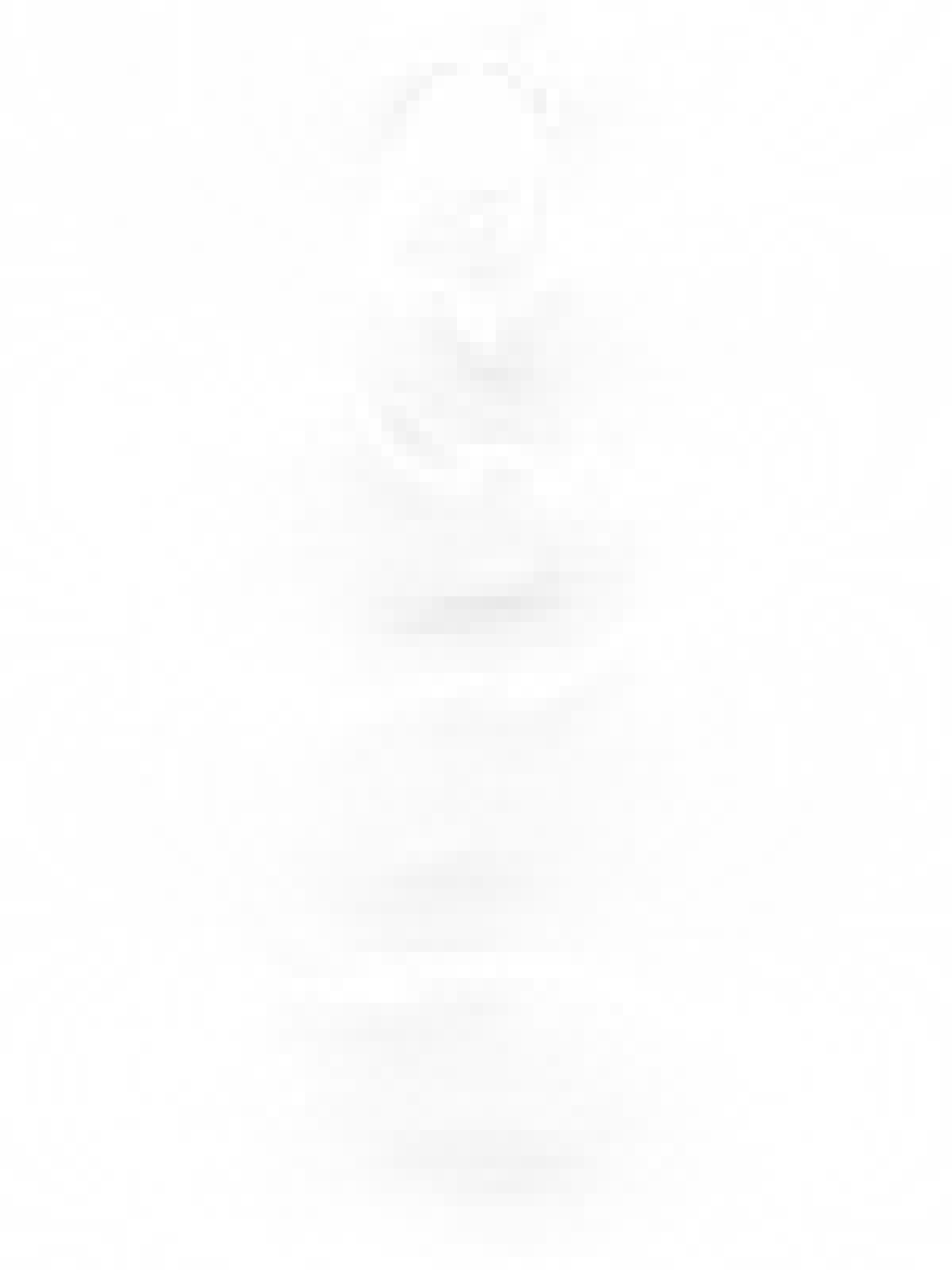} &    
    \includegraphics[width=0.32\columnwidth, angle=90]{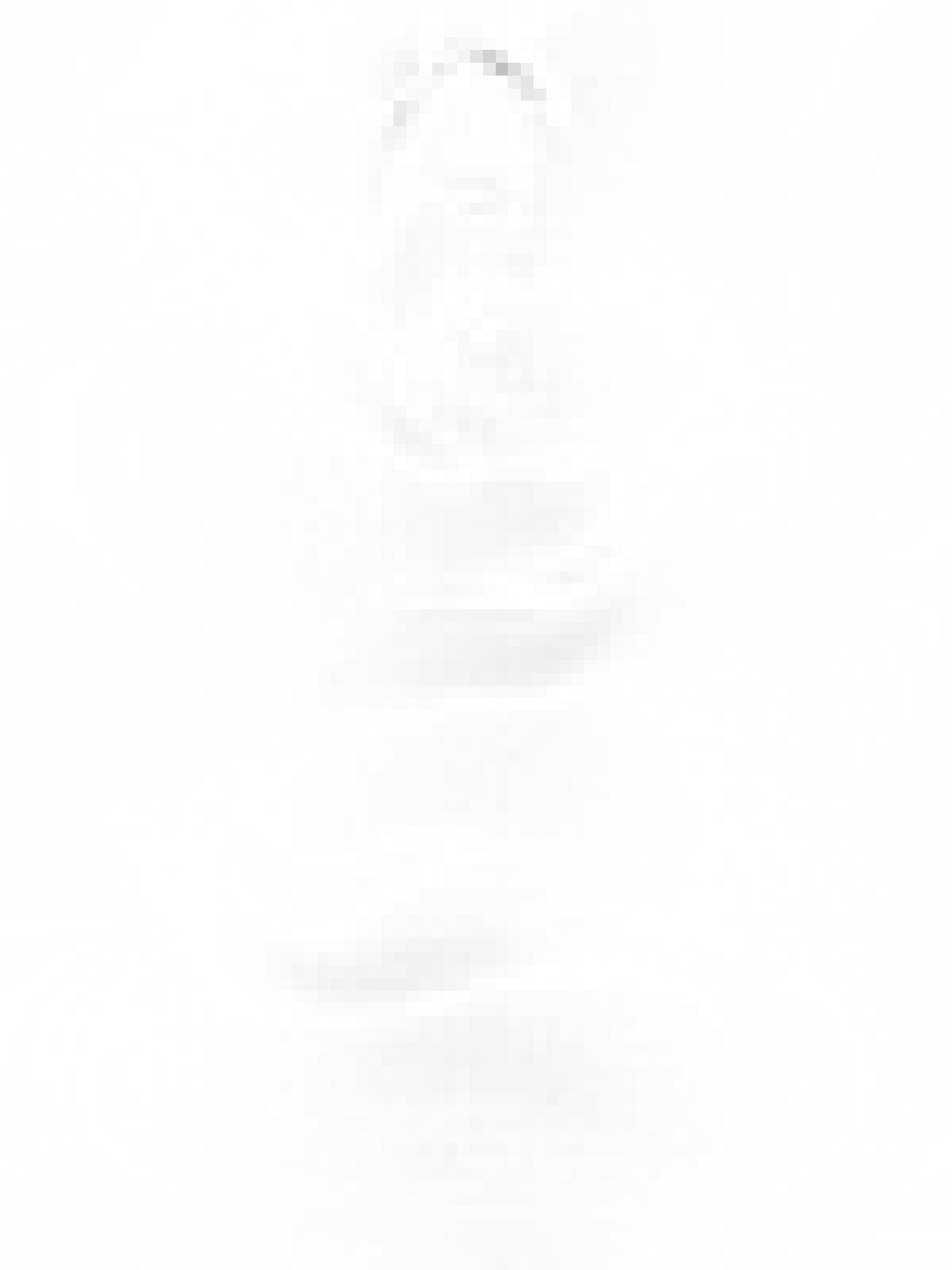} &       
    \includegraphics[width=0.32\columnwidth, angle=90]{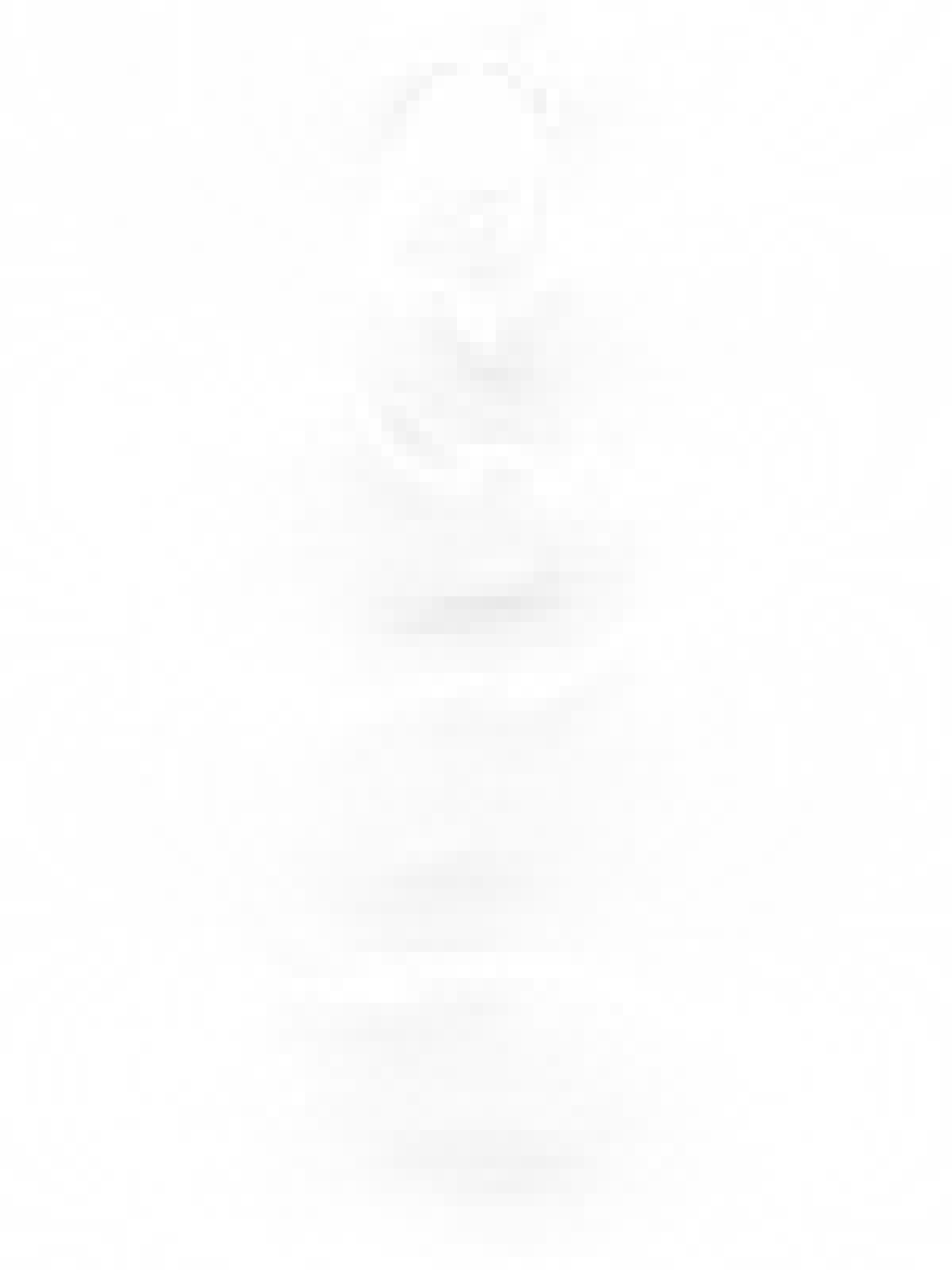} &     
    \includegraphics[width=0.32\columnwidth, angle=90]{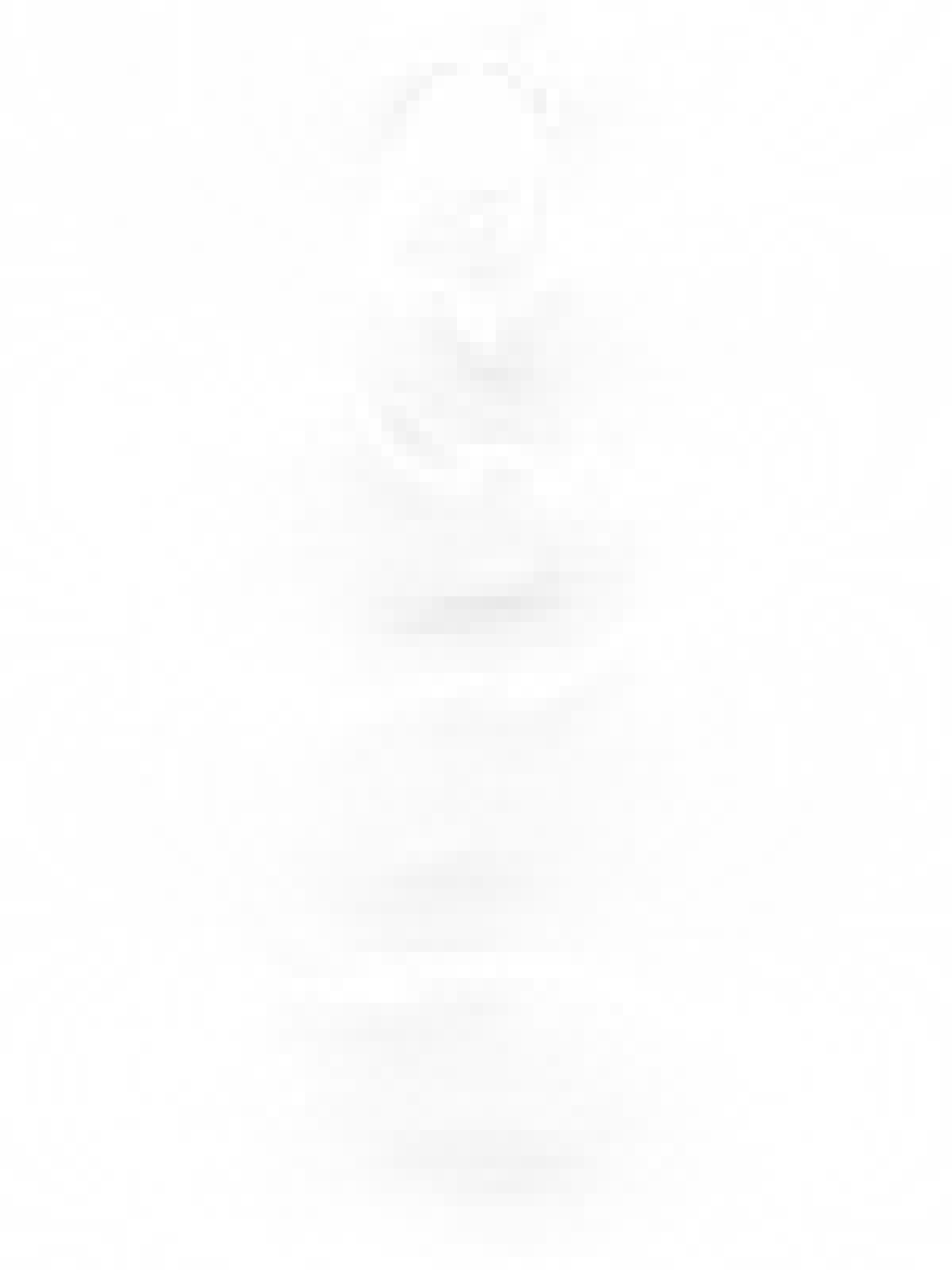}     
\\  
    \multirow{-2}{*}{\rotatebox{90}{\textcolor{black}{Trilinear}}} &
    \includegraphics[width=0.32\columnwidth, angle=90]{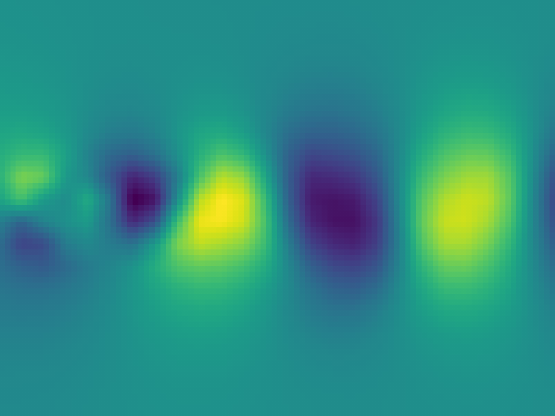} &    
    \includegraphics[width=0.32\columnwidth, angle=90]{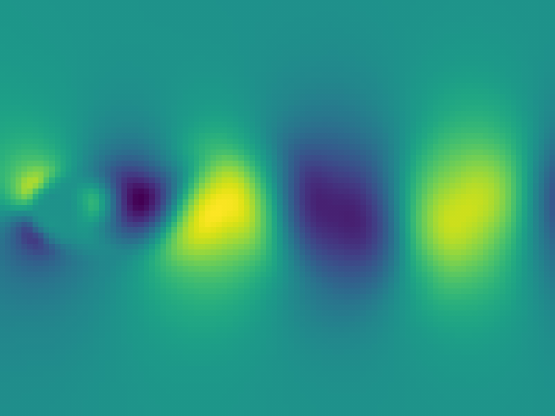} &     
    \includegraphics[width=0.32\columnwidth, angle=90]{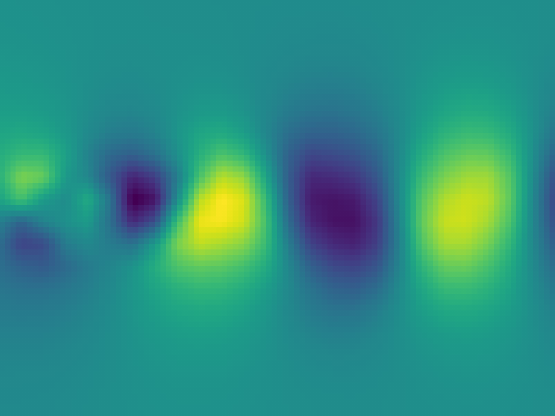} &     
    \includegraphics[width=0.32\columnwidth, angle=90]{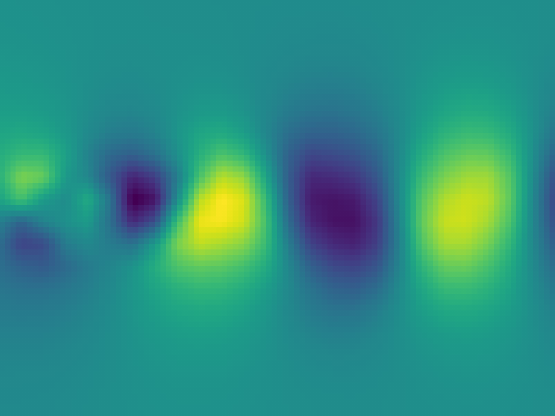}     
\\  
    &
    \includegraphics[width=0.32\columnwidth, angle=90]{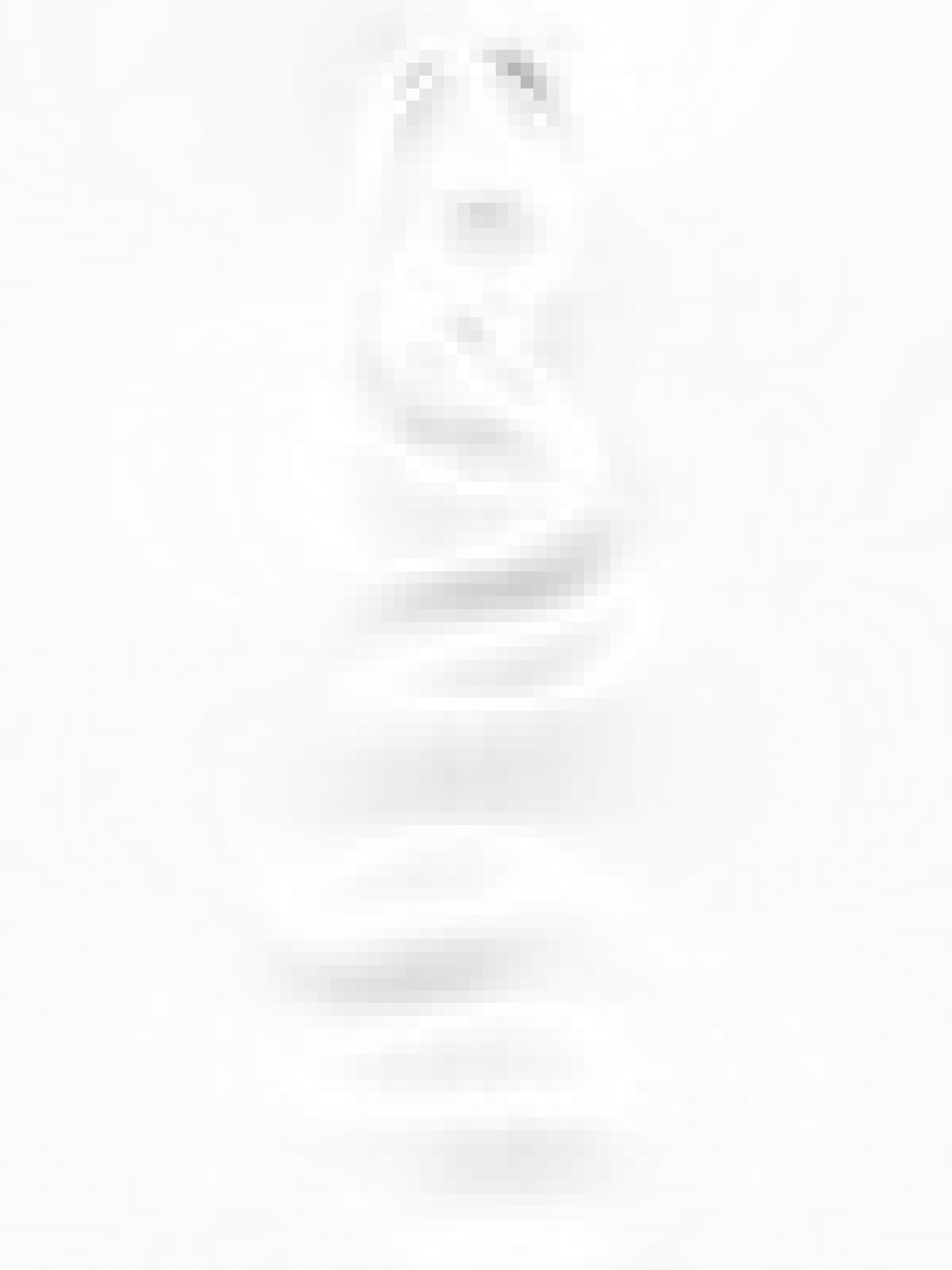} &   
    \includegraphics[width=0.32\columnwidth, angle=90]{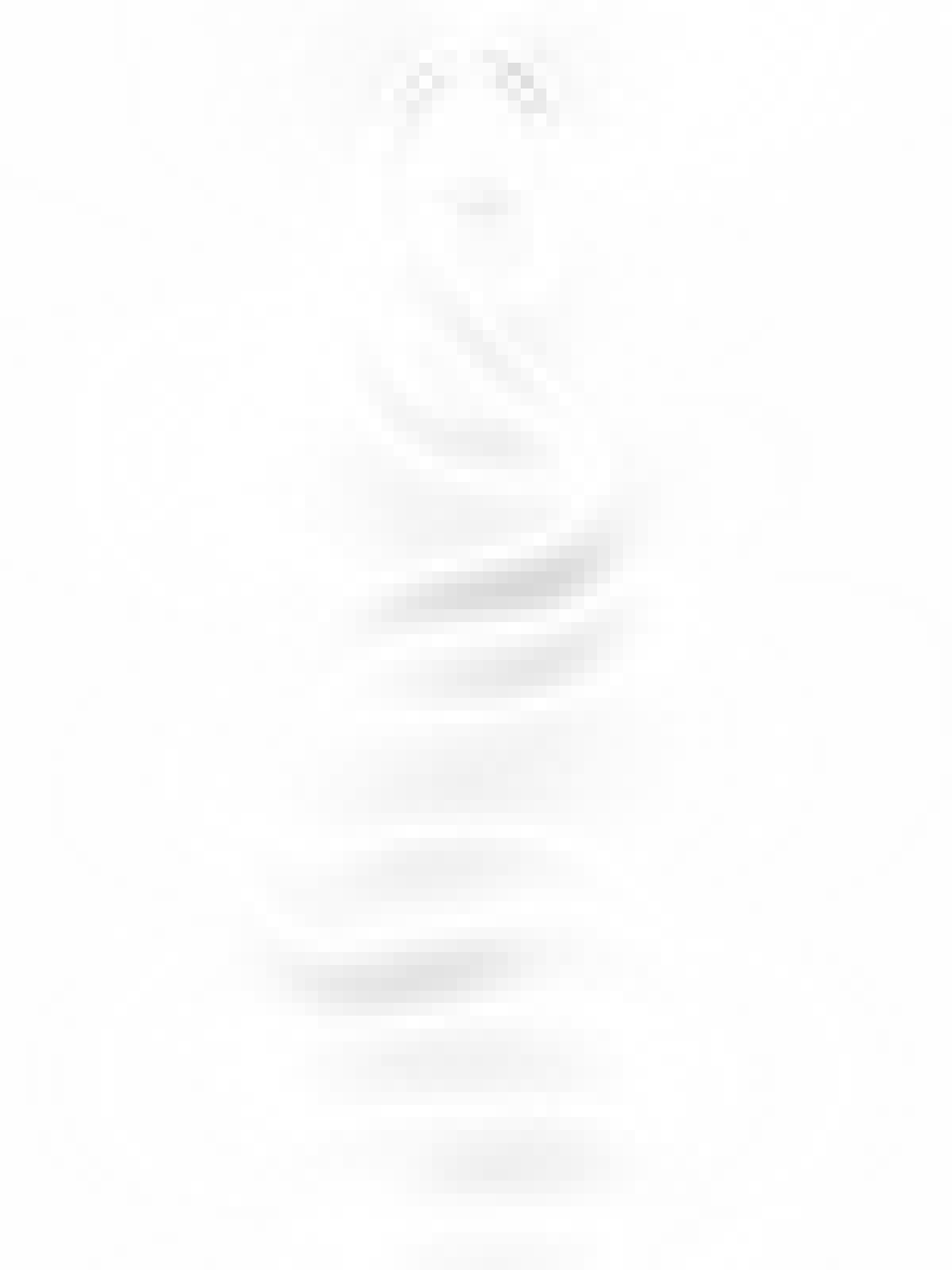} &      
    \includegraphics[width=0.32\columnwidth, angle=90]{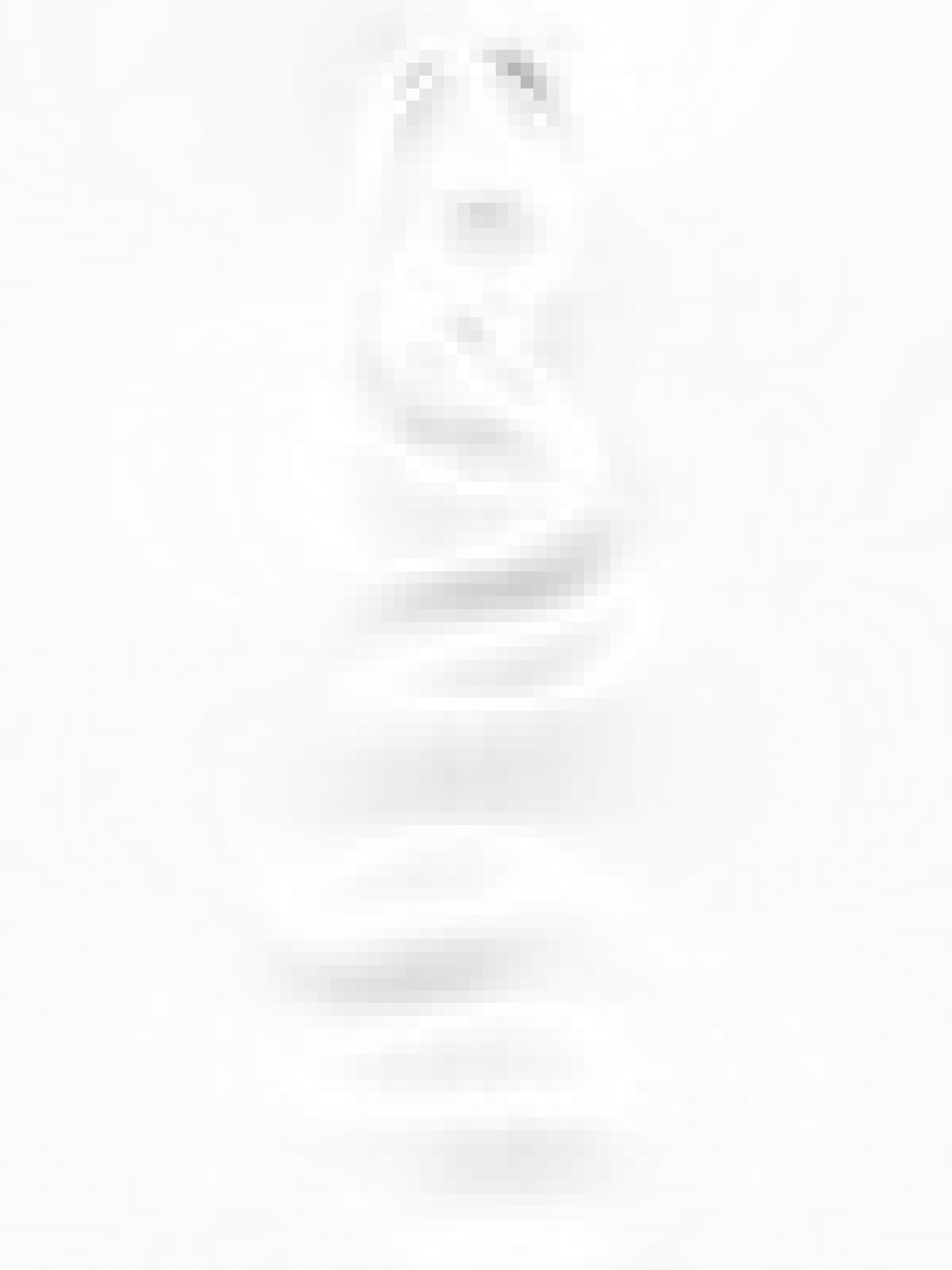} &     
    \includegraphics[width=0.32\columnwidth, angle=90]{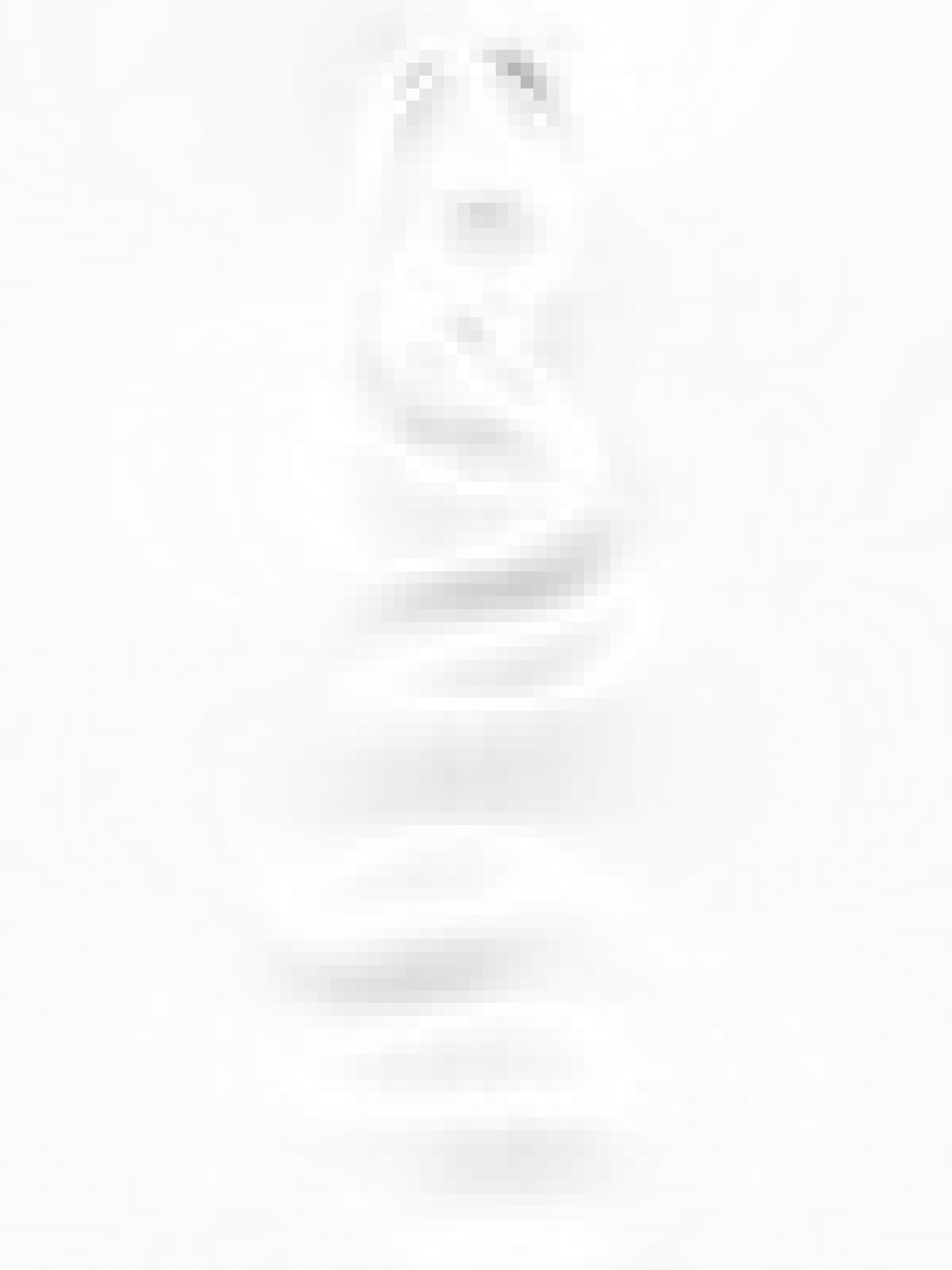} 
\\ &
(1) S$\times$4, T$\times$2 & (2) S$\times$2, T$\times$2& (3) S$\times$4, T$\times$4&(4) S$\times$4, T$\times$8
\\
\end{tabular}
\caption{Visualization and error comparison for extended resolution (Cylinder). Top to bottom: FFEINR, Trilinear. We train the model at a factor of $(S\times4, T\times2)$. For scale factors outside the training setting, FFEINR still achieves better visualization results than Trilinear at the edge of cylinder. Besides, from another point of view, we can see that the visualization error of FFEINR is very small, while the error of the Trilinear is large in the vortex street region.}
\label{fig:qual4extend1}
\end{figure*}
\begin{figure*}[!h]
\setlength{\abovecaptionskip}{0.cm}
\setlength{\belowcaptionskip}{-0.8cm}
\begin{tabular}{ccccc}
\centering
    \multirow{-2}{*}{\rotatebox{90}{\textcolor{black}{FFEINR}}} &
    \includegraphics[width=0.432\columnwidth]{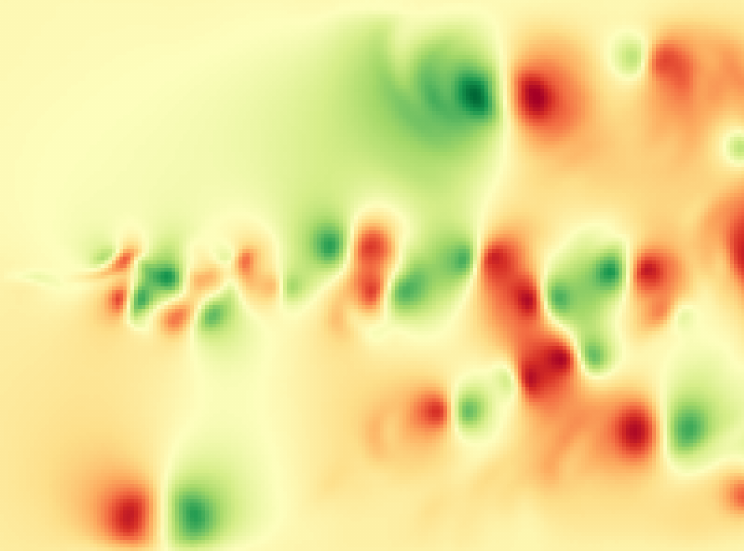} &     
    \includegraphics[width=0.432\columnwidth]{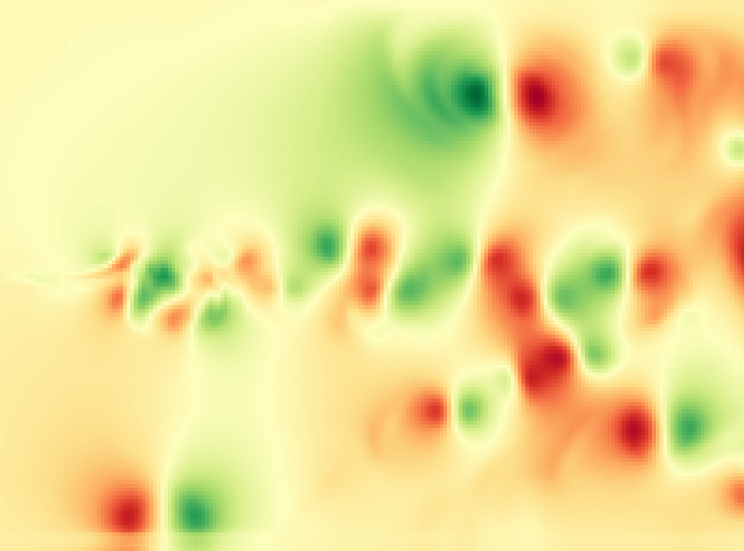} &     
    \includegraphics[width=0.432\columnwidth]{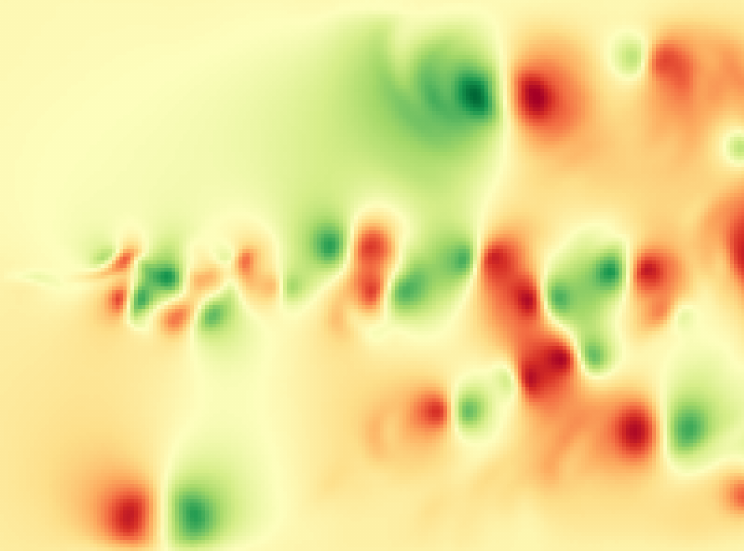} &     
    \includegraphics[width=0.432\columnwidth]{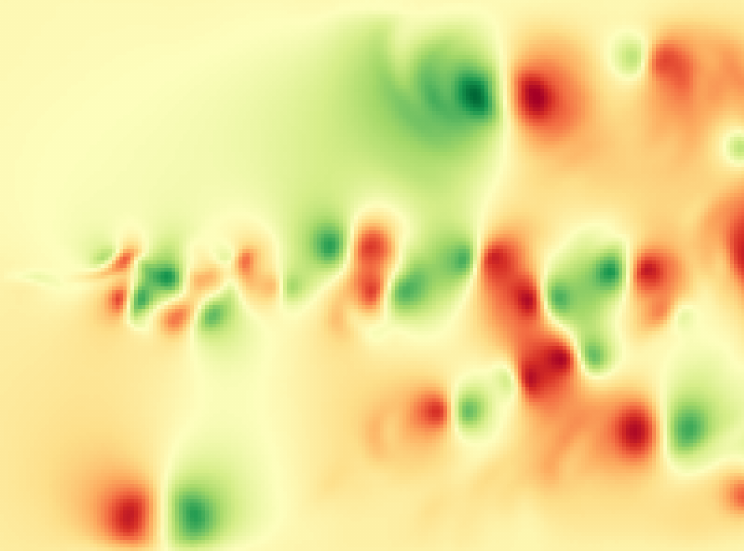}     
\\  
    &
    \includegraphics[width=0.432\columnwidth]{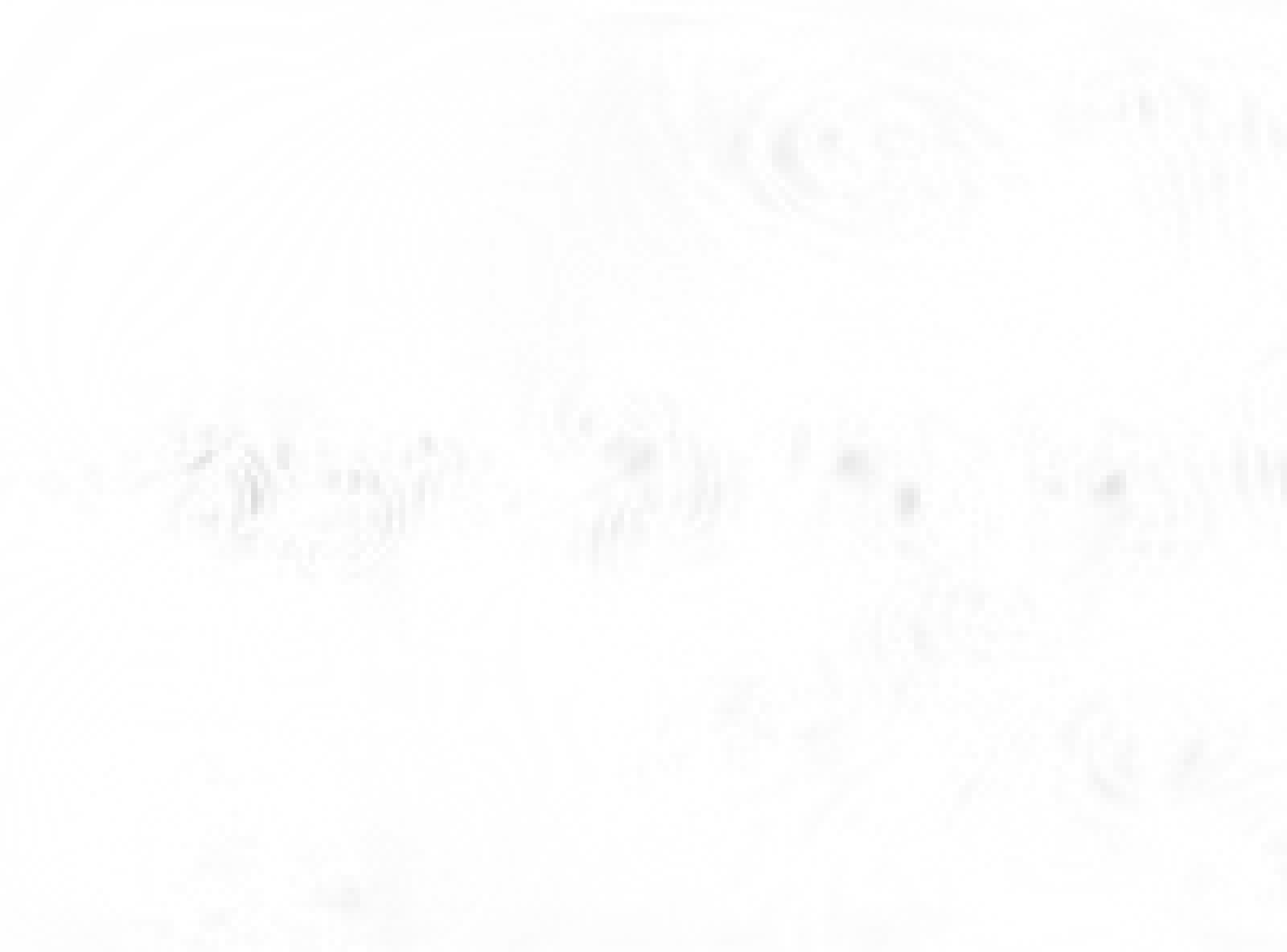} &    
    \includegraphics[width=0.432\columnwidth]{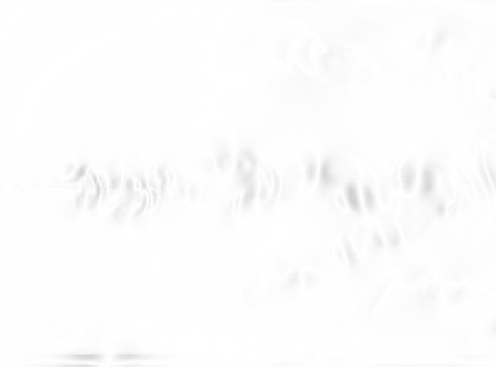} &       
    \includegraphics[width=0.432\columnwidth]{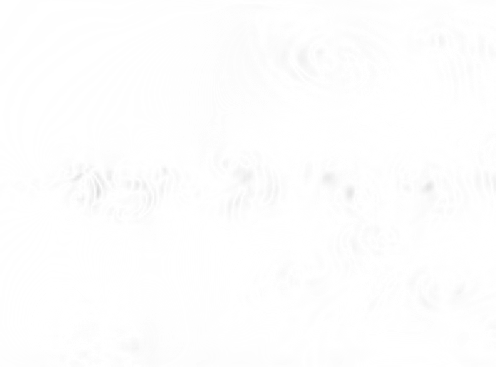} &     
    \includegraphics[width=0.432\columnwidth]{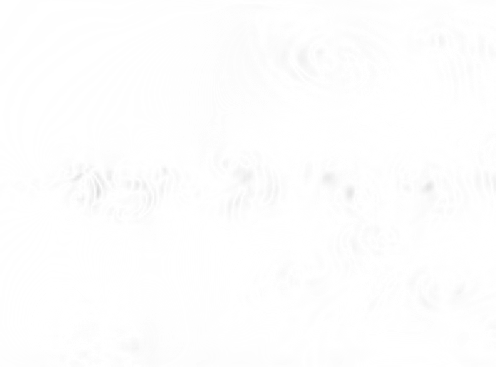}     
\\  
    \multirow{-2}{*}{\rotatebox{90}{\textcolor{black}{Trilinear}}} &
    \includegraphics[width=0.432\columnwidth]{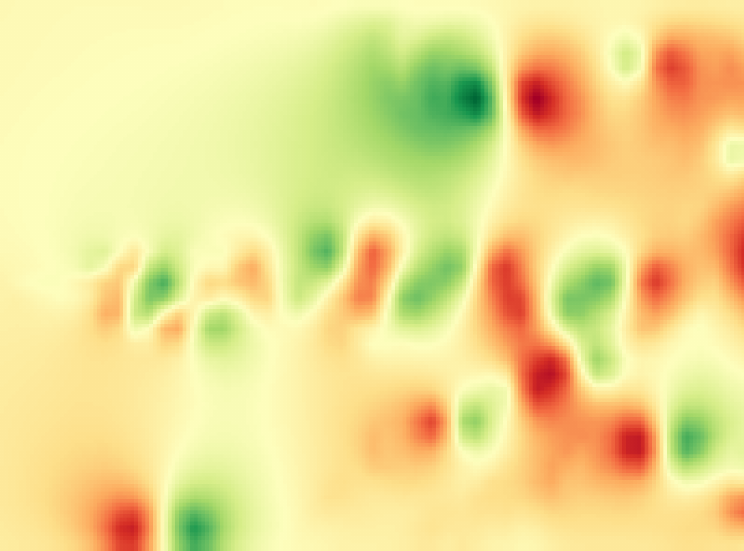} &    
    \includegraphics[width=0.432\columnwidth]{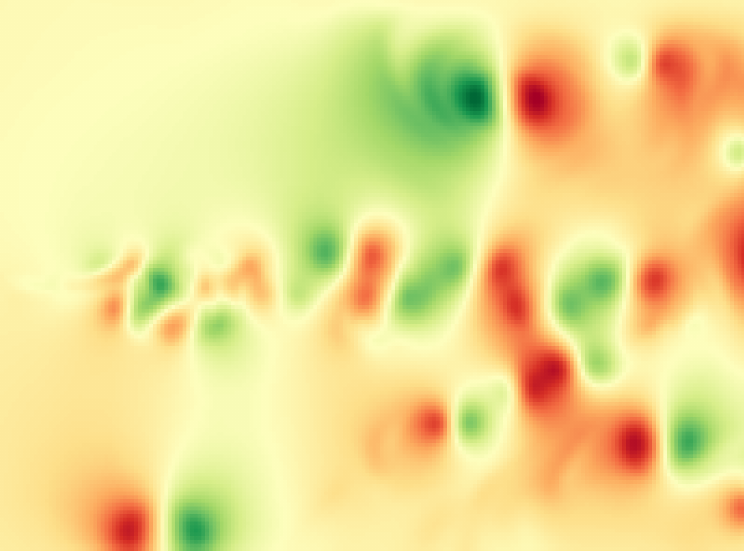} &     
    \includegraphics[width=0.432\columnwidth]{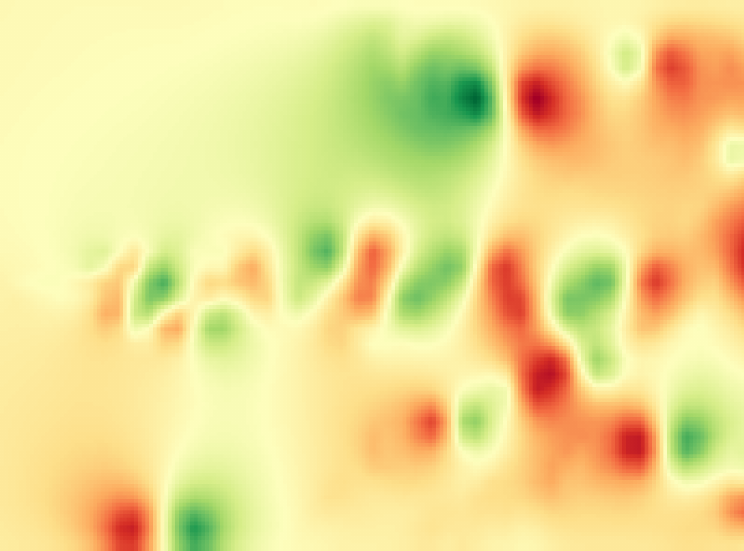} &     
    \includegraphics[width=0.432\columnwidth]{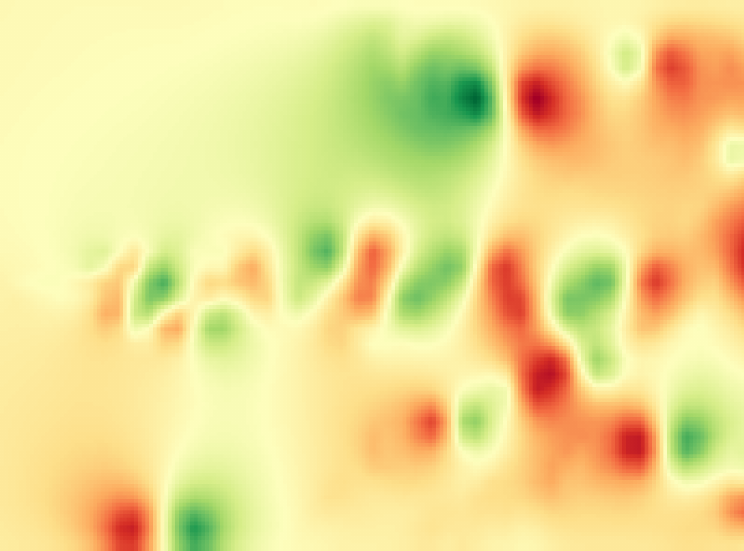}     
\\  
    &
    \includegraphics[width=0.432\columnwidth]{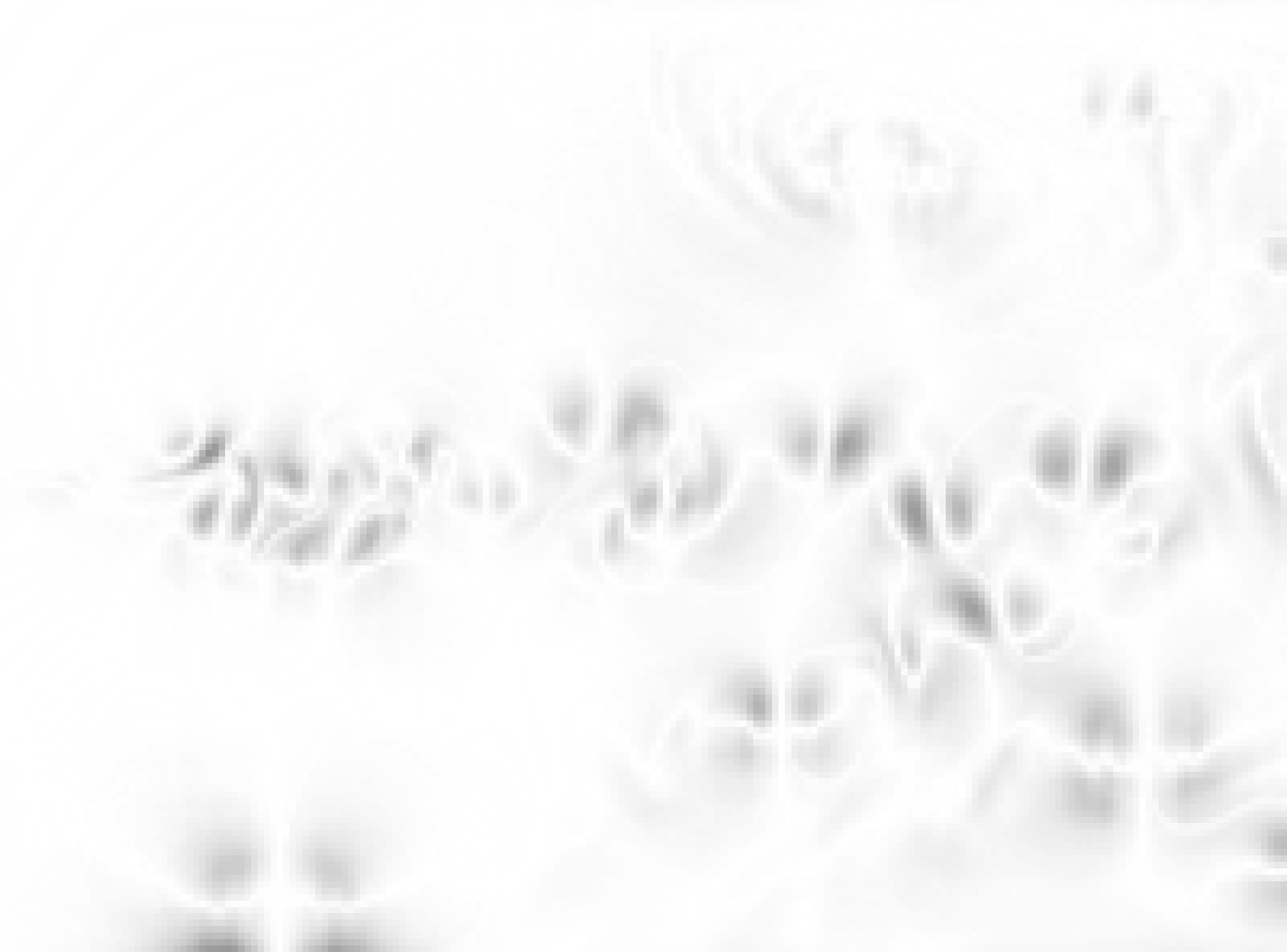} &   
    \includegraphics[width=0.432\columnwidth]{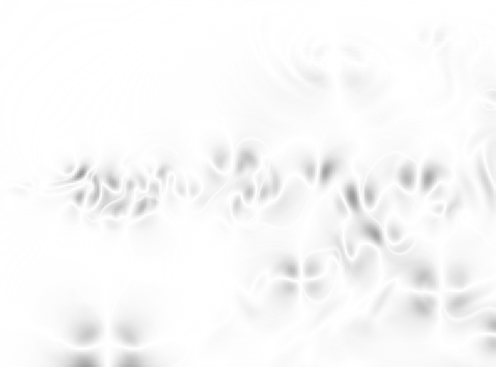} &      
    \includegraphics[width=0.432\columnwidth]{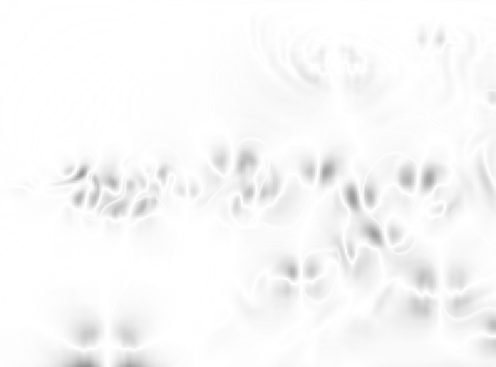} &     
    \includegraphics[width=0.432\columnwidth]{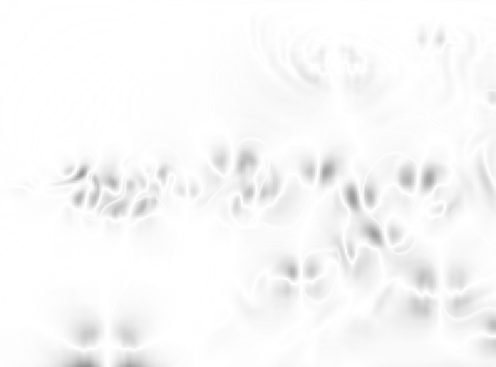}     
\\ &
(1) S$\times$4, T$\times$2 & (2) S$\times$2, T$\times$2& (3) S$\times$4, T$\times$4&(4) S$\times$4, T$\times$8
\\
\end{tabular}
\caption{Visualization and error comparison for extended resolution (HeatedCylinder). Top to bottom: FFEINR, Trilinear. We train the model at a factor of $(S\times4, T\times2)$. For scale factors outside the training setting, FFEINR can still maintain the tiny vortices in the flow field visualization results, but this detail is partially lost in the Trilinear interpolation method. A similar situation can be seen from the results of visualizing the errors. The error of Trilinear is much larger than that of FFEINR.}
\label{fig:qual4extend2}
\end{figure*}

\textbf{Quantitative and qualitative analysis for extended super-resolution.}
We have mentioned earlier that traditional convolutional based super-resolution networks require a fixed scale factor to be determined before training, as this directly determines the number of upsampling layers in the network. 
The trained model only supports this upsampling factor during inference. 
In contrast, implicit neural representations have the natural advantage of supporting more flexible resolutions,  
as we can input any query point into the neural representations and output the corresponding flow field data. 
We can use this continuous representation to obtain super-resolution data when we set the query points on a grid with higher resolution.
However, since we need to downsample the high-resolution data to obtain low-resolution model inputs, 
we still need to determine a specific data downsampling factor, which is also the upsampling factor during network training. 
Can implicit representations with a fixed scale factor during training support super-resolution tasks of arbitrary resolution during inference as theoretically possible? 
To answer this question, we then conduct experiments within a certain range, and the quantitative results are shown in Table~\ref{tab:quan4extended} and \textcolor{black}{Table~\ref{tab:quan4extended2} (in Appendix~\ref{apndx:factors})}. 
\textcolor{black}{We first train the model at a fixed factor, e.g., $SR=(Spatial\times4, Temporal\times2)$, and use it in the inference stage for extended super-resolution tasks, e.g., $ESR=(S\times2, T\times2), (S\times4, T\times4)$, and $(S\times4, T\times8)$. }
Our method, FFEINR, achieves higher PSNR, SSIM and RMSE than the baseline method in most experiments. 
From a qualitative perspective, we present the results of super-resolution for different datasets in Fig.~\ref{fig:qual4extend1} and Fig.~\ref{fig:qual4extend2}. 
For different scaling factors, FFEINR is able to recover features of high-resolution data to the maximum extent, 
and the visualization error of the data is very small. 
In contrast, the trilinear interpolation method shows very blurred edges in the visualization results, and significant differences in visualization errors are observed. 
\textcolor{black}{In order to further illustrate the impact of the fixed scale factors on network performance, we conduct supplementary experiments on the Cylinder dataset (Table~\ref{tab:quan4extended2} in Appendix~\ref{apndx:factors}). Different fixed scale factors are set during training, and the results of the extended resolution are tested during the inference phase. 
A more detailed discussion can be found in Appendix~\ref{apndx:factors}.} 
Therefore, we can conclude that although we set a fixed factor during training, the model supports multiple \textcolor{black}{scale} factors during the inference phase and can flexibly generate flow field data with extended resolution.


\settowidth\rotheadsize{\theadfont FFEINR}

\textcolor{black}{
\textbf{Quantitative and qualitative comparison with state-of-the-arts.} 
To illustrate the competitive performances of our method in terms of model size and performance metrics, we compare FFEINR with the state of the art method. Firstly, we compared it with the state of the art spatio-temporal super-resolution method, TMNet~\cite{xu_temporal_2021}. 
TMNet needs to pre-train the $(S\times4, T\times2)$ model before the $(S\times4, T\times8)$ model training. 
As shown in Table~\ref{tab:quan4STSR}, from a quantitative perspective, FFEINR has smaller storage overhead and faster inference speed, although the training time is slightly longer compared to TMNet. As for the training results, FFEINR performs better than TMNet with the set of scale factor used for training. In addition, FFEINR can achieve higher PSNR in out-of-distribution tasks $(S\times4, T\times4)$.
Next, we compare it with the lossy compression algorithm. As shown in Fig.~\ref{fig:qual4SZ2}, when the PSNR is the same, although the compression rate of FFEINR (23.71:1) is lower than SZ (201.26:1)~\cite{liang_error-controlled_2018,zhao_significantly_2020,zhao_optimizing_2021}, the visualization effect is better and the features of  streamlines can be retained to a greater extent. When applying neural network based super-resolution techniques to data reduction, the compression rate may be reduced due to the storage space required by the model itself. This problem can be alleviated by model compression and architecture optimization.}

\begin{table}[!t]
\setlength{\abovecaptionskip}{0.5cm}
\setlength{\belowcaptionskip}{-1.cm}
\caption{\textcolor{black}{Quantitative comparison with spatio-temporal super-resolution model (PipeCylinder).}}
\label{tab:quan4STSR}
\setlength\tabcolsep{3pt} 
\begin{tabular}{c|c|cc|cc}
\hline
\textbf{Method} & \textbf{\begin{tabular}[c]{@{}c@{}}Model Size\\(MB)\end{tabular}} & \textbf{\begin{tabular}[c]{@{}c@{}}Training\\Time (h)\end{tabular}} & \textbf{\begin{tabular}[c]{@{}c@{}}Inference\\Time(s)\end{tabular}} & \multicolumn{1}{c}{\begin{tabular}[c]{@{}c@{}}\textbf{SR}\\ Sx4, Tx8\end{tabular}} & 
\multicolumn{1}{c}{\begin{tabular}[c]{@{}c@{}}\textbf{ESR}\\ Sx4, Tx4\end{tabular}} \\ \hline
FFEINR & \textbf{43.2} & 1.92 & \textbf{0.2793} & \textbf{50.93} & 
\textbf{51.10} \\ \hline
TMNet & 46.9 & 1.18 & 0.3385 & 48.18 & 
48.99 \\ \hline
Trilinear & -- & -- & \textcolor{red}{0.0017} & 35.89 & 
35.71 \\ \hline
\end{tabular}%
\end{table}


\begin{figure}[!h]
\setlength{\abovecaptionskip}{0.cm}
\setlength{\belowcaptionskip}{-0.8cm}
\begin{tabular}{c}

\includegraphics[width=0.32\columnwidth, angle=-90]{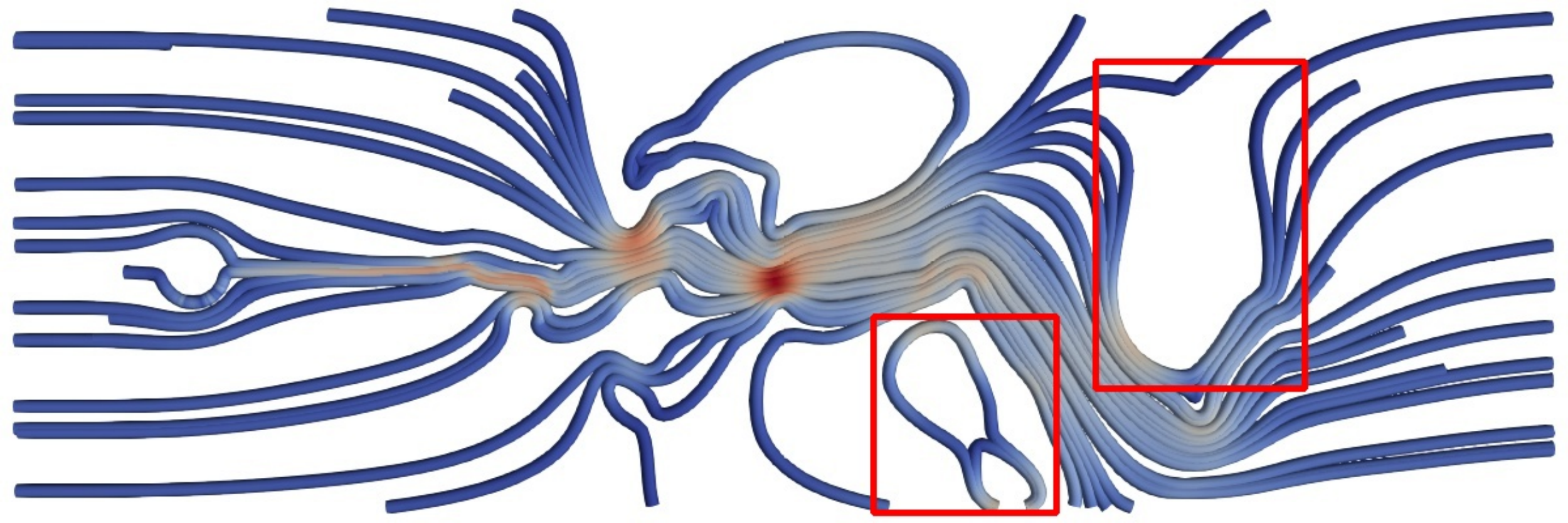} \\   
(1) Ground Truth \\
\includegraphics[width=0.32\columnwidth, angle=-90]{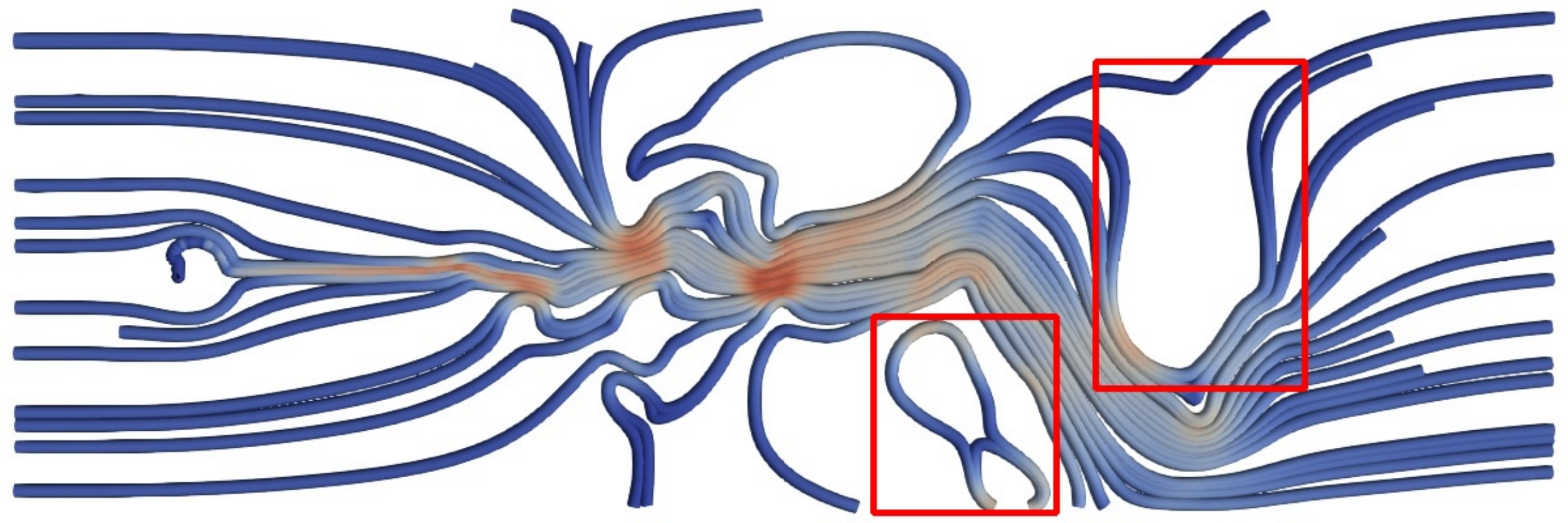}  \\    
(2) FFEINR \\
\includegraphics[width=0.32\columnwidth, angle=-90]{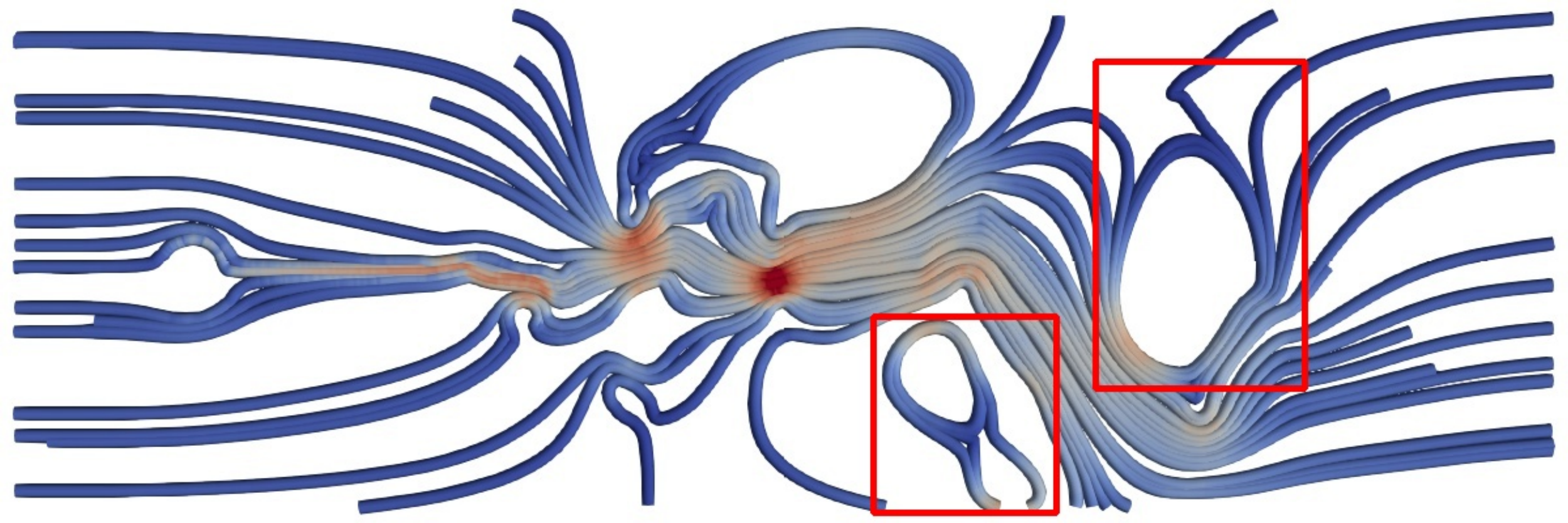} \\
(3) SZ~\cite{liang_error-controlled_2018,zhao_significantly_2020,zhao_optimizing_2021}
\end{tabular}
\caption{\textcolor{black}{Qualitative comparison with lossy compression (HeatedCylinder). The PSNR is 43.9 for both the predicted and decompressed data. The compression rates of FFEINR and SZ are 23.71 and 201.26, respectively.}}
\label{fig:qual4SZ2}
\end{figure}

\textbf{One-stage-training VS two-stage-training.}
The training process of our model involves two options: two-stage training and one-stage training. 
The two-stage mode involves training the super-resolution model with a fixed scale factor, followed by fine-tuning with uniformly distributed upsampling factors. 
The original intention is to expand the receptive field of the implicit representation network, so that it can learn features at different scales, and potentially improve the performance of the model in arbitrary resolution tasks in theory. 
However, we choose one-stage-training and only train the model with a fixed scale factor. 
As mentioned earlier, we have demonstrated that this mode can be extended to multiple super-resolution during the inference stage. 
In this part, we test and discuss the effectiveness of using two-stage training. 
Table~\ref{tab:quan4stage} shows the results of using FFEINR (1-stage) with one-stage training and \textcolor{black}{VideoINR} (2-stage) with fixed upsampling factors for 10000 iterations and multiple upsampling factors for 10,000 iterations. 
From a statistical point of view, the one-stage training achieves the best results, while the two-stage training does not achieve the expected results in fact. 
The possible reason is that setting arbitrary scale factors distracts the attention of network training, 
and greedy optimization goals make the optimization process of the network more difficult, 
resulting in a negative impact on the training results. 
Nevertheless, INR-based methods outperform the baseline method in all three indicators.
\begin{table}[t]
\centering
\setlength{\abovecaptionskip}{0.5cm}
\caption{Quantitative comparison for one-stage training VS two-stage training. The results of \textcolor{black}{VideoINR} (2-stage) are worse than those of FFEINR (1-stage). 
Nevertheless, \textcolor{black}{the INR-based methods} outperform the baseline method in all three indicators.}
\label{tab:quan4stage}
\renewcommand\arraystretch{1.0}
\setlength\tabcolsep{3pt} 
\begin{tabular}{cccc}
\cline{1-4}
\multicolumn{1}{c|}{\textbf{Method}} &
  \multicolumn{1}{c}{\textbf{PSNR}} &
  \multicolumn{1}{c}{\textbf{SSIM}} &
  \multicolumn{1}{c}{\textbf{\begin{tabular}[c]{@{}c@{}}RMSE\\ $u_x$/$u_y$\end{tabular}}}
\\ \cline{1-4}
\multicolumn{1}{c|}{\textbf{FFEINR (1-stage)}} &
  \multicolumn{1}{c}{\textbf{ \textcolor{black}{46.68}}} &
  \multicolumn{1}{c}{\textbf{ \textcolor{black}{0.994}}} &
  \multicolumn{1}{c}{\textbf{ \textcolor{black}{0.050/0.069}}}\\ \hline
\multicolumn{1}{c|}{\textcolor{black}{VideoINR} (2-stage)} &
  \multicolumn{1}{c}{\textcolor{black}{ 42.30}} &
  \multicolumn{1}{c}{\textcolor{black}{ 0.991}} &
  \multicolumn{1}{c}{\textcolor{black}{ 0.101/0.099}}\\ \hline
\multicolumn{1}{c|}{Trilinear} &
  \multicolumn{1}{c}{\textcolor{black}{ 35.59}} &
  \multicolumn{1}{c}{\textcolor{black}{ 0.986}} &
  \multicolumn{1}{c}{\textcolor{black}{ 0.194/0.222}}\\ \cline{1-4}
\end{tabular}%
\end{table}

\subsection{Discussion and Limitations}\label{ssec:disandlimit}
\subsubsection{Advantages}\label{sssec:advantages}
The FFEINR model studied in this paper has the following advantages. 
\textcolor{black}{Firstly, our method can achieve extended resolution, with a fixed scale factor during training and multiple scale factors during inference. 
We fully leverage the powerful representation capabilities of implicit representations for continuous spatio-temporal domains. 
By inputting continuous coordinates at any resolution, the model can output flow field data at the corresponding resolution, 
solving the problem of traditional convolutional based super-resolution networks that only supporting one upsampling factor during inference. 
Secondly, we achieve the super-resolution task for flow field data using a simpler and lighter model. 
With low-resolution data as input, high-resolution data can be obtained through the decoding process of SpatialINR, TemporalINR, and Decoder. 
All three decoders are based on SIREN~\cite{sitzmann_implicit_2020} and we do not construct residual blocks based on SIREN and use a deeper network architecture as suggested by~\cite{han_coordnet_2022}. 
Compared to the state of the art spatio-temporal super-resolution method, our model outperforms the compared methods in model size, inference speed, and results metrics.}
Finally, our method can be used for data compression. Thanks to upsampling at multiple spatio-temporal resolution, \textcolor{black}{we can compress the data} and recover it flexibly to high-resolution flow field data as needed.

\subsubsection{Limitations}\label{sssec:limitations}
While FFEINR achieves excellent visualization and competitive metrics, it still has limitations.
Firstly, we currently need to perform separate training on each flow field data. 
Due to the limited amount of open-source flow field data available, we are unable to construct a large Benchmark dataset similar to the computer vision domain for training. 
Thus, further tests and improvements are necessary to enhance the generalization ability of the model.
Secondly, the performance of the model on complex flow field data needs to be improved.
Of the three datasets, HeatedCylinder has the most complex flow field motion patterns. In our experiment, the advantage of the model on the HeatedCylinder dataset is not as pronounced as the other two datasets.
This indicates that there is still room for improvement in the modeling capabilities of FFEINR for more complex flow field features. 
Finally, as FFEINR is a data-driven model, it may not be sufficient for flow field data with physical prior knowledge. 
To enhance the physical interpretability of the super-resolution results, it is necessary to include physics-informed modules in the design of the loss function or the input phase.

%% file: body/6-Conclusions.tex
\section{Conclusions}\label{sec:conclusions}

In this paper, we study a flow feature-enhanced implicit neural representation, which is an adaptive optimization of the implicit representation for spatio-temporal super-resolution problems, while solving the fixed scale factors problem in traditional convolutional super-resolution networks. 
By representing the flow field as a continuous implicit function, 
it can support inputting arbitrary spatio-temporal coordinates to obtain the corresponding physical field data. 
With feature enhancement at the input layer, assistance from advanced network modules, 
and optimization of the network training process, 
we achieve significantly better results than the trilinear interpolation method. 
From a quantitative point of view, our results are advantageous in terms of PSNR, SSIM, and RMSE metrics. 
The test results on the Cylinder dataset with von Karman vortex street achieve 45dB+ PSNR, with some results having RMSE even one order of magnitude lower than the trilinear interpolation. 
From a qualitative point of view, FFEINR preserves the sharp features of the flow field in regions such as the edges and the interior of a vortex during super-resolution upsampling, 
and the visualization error is significantly smaller than the trilinear method. 
In addition, we investigate the scalability of the network to multiple super-resolution. 
The model trained in limited $(S\times4, T\times2)$ mode still achieves better results than the trilinear baseline method under the settings of $S\times2$, $T\times4$, and $T\times8$. 
This demonstrates the powerful potential of such infinite-resolution implicit representations for solving spatio-temporal super-resolution problems. 

In the future, we will add physical governing equations (such as Navier-Stokes Eqution, etc.) to the design of loss function and apply soft or hard constraints to the physical characteristics of the model in the optimization process, 
which allows the final implicit representation to generate data that is more amenable to physical theorems and enhances the physical interpretability of the model.
Additionally, we consider using the physics-informed implicit representations for out-of-frame interpolation or prediction problems to facilitate the study of PINN.

%% file: body/7-Acknowledgments.tex
\acknowledgments{%

This work was partially supported by the National Key R\&D Program of China under Grand No. 2021YFE0108400, 
partly supported by National Natural Science Foundation of China under Grant No. 62172294.

}

%% file: body/8-Appendices.tex
\clearpage    

\appendix 

\section{Supplementary experiments}\label{appendix}

\subsection{Performance on different timesteps}\label{apndx:timestep}
\begin{figure}[!htb]
\centering
\setlength{\abovecaptionskip}{0.5cm}
\setlength{\belowcaptionskip}{-0.5cm}
\begin{tabular}{ccc}
\centering
\includegraphics[width=0.28\columnwidth]{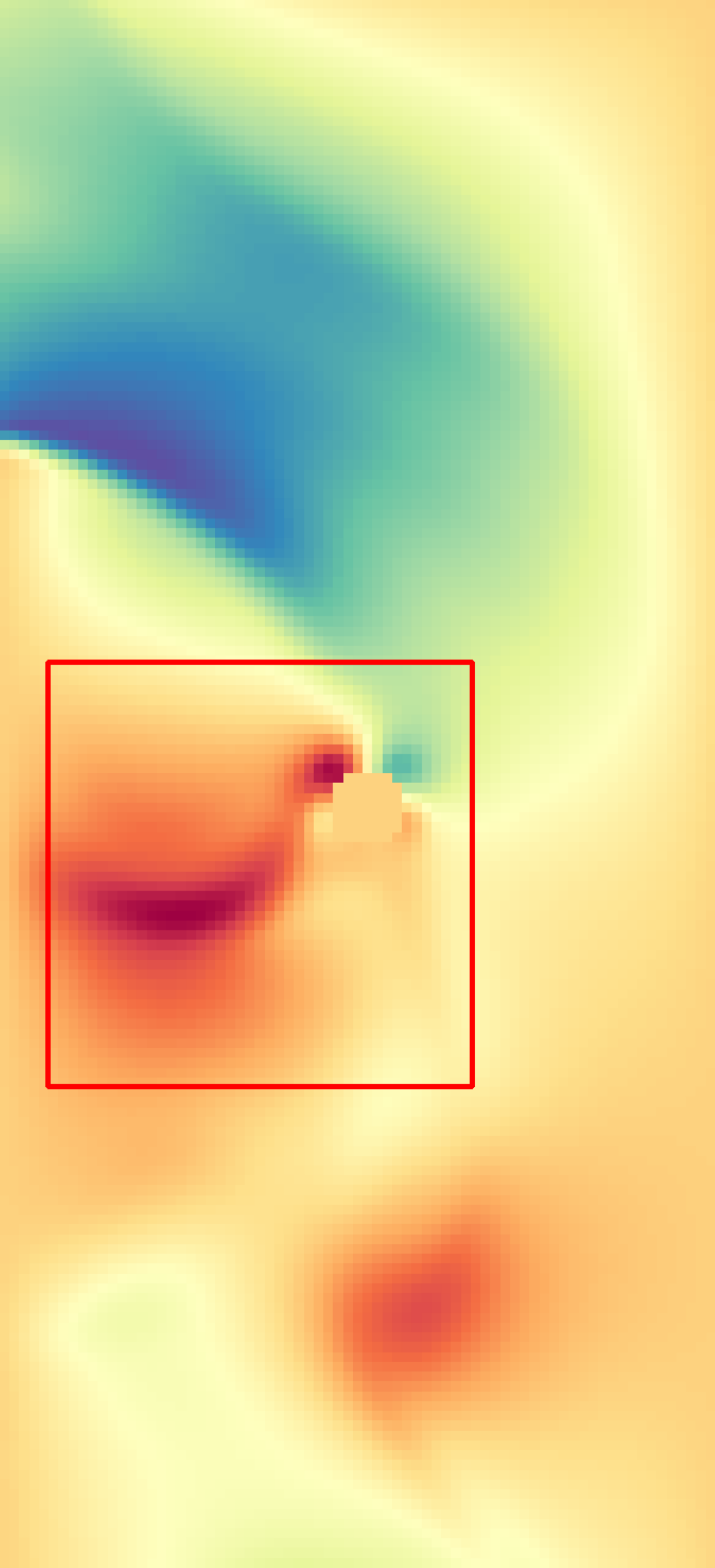} &     
\includegraphics[width=0.28\columnwidth]{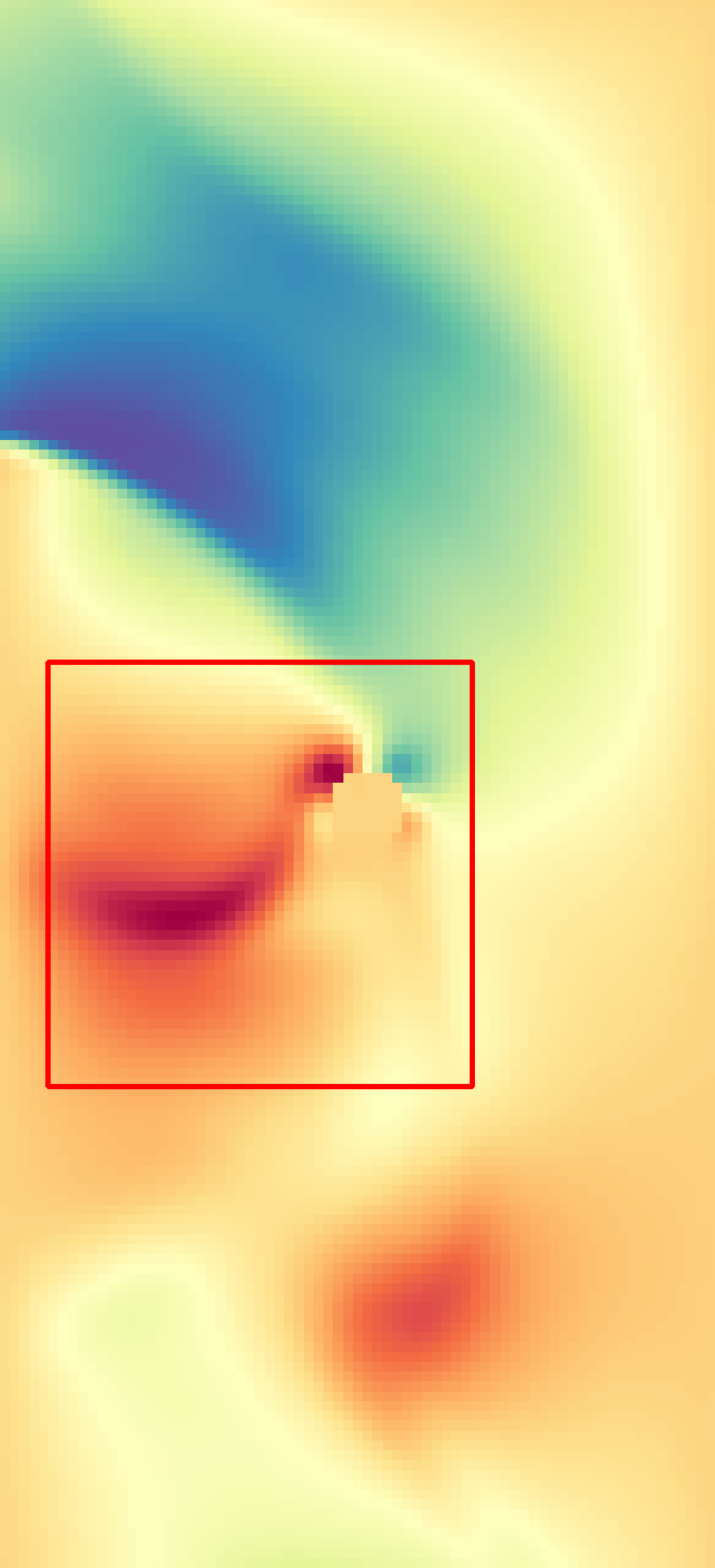} &     
\includegraphics[width=0.28\columnwidth]{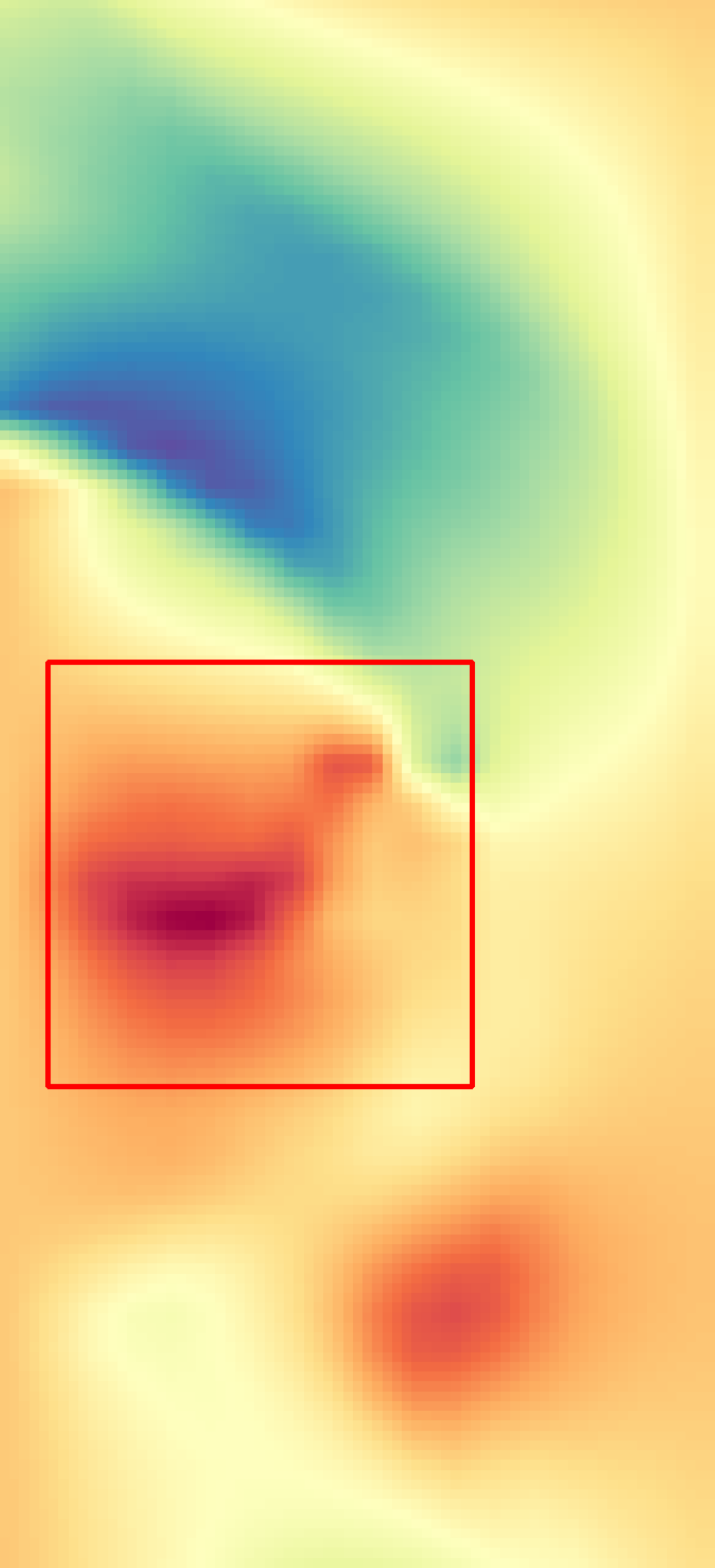} 

\\
\includegraphics[width=0.28\columnwidth]{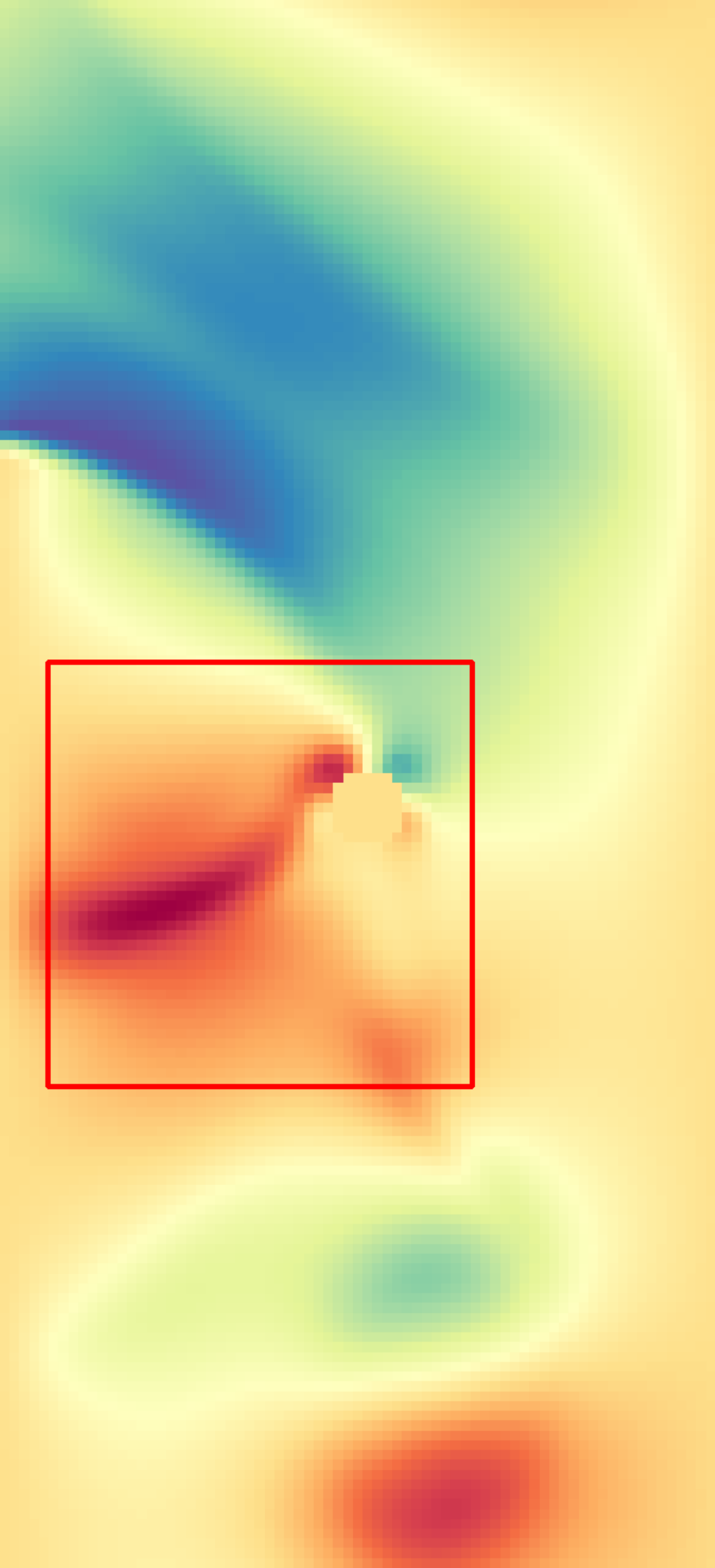} &     
\includegraphics[width=0.28\columnwidth]{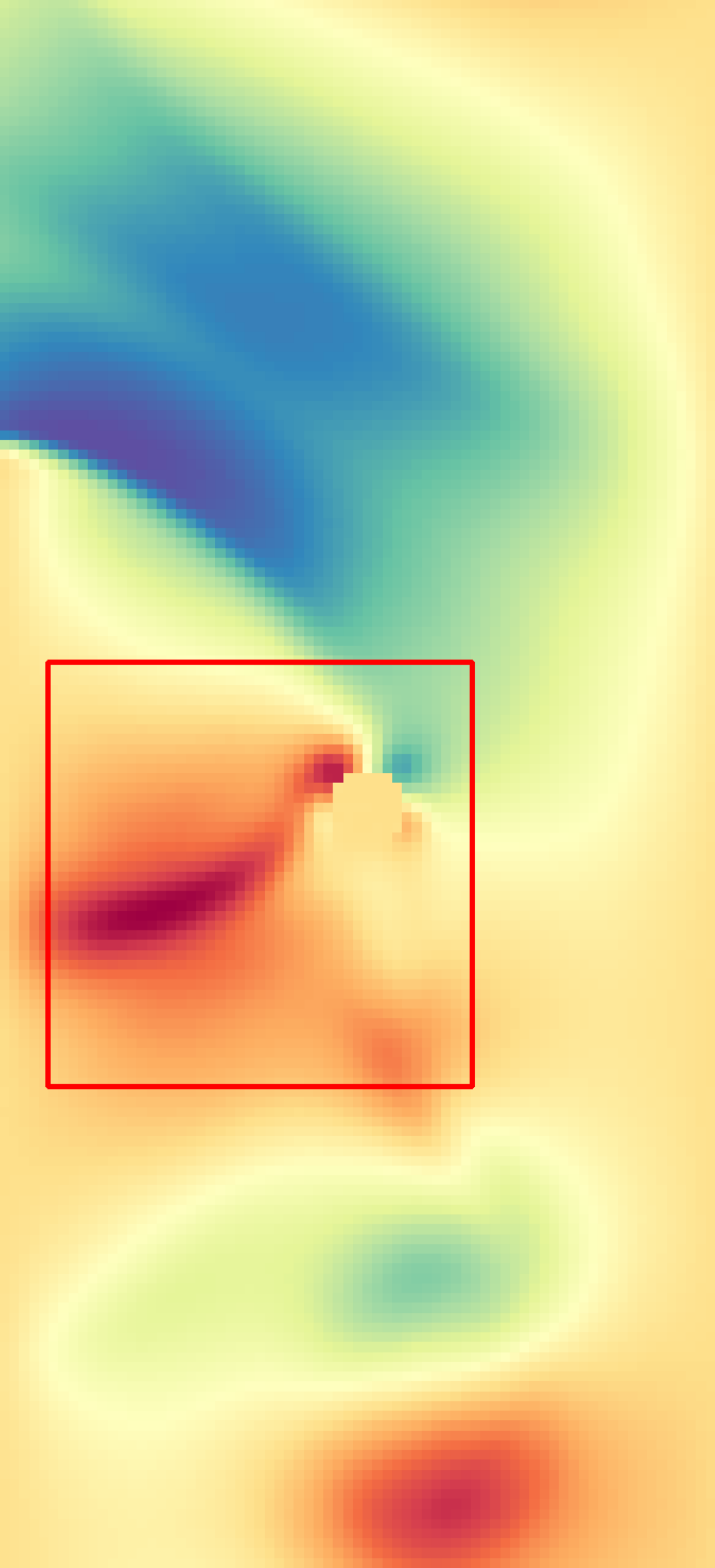} &     
\includegraphics[width=0.28\columnwidth]{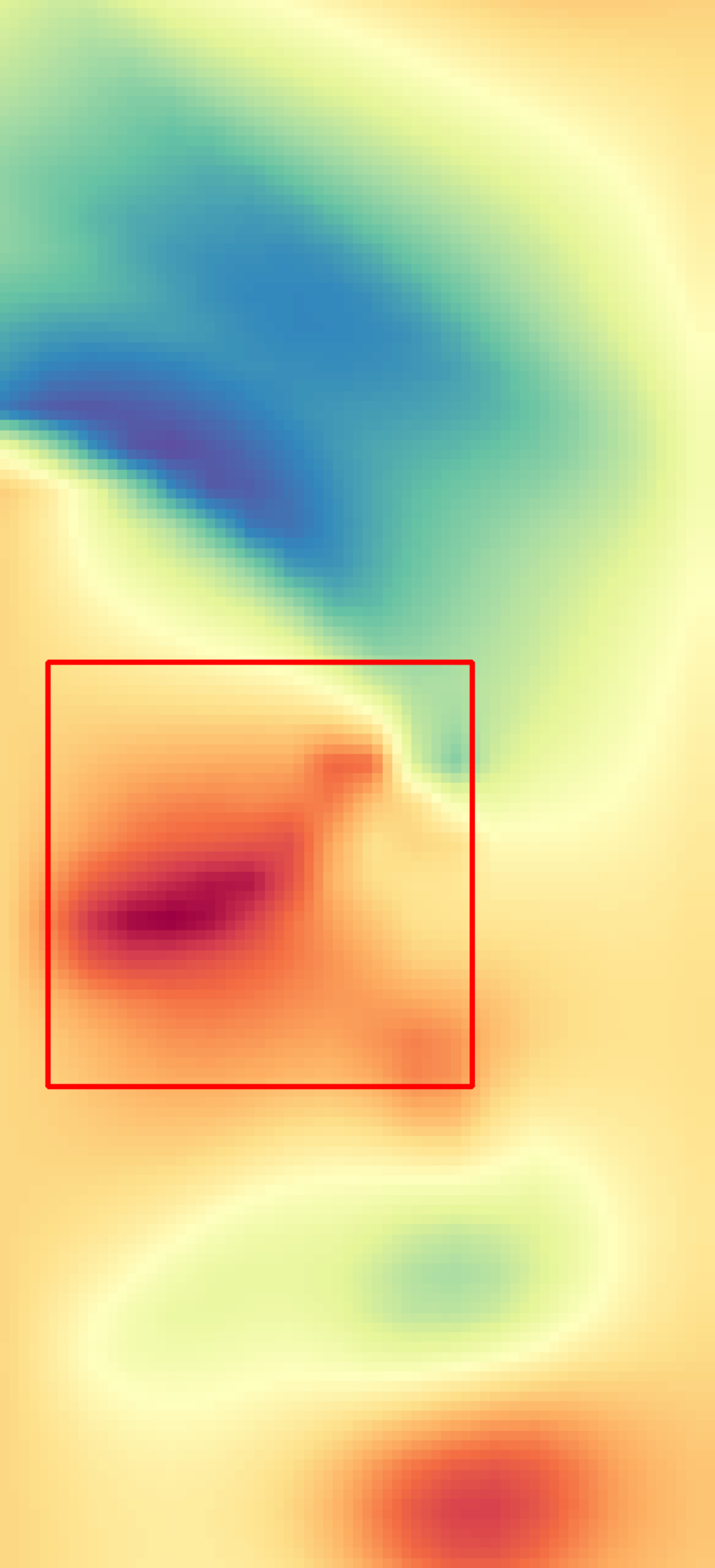} 
\\
\includegraphics[width=0.28\columnwidth]{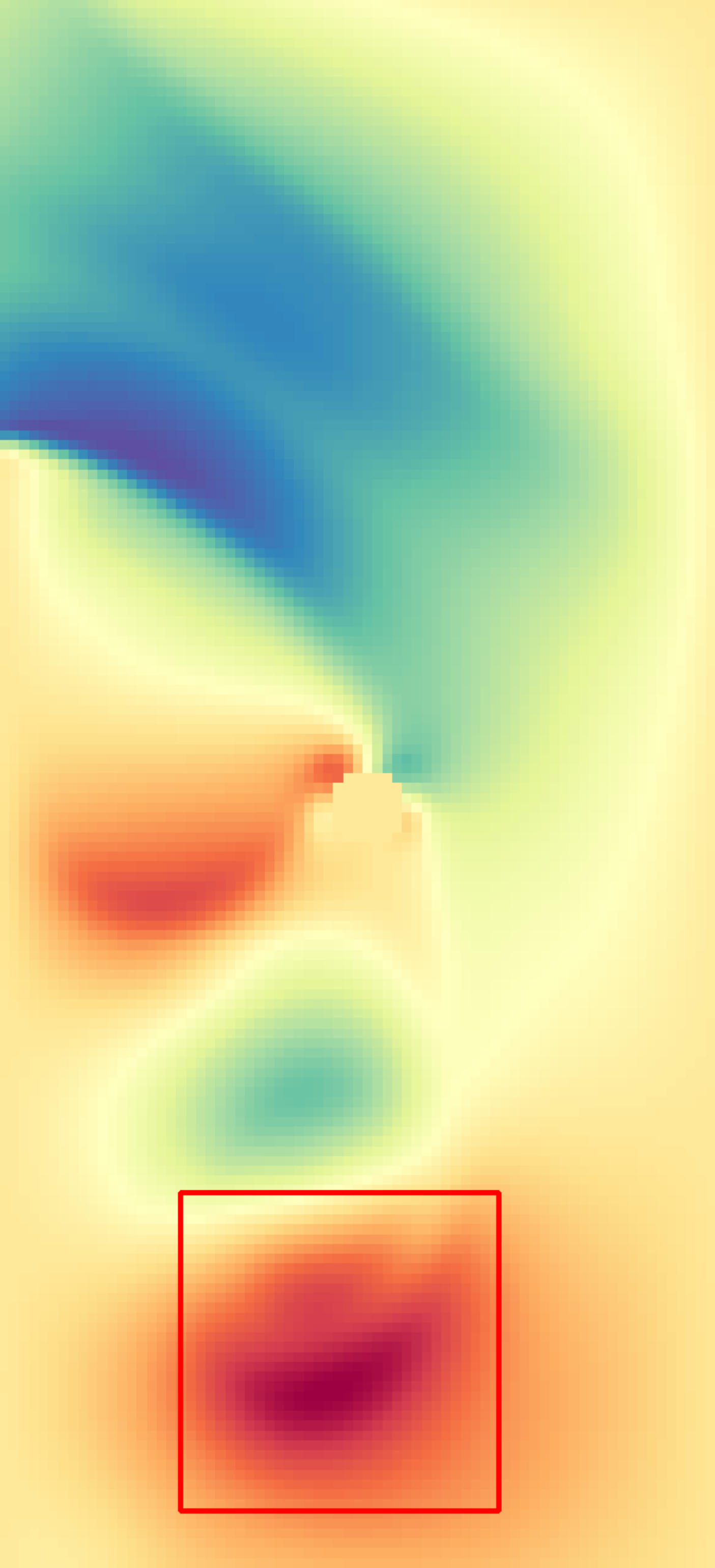} &     
\includegraphics[width=0.28\columnwidth]{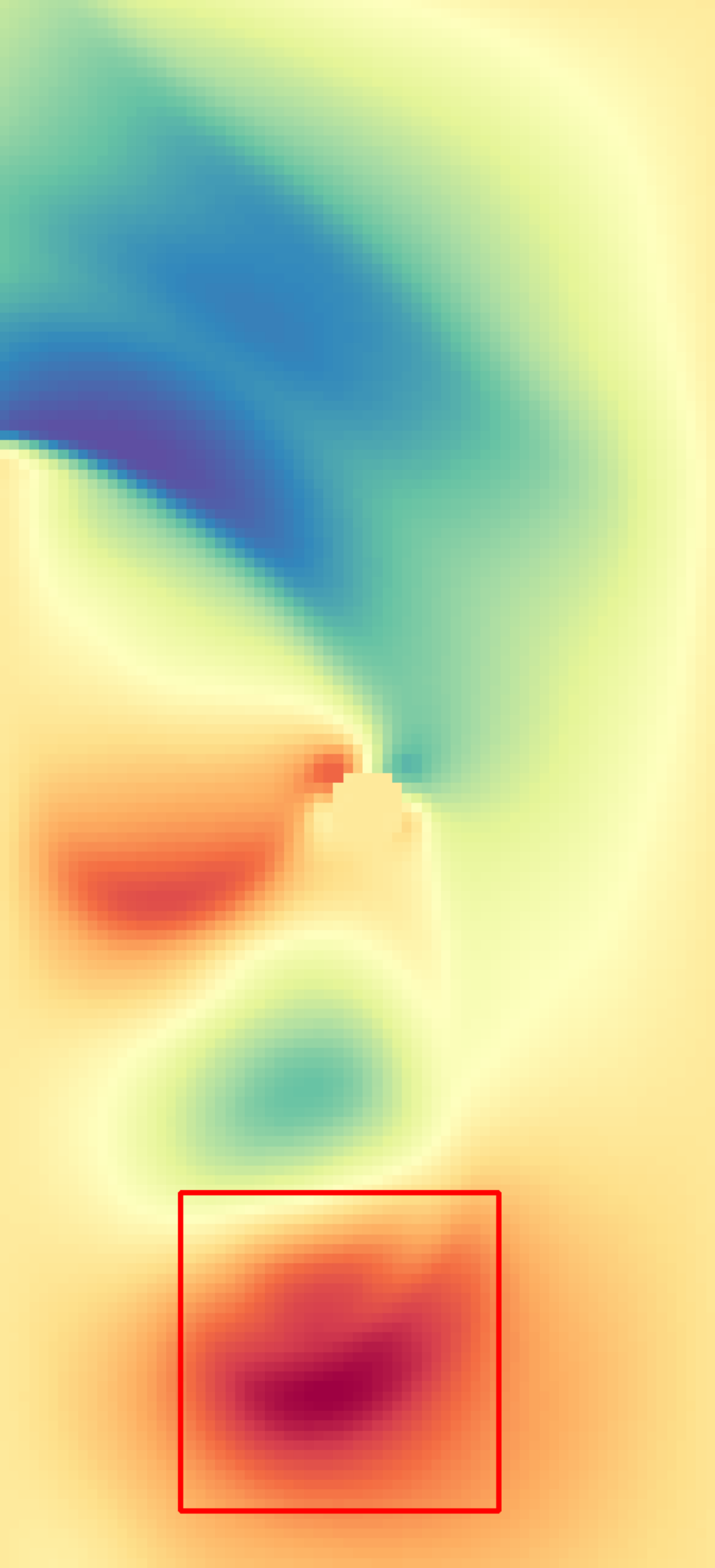} &     
\includegraphics[width=0.28\columnwidth]{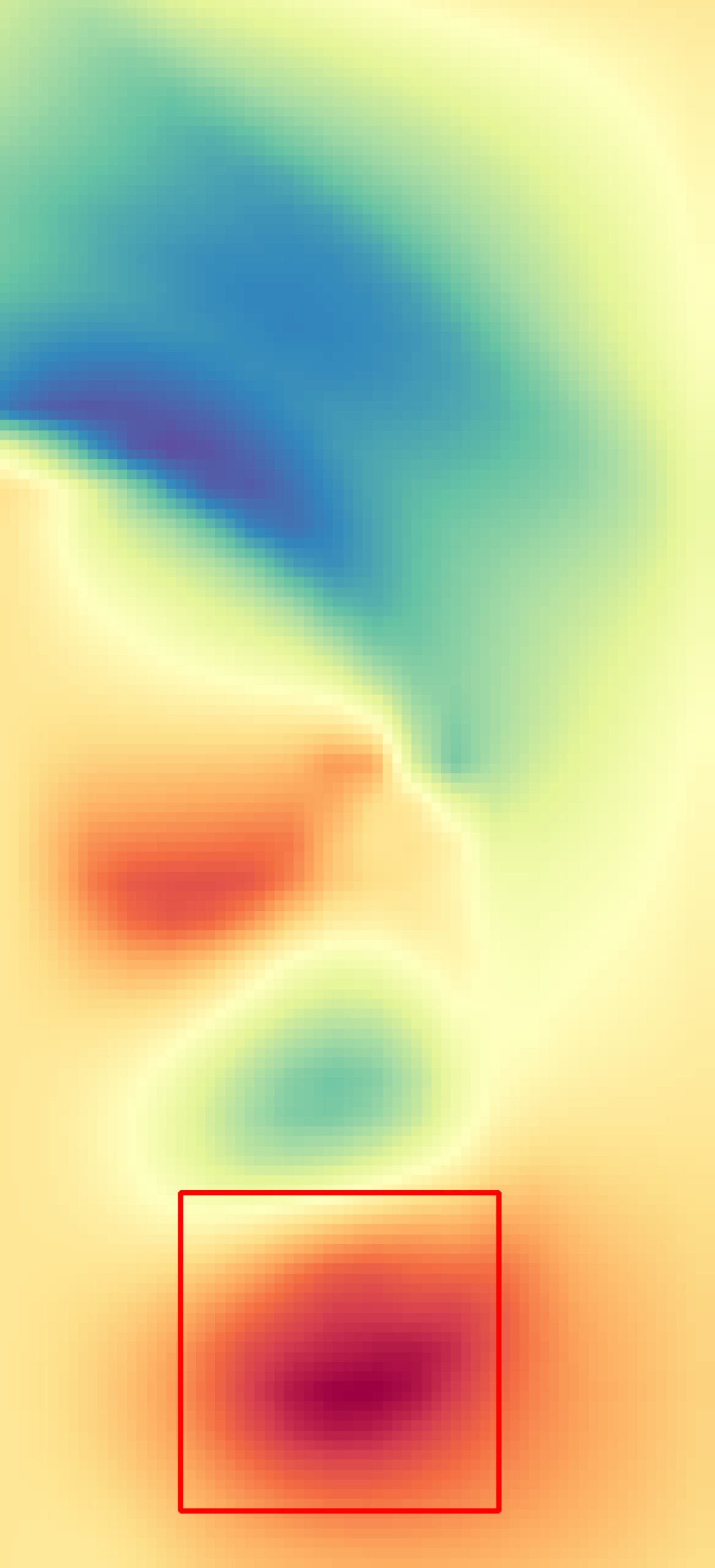} 
\\
(1) Ground Truth & (2) FFEINR & (3) Trilinear
\end{tabular}
\caption{Qualitative comparison at different timesteps. Top to bottom: $t=1070, 1366, 1430$. FFEINR outperforms Trilinear in predicting data at different timesteps. \textcolor{black}{Specifically, FFEINR achieves more accurate results than interpolation methods in the region of the flow field around obstacles or in the interior of vortices across timesteps.}}
\label{fig:qual4timestep}
\end{figure}

\textcolor{black}{
As shown in Fig.~\ref{fig:qual4timestep}, we compare the visualization results of PipeCylinder dataset at different timesteps.
FFEINR achieves better results than the baseline method in the region of the flow field around obstacles or in the interior of vortices across different timesteps.}

\subsection{Performance with different fixed scale factors}\label{apndx:factors}
\textcolor{black}{In order to illustrate the impact of fixed scale factors on network performance, we conduct supplementary experiments on the Cylinder dataset. As shown in Table~\ref{tab:quan4extended2}, the scale factor can affect training time and model performance. 
The larger the scale factor in the temporal dimension during training, the longer the training time. 
In the experiments with the factor of $(S\times4, T\times4)$ and $(S\times4, T\times8)$, the model performance decreases significantly compared to the factor $(S\times4, T\times2)$, but the effects of extended resolution inference is roughly the same, 
which means that the out-of-distribution performance of the model is relatively stable. 
However, in our experiment, if the scale factor in the spatial dimension during training is too small $(S\times2)$ or too large $(S\times8)$, the model performs poorly in extended resolution tasks.
The reason may be that the factor of $S\times4$ achieves a balance between the degree of low-resolution data loss and the interpolation performance of the model.}

\begin{table}[!h]
\centering
\caption{\textcolor{black}{Quantitative comparison of the extended resolution with different training scale factors. Except for $(S\times2, T\times2)$, FFEINR outperforms Trilinear in most indicators. Considering the trade-off between training time and effectiveness, setting the scaling factor of $(S\times4, T\times2)$ during training can achieve the optimal results.}}
\label{tab:quan4extended2}
\setlength\tabcolsep{3pt} 
\setlength{\abovecaptionskip}{0.5cm}
\setlength{\belowcaptionskip}{-0.8cm}
\begin{tabular}{c|c|c|ccc}
\hline
\textbf{Method} & \textbf{\begin{tabular}[c]{@{}c@{}}Training SR \& \\ Training time\end{tabular}} & \textbf{\begin{tabular}[c]{@{}c@{}}Inference\\ESR\end{tabular}} & \textbf{PSNR} & \textbf{SSIM} & \textbf{\begin{tabular}[c]{@{}c@{}}RMSE\\ $u_x$/$u_y$\end{tabular}} \\ \hline
FIFENR & \multicolumn{1}{c|}{\multirow{8}{*}{\begin{tabular}[c]{@{}c@{}}Sx4, Tx2\\ 3960s\end{tabular}}} & \multicolumn{1}{c|}{\multirow{2}{*}{Sx4, Tx2}} & \textbf{46.68 } & \textbf{0.994 } & \textbf{0.050/0.069} \\
\cline{1-1}    Trilinear &       &       & 35.59  & 0.986  & 0.194/0.222 \\
\cline{1-1}\cline{3-6}    FIFENR &       & \multicolumn{1}{c|}{\multirow{2}{*}{Sx2, Tx2}} & \textbf{40.04 } & \textbf{0.993 } & 0.152/\textbf{0.104} \\
\cline{1-1}    Trilinear &       &       & 38.82  & 0.992  & 0.104/0.178 \\
\cline{1-1}\cline{3-6}    FIFENR &       & \multicolumn{1}{c|}{\multirow{2}{*}{Sx4, Tx4}} & \textbf{38.50 } & \textbf{0.984 } & \textbf{0.122/0.234} \\
\cline{1-1}    Trilinear &       &       & 34.08  & 0.979  & 0.209/0.286 \\
\cline{1-1}\cline{3-6}    FIFENR &       & \multicolumn{1}{c|}{\multirow{2}{*}{Sx4, Tx8}} & \textbf{36.79 } & 0.978  & \textbf{0.143}/0.282 \\
\cline{1-1}    Trilinear &       &       & 34.92  & 0.982  & 0.200/0.253 \\
    \hline
    FIFENR & \multicolumn{1}{c|}{\multirow{8}{*}{\begin{tabular}[c]{@{}c@{}}Sx2, Tx2\\ 5880s\end{tabular}}} & \multicolumn{1}{c|}{\multirow{2}{*}{Sx2, Tx2}} & \textbf{43.11 } & \textbf{0.993 } & \textbf{0.081/0.097} \\
\cline{1-1}    Trilinear &       &       & 38.82  & 0.992  & 0.104/0.178 \\
\cline{1-1}\cline{3-6}    FIFENR &       & \multicolumn{1}{c|}{\multirow{2}{*}{Sx4, Tx2}} & 35.51  & 0.985  & 0.255/\textbf{0.173} \\
\cline{1-1}    Trilinear &       &       & 35.59  & 0.986  & 0.194/0.222 \\
\cline{1-1}\cline{3-6}    FIFENR &       & \multicolumn{1}{c|}{\multirow{2}{*}{Sx2, Tx4}} & \textbf{37.12 } & 0.983  & \textbf{0.137/0.246} \\
\cline{1-1}    Trilinear &       &       & 35.86  & 0.985  & 0.138/0.266 \\
\cline{1-1}\cline{3-6}    FIFENR &       & \multicolumn{1}{c|}{\multirow{2}{*}{Sx2, Tx8}} & 35.86  & 0.979  & 0.155/0.290 \\
\cline{1-1}    Trilinear &       &       & 37.10  & 0.988  & 0.123/0.229 \\
    \hline
    FIFENR & \multicolumn{1}{c|}{\multirow{8}{*}{\begin{tabular}[c]{@{}c@{}}Sx4, Tx4\\ 9360s\end{tabular}}} & \multicolumn{1}{c|}{\multirow{2}{*}{Sx4, Tx4}} & \textbf{41.41 } & \textbf{0.994 } & \textbf{0.128/0.085} \\
\cline{1-1}    Trilinear &       &       & 34.08  & 0.979  & 0.209/0.286 \\
\cline{1-1}\cline{3-6}    FIFENR &       & \multicolumn{1}{c|}{\multirow{2}{*}{Sx2, Tx4}} & \textbf{36.91 } & \textbf{0.992 } & 0.219/\textbf{0.129} \\
\cline{1-1}    Trilinear &       &       & 35.43  & 0.985  & 0.138/0.266 \\
\cline{1-1}\cline{3-6}    FIFENR &       & \multicolumn{1}{c|}{\multirow{2}{*}{Sx4, Tx2}} & \textbf{41.50 } & \textbf{0.994 } & \textbf{0.127/0.083} \\
\cline{1-1}    Trilinear &       &       & 35.59  & 0.986  & 0.194/0.222 \\
\cline{1-1}\cline{3-6}    FIFENR &       & \multicolumn{1}{c|}{\multirow{2}{*}{Sx4, Tx8}} & \textbf{41.23 } & \textbf{0.994 } & \textbf{0.129/0.090} \\
\cline{1-1}    Trilinear &       &       & 34.92  & 0.982  & 0.200/0.253 \\
    \hline
    FIFENR & \multicolumn{1}{c|}{\multirow{8}{*}{\begin{tabular}[c]{@{}c@{}}Sx4, Tx8\\ 13980s\end{tabular}}} & \multicolumn{1}{c|}{\multirow{2}{*}{Sx4, Tx8}} & \textbf{39.24 } & \textbf{0.993 } & \textbf{0.167/0.104} \\
\cline{1-1}    Trilinear &       &       & 34.92  & 0.982  & 0.200/0.253 \\
\cline{1-1}\cline{3-6}    FIFENR &       & \multicolumn{1}{c|}{\multirow{2}{*}{Sx2, Tx8}} & 35.54  & \textbf{0.993 } & 0.259/\textbf{0.145} \\
\cline{1-1}    Trilinear &       &       & 37.10  & 0.988  & 0.123/0.229 \\
\cline{1-1}\cline{3-6}    FIFENR &       & \multicolumn{1}{c|}{\multirow{2}{*}{Sx4, Tx2}} & \textbf{39.28 } & \textbf{0.993 } & \textbf{0.167/0.102} \\
\cline{1-1}    Trilinear &       &       & 35.59  & 0.986  & 0.194/0.222 \\
\cline{1-1}\cline{3-6}    FIFENR &       & \multicolumn{1}{c|}{\multirow{2}{*}{Sx4, Tx4}} & \textbf{39.25 } & \textbf{0.993 } & \textbf{0.167/0.104} \\
\cline{1-1}    Trilinear &       &       & 34.08  & 0.979  & 0.209/0.286 \\ \hline
\end{tabular}%
\end{table}

%% file: template.bbl
\begin{thebibliography}{10}

\bibitem{BaezaRojo19SciVisa}
I.~Baeza~Rojo and T.~G{\"u}nther.
\newblock Vector field topology of time-dependent flows in a steady reference
  frame.
\newblock {\em IEEE Transactions on Visualization and Computer Graphics}, 2019.

\bibitem{barron_mip-nerf_nodate}
J.~T. Barron, B.~Mildenhall, M.~Tancik, P.~Hedman, R.~Martin-Brualla, and P.~P.
  Srinivasan.
\newblock Mip-nerf: A multiscale representation for anti-aliasing neural
  radiance fields.
\newblock In {\em Proceedings of IEEE/CVF International Conference on Computer
  Vision}, pp. 5835--5844, 2021.

\bibitem{bashir_comprehensive_2021}
S.~M.~A. Bashir, Y.~Wang, M.~Khan, and Y.~Niu.
\newblock {A Comprehensive Review of Deep Learning-Based Single Image
  Super-Resolution}.
\newblock {\em PeerJ Computer Science}, 7:1--56, 2021.

\bibitem{cai_physics-informed_2021}
S.~Cai, Z.~Mao, Z.~Wang, M.~Yin, and G.~E. Karniadakis.
\newblock Physics-informed neural networks ({PINNs}) for fluid mechanics: a
  review.
\newblock {\em Acta Mechanica Sinica}, 37(12):1727--1738, 2021.

\bibitem{chen_mvsnerf_2021}
A.~Chen, Z.~Xu, F.~Zhao, X.~Zhang, F.~Xiang, J.~Yu, and H.~Su.
\newblock {MVSNeRF}: Fast generalizable radiance field reconstruction from
  multi-view stereo.
\newblock In {\em Proceedings of {IEEE}/{CVF} International Conference on
  Computer Vision}, pp. 14104--14113, 2021.

\bibitem{chen_hnerv_2023}
H.~Chen, M.~Gwilliam, S.-N. Lim, and A.~Shrivastava.
\newblock {HN}e{RV}: Neural representations for videos.
\newblock In {\em Proceedings of {IEEE}/{CVF} Conference on Computer Vision and
  Pattern Recognition}, 2023.

\bibitem{chen_nerv_nodate}
H.~Chen, B.~He, H.~Wang, Y.~Ren, S.-N. Lim, and A.~Shrivastava.
\newblock {NeRV: Neural Representations for Videos}.
\newblock In {\em Proceedings of Advances in Neural Information Processing
  Systems}, 2021.

\bibitem{chen_videoinr_2022}
Z.~Chen, Y.~Chen, J.~Liu, X.~Xu, V.~Goel, Z.~Wang, H.~Shi, and X.~Wang.
\newblock {VideoINR}: Learning video implicit neural representation for
  continuous space-time super-resolution.
\newblock In {\em Proceedings of {IEEE}/{CVF} Conference on Computer Vision and
  Pattern Recognition}, pp. 2037--2047, 2022.

\bibitem{chu_physics_2022}
M.~Chu, L.~Liu, Q.~Zheng, E.~Franz, H.-P. Seidel, C.~Theobalt, and R.~Zayer.
\newblock Physics informed neural fields for smoke reconstruction with sparse
  data.
\newblock {\em {ACM} Transactions on Graphics}, 41(4):1--14, 2022.

\bibitem{fukami_machine-learning-based_2021}
K.~Fukami, K.~Fukagata, and K.~Taira.
\newblock Machine-learning-based spatio-temporal super resolution
  reconstruction of turbulent flows.
\newblock {\em Journal of Fluid Mechanics}, 909:A9, 2021.

\bibitem{Guenther17}
T.~G{\"u}nther, M.~Gross, and H.~Theisel.
\newblock Generic objective vortices for flow visualization.
\newblock {\em ACM Transactions on Graphics}, 36(4):141:1--141:11, 2017.

\bibitem{guo_ssr-vfd_2020}
L.~Guo, S.~Ye, J.~Han, H.~Zheng, H.~Gao, D.~Z. Chen, J.~X. Wang, and C.~Wang.
\newblock {SSR-VFD: Spatial Super-Resolution for Vector Field Data Analysis and
  Visualization}.
\newblock In {\em Proceedings of the {IEEE} Pacific Visualization Symposium},
  pp. 71--80, 2020.

\bibitem{han_ssr-tvd_2020}
J.~Han and C.~Wang.
\newblock {SSR-TVD: Spatial Super-Resolution for Time-Varying Data Analysis and
  Visualization}.
\newblock {\em IEEE Transactions on Visualization and Computer Graphics},
  28(6):2445--2456, 2020.

\bibitem{han_tsr-tvd_2020}
J.~Han and C.~Wang.
\newblock {TSR-TVD: Temporal Super-Resolution for Time-Varying Data Analysis
  and Visualization}.
\newblock {\em IEEE Transactions on Visualization and Computer Graphics},
  26(1):205--215, 2020.

\bibitem{han_coordnet_2022}
J.~Han and C.~Wang.
\newblock {CoordNet}: Data generation and visualization generation for
  time-varying volumes via a coordinate-based neural network.
\newblock {\em {IEEE} Transactions on Visualization and Computer Graphics}, pp.
  1--12, 2022.

\bibitem{han_tsr-vfd_2022}
J.~Han and C.~Wang.
\newblock {TSR-VFD: Generating temporal super-resolution for unsteady vector
  field data}.
\newblock {\em Computers \& Graphics}, 103:168--179, 2022.

\bibitem{han_stnet_2022}
J.~Han, H.~Zheng, D.~Z. Chen, and C.~Wang.
\newblock {STNet}: An end-to-end generative framework for synthesizing
  spatiotemporal super-resolution volumes.
\newblock {\em {IEEE} Transactions on Visualization and Computer Graphics},
  28(1):270--280, 2022.

\bibitem{jiao_esrgan-based_2023}
C.~Jiao, C.~Bi, L.~Yang, Z.~Wang, Z.~Xia, and K.~Ono.
\newblock {ESRGAN}-based visualization for large-scale volume data.
\newblock {\em Journal of Visualization}, 26(3):649--665, 2023.

\bibitem{jin_nsfnets_2021}
X.~Jin, S.~Cai, H.~Li, and G.~E. Karniadakis.
\newblock {NSFnets} (navier-stokes flow nets): Physics-informed neural networks
  for the incompressible navier-stokes equations.
\newblock {\em Journal of Computational Physics}, 426:109951, 2021.

\bibitem{karniadakis_physics-informed_2021}
G.~E. Karniadakis, I.~G. Kevrekidis, L.~Lu, P.~Perdikaris, S.~Wang, and
  L.~Yang.
\newblock Physics-informed machine learning.
\newblock {\em Nature Reviews Physics}, 3(6):422--440, 2021.

\bibitem{li_data_2018}
S.~Li, N.~Marsaglia, C.~Garth, J.~Woodring, J.~P. Clyne, and H.~Childs.
\newblock {Data Reduction Techniques for Simulation, Visualization and Data
  Analysis}.
\newblock {\em Computer Graphics Forum}, 37(6):422--447, 2018.

\bibitem{avidan_e-nerv_2022}
Z.~Li, M.~Wang, H.~Pi, K.~Xu, J.~Mei, and Y.~Liu.
\newblock E-{NeRV}: Expedite neural video representation with disentangled
  spatial-temporal context.
\newblock In {\em Proceedings of European Conference on Computer Vision}, vol.
  13695, pp. 267--284, 2022.

\bibitem{liang_error-controlled_2018}
X.~Liang, S.~Di, D.~Tao, S.~Li, S.~Li, H.~Guo, Z.~Chen, and F.~Cappello.
\newblock Error-controlled lossy compression optimized for high compression
  ratios of scientific datasets.
\newblock In {\em Proceedings of {IEEE} International Conference on Big Data
  (Big Data)}, pp. 438--447, 2018.

\bibitem{liu_video_2022}
H.~Liu, Z.~Ruan, P.~Zhao, C.~Dong, F.~Shang, Y.~Liu, L.~Yang, and R.~Timofte.
\newblock Video super-resolution based on deep learning: a comprehensive
  survey.
\newblock {\em Artificial Intelligence Review}, 55(8):5981--6035, 2022.

\bibitem{lu_compressive_2021}
Y.~Lu, K.~Jiang, J.~A. Levine, and M.~Berger.
\newblock {Compressive Neural Representations of Volumetric Scalar Fields}.
\newblock {\em Computer Graphics Forum}, 40(3):135--146, 2021.

\bibitem{mildenhall_nerf_nodate}
B.~Mildenhall, P.~P. Srinivasan, M.~Tancik, J.~T. Barron, R.~Ramamoorthi, and
  R.~Ng.
\newblock {NeRF}: Representing scenes as neural radiance fields for view
  synthesis.
\newblock In {\em Proceedings of European Conference on Computer Vision}, pp.
  405--421, 2020.

\bibitem{nguyen-phuoc_snerf_2022}
T.~Nguyen-Phuoc, F.~Liu, and L.~Xiao.
\newblock Snerf: Stylized neural implicit representations for 3d scenes.
\newblock {\em {ACM} Transactions on Graphics}, 41(4), 2022.

\bibitem{pandey_perspective_2020}
S.~Pandey, J.~Schumacher, and K.~R. Sreenivasan.
\newblock A perspective on machine learning in turbulent flows.
\newblock {\em Journal of Turbulence}, 21(9):567--584, 2020.

\bibitem{gerrisflowsolver}
S.~Popinet.
\newblock Free computational fluid dynamics.
\newblock {\em ClusterWorld}, 2(6), 2004.

\bibitem{raissi_physics-informed_2019}
M.~Raissi, P.~Perdikaris, and G.~Karniadakis.
\newblock Physics-informed neural networks: A deep learning framework for
  solving forward and inverse problems involving nonlinear partial differential
  equations.
\newblock {\em Journal of Computational Physics}, 378:686--707, 2019.

\bibitem{raissi_hidden_2020}
M.~Raissi, A.~Yazdani, and G.~E. Karniadakis.
\newblock Hidden fluid mechanics: Learning velocity and pressure fields from
  flow visualizations.
\newblock {\em Science}, 367(6481):1026--1030, 2020.

\bibitem{goos_increasing_2002}
E.~Shechtman, Y.~Caspi, and M.~Irani.
\newblock {Increasing Space-Time Resolution in Video}.
\newblock In {\em Proceedings of European Conference on Computer Vision}, pp.
  753--768, 2002.

\bibitem{sitzmann_implicit_2020}
V.~Sitzmann, J.~N. Martel, A.~W. Bergman, D.~B. Lindell, and G.~Wetzstein.
\newblock Implicit neural representations with periodic activation functions.
\newblock In {\em Proceedings of Advances in Neural Information Processing
  Systems}, 2020.

\bibitem{takamoto_pdebench_2022}
M.~Takamoto, T.~Praditia, R.~Leiteritz, D.~MacKinlay, F.~Alesiani,
  D.~Pfl{\"u}ger, and M.~Niepert.
\newblock {PDEBENCH}: An extensive benchmark for scientific machine learning.
\newblock In {\em Proceedings of ICLR 2023 Workshop on Physics for Machine
  Learning}, 2023.

\bibitem{wang_dl4scivis_2022}
C.~Wang and J.~Han.
\newblock {DL4SciVis: A State-of-the-Art Survey on Deep Learning for Scientific
  Visualization}.
\newblock {\em IEEE Transactions on Visualization and Computer Graphics},
  Accept.

\bibitem{wang_when_2022}
S.~Wang, X.~Yu, and P.~Perdikaris.
\newblock When and why {PINNs} fail to train: A neural tangent kernel
  perspective.
\newblock {\em Journal of Computational Physics}, 449:110768, 2022.

\bibitem{wang_esrgan_2019}
X.~Wang, K.~Yu, S.~Wu, J.~Gu, Y.~Liu, C.~Dong, Y.~Qiao, and C.~C. Loy.
\newblock {ESRGAN: Enhanced Super-Resolution Generative Adversarial Networks}.
\newblock In {\em Proceedings of Computer Vision - {ECCV} 2018 Workshops}, pp.
  63--79, 2018.

\bibitem{xiang_zooming_2020}
X.~Xiang, Y.~Tian, Y.~Zhang, Y.~Fu, J.~P. Allebach, and C.~Xu.
\newblock Zooming slow-mo: Fast and accurate one-stage space-time video
  super-resolution.
\newblock In {\em Proceedings of {IEEE}/{CVF} Conference on Computer Vision and
  Pattern Recognition}, pp. 3367--3376, 2020.

\bibitem{xu_temporal_2021}
G.~Xu, J.~Xu, Z.~Li, L.~Wang, X.~Sun, and M.-M. Cheng.
\newblock Temporal modulation network for controllable space-time video
  super-resolution.
\newblock In {\em Proceedings of {IEEE}/{CVF} Conference on Computer Vision and
  Pattern Recognition ({CVPR})}, pp. 6384--6393, 2021.

\bibitem{yu_pixelnerf_2021}
A.~Yu, V.~Ye, M.~Tancik, and A.~Kanazawa.
\newblock {pixelNeRF}: Neural radiance fields from one or few images.
\newblock In {\em Proceedings of IEEE Conference on Computer Vision and Pattern
  Recognition}, 2021.

\bibitem{zhao_optimizing_2021}
K.~Zhao, S.~Di, M.~Dmitriev, T.-L.~D. Tonellot, Z.~Chen, and F.~Cappello.
\newblock Optimizing error-bounded lossy compression for scientific data by
  dynamic spline interpolation.
\newblock In {\em Proceedings of {IEEE} International Conference on Data
  Engineering}, pp. 1643--1654, 2021.

\bibitem{zhao_significantly_2020}
K.~Zhao, S.~Di, X.~Liang, S.~Li, D.~Tao, Z.~Chen, and F.~Cappello.
\newblock Significantly improving lossy compression for {HPC} datasets with
  second-order prediction and parameter optimization.
\newblock In {\em Proceedings of International Symposium on High-Performance
  Parallel and Distributed Computing}, pp. 89--100, 2020.

\end{thebibliography}
